%% file: EXOT-2016-23-PAPER.tex
\pdfoutput=1
\newcommand*{\ATLASLATEXPATH}{latex/}
\documentclass[UKenglish,cernpreprint,texlive=2016,LANGEDIT=false]{\ATLASLATEXPATH atlasdoc}
\LEcontact{Pat Ward cpw1@hep.phy.cam.ac.uk}
\usepackage{multirow}
\usepackage{xspace}
\usepackage{verbatim}
\usepackage{siunitx}

\usepackage[subfigure=true]{\ATLASLATEXPATH atlaspackage}
\usepackage{\ATLASLATEXPATH atlasbiblatex}
\usepackage{\ATLASLATEXPATH atlasphysics}

\usepackage{\ATLASLATEXPATH atlascontribute}

\graphicspath{{logos/}{figures/}}

\usepackage{EXOT-2016-23-PAPER-defs}

\newcommand\mjj{\ensuremath{m_{jj}}}

\def\met{\ensuremath{E_{\mathrm{T}}^{\mathrm{miss}}}\xspace}

\def\metnolep{\ensuremath{E_{\mathrm{T}}^{\mathrm{miss (no\ lepton)}}}}
\def\mptnolep{\ensuremath{p_{\mathrm{T}}^{\mathrm{miss  (no\ lepton)}}}}
\def\mpt{\ensuremath{p_{\mathrm{T}}^{\mathrm{miss}}}}

\def\metvec{\ensuremath{{\vec{E}_{\mathrm{T}}^{\mathrm{miss}}}}}

\def\mptvec{\ensuremath{{\vec{p}_{\mathrm{T}}^{\mathrm{miss}}}}}

\def\sigvis{\ensuremath{\sigma_{\mathrm{vis}}}\xspace}
\def\sigvisw{\ensuremath{\sigma_{\mathrm{vis,}\,\emph{W}\mathrm{+DM}}}\xspace}
\def\sigvisz{\ensuremath{\sigma_{\mathrm{vis,}\,\emph{Z}\mathrm{+DM}}}\xspace}
\def\pt{\ensuremath{p_{\mathrm{T}}}}

\def\ptv{\ensuremath{p_{\mathrm{T}}^V}}

\def\vec#1{{\mbox{$\boldsymbol{#1}$}}}
\def\ifb{\ensuremath{\mathrm{fb}^{-1}}}

\def\pt{\ensuremath{p_{\mathrm{T}}}}

\hypersetup{pdftitle={ATLAS draft},pdfauthor={The ATLAS Collaboration}}
\input{EXOT-2016-23-PAPER-metadata}

\begin{document}

\maketitle

\section{Introduction}
\label{sec:intro}
\input{introduction}

\section{ATLAS detector}
\label{sec:exp}
\input{experiment}

\section{Signal models}
\label{sec:models}
\input{models}

\section{Simulated signal and background samples}
\label{sec:mc}
\input{mc}

\section{Object reconstruction and identification}
\label{sec:objects}
\input{objects}

\section{Event selection and categorization}
\label{sec:selection}
\input{selection}

\section{Background estimation}
\label{sec:BG}
\input{background}

\section{Systematic uncertainties}
\label{sec:sys}
\input{systematics}

\section{Results}
\label{sec:results}
\input{results}
\FloatBarrier
\section{Summary}
\label{sec:summary}
\input{summary}

\section*{Acknowledgements}
\input{acknowledgements/Acknowledgements}

\newpage
\printbibliography

\clearpage
\input{atlas_authlist}

\end{document}

%% file: EXOT-2016-23-PAPER-metadata.tex
\AtlasTitle{
Search for dark matter in events with a hadronically decaying vector boson and missing transverse momentum in $pp$ collisions at $\sqrt{s} = 13$~TeV with the ATLAS detector}

\author{The ATLAS Collaboration}

\AtlasRefCode{EXOT-2016-23}

\PreprintIdNumber{CERN-EP-2018-083}

\AtlasJournal{JHEP}

\AtlasAbstract{% 
A search for dark matter (DM) particles produced in association with a hadronically decaying vector boson is performed using $pp$ collision data at a centre-of-mass energy of $\sqrt{s}=13$~\TeV\ corresponding to an integrated luminosity of 36.1 fb$^{-1}$, recorded by the ATLAS detector at the Large Hadron Collider. This analysis improves on previous searches for processes with hadronic decays of $W$ and $Z$ bosons in association with large missing transverse momentum (mono-$W/Z$ searches) due to the larger dataset and further optimization of the event selection and signal region definitions. In addition to the mono-$W/Z$ search, the as yet unexplored hypothesis of a new vector boson $Z'$ produced in association with dark matter is considered (mono-$Z'$~search). No significant excess over the Standard Model prediction is observed. The results of the mono-$W/Z$ search are interpreted in terms of limits on invisible Higgs boson decays into dark matter particles, constraints on the parameter space of the simplified vector-mediator model and generic upper limits on the visible cross sections for $W/Z$+DM production.  The results of the mono-$Z'$ search are shown in the framework of several simplified-model scenarios involving DM production in association with the $Z'$ boson.}

%% file: introduction.tex
Numerous cosmological observations indicate that a large part of the mass of the universe is composed of dark matter (DM), yet its exact, possibly particle, nature and its connection to the Standard Model (SM) of particle physics remain unknown. Discovery of DM particles and understanding their interactions with SM particles is one of the greatest quests in particle physics and cosmology today. Several different experimental approaches are being exploited. Indirect detection experiments search for signs of DM annihilation or decays in outer space, while direct detection experiments are sensitive to low-energy recoils of nuclei induced by interactions with DM particles from the galactic halo. The interpretation of these searches is subject to astrophysical uncertainties in DM abundance and composition.
Searches at particle colliders, for which these uncertainties are irrelevant, are complementary if DM candidates can be produced in particle collisions.
Weakly interacting massive particles (WIMPs), one of the leading DM candidates, could be produced in proton--proton ($pp$) collisions at the Large Hadron Collider (LHC) and detected by measuring the momentum imbalance associated with the recoiling SM particles.

A typical DM signature which can be detected by the LHC experiments is a large overall missing transverse momentum $\MET$ from a pair of DM particles which are recoiling against one or more SM particles. Several searches for such signatures performed with LHC $pp$ collision data at centre-of-mass energies of 7, 8 and 13~\TeV\ observed 
no deviations from SM predictions and set limits on various DM particle models. 
%, probing the DM production in association with a hadronic jet~\cite{EXOT-2016-27,CMS-EXO-16-048, EXOT-2013-13,CMS-EXO-12-048}, heavy-flavor quarks~\cite{SUSY-2016-18, CMS-EXO-16-051, CMS-EXO-16-005, TOPQ-2013-11, EXOT-2014-01, CMS-EXO-14-004}, a photon~\cite{EXOT-2016-32, CMS-EXO-16-039, EXOT-2014-06, CMS-EXO-12-047}, a hadronically~\cite{EXOT-2015-08, CMS-EXO-16-048, EXOT-2012-27, CMS-EXO-12-055} or leptonically decaying $W$ or $Z$ boson~\cite{atlas_Hinv_Zll13,CMS-EXO-16-052, EXOT-2012-26, CMS-EXO-12-060,CMS-EXO-12-054}, and a Higgs boson~\cite{atlasMonoHbb16, HIGG-2016-18, CMS-EXO-16-012,  EXOT-2014-20, HIGG-2014-05, EXOT-2014-20}. 
Measurements include those probing DM production in association with a hadronically decaying $W$ or $Z$ boson~\cite{EXOT-2015-08, CMS-EXO-16-048, EXOT-2012-27, CMS-EXO-12-055} and dedicated searches for the so-called invisible decays of the Higgs boson into a pair of DM particles, targeting Higgs boson production in association with a hadronically decaying vector boson~\cite{HIGG-2015-03, HIGG-2014-07, CMS-HIG-13-030}.
%via vector boson fusion or in association with a leptonically or hadronically decaying vector boson~\cite{CMS-HIG-16-016, HIGG-2015-03, atlas_Hinv_Zll13, HIGG-2013-16, HIGG-2014-07, HIGG-2013-03, CMS-HIG-13-030}. 
In the SM, the invisible Higgs boson decays occur through the $H \rightarrow ZZ^{\star} \rightarrow \nu\nu\nu\nu$ process with a branching ratio $\mathcal{B}^{\textrm{SM}}_{H \rightarrow \textrm{inv.}}$ of $1.06 \times 10^{-3}$ for a Higgs boson mass $m_{H}=125$~\GeV~\cite{Higgs_YR4}.
Some extensions of the SM allow invisible decays of the Higgs boson into DM or
neutral long-lived massive particles~\cite{hinv_ref1, hinv_ref2, hinv_ref3, hinv_ref4, hinv_ref5} with a significantly larger branching ratio  \BHinv . 
In this case $H$ is required to have properties similar to those of a SM Higgs boson and is assumed to be the Higgs boson with mass of 125~GeV that was discovered at the LHC. 
At present, the most stringent upper limit on \BHinv is about 23\% at 95\%
confidence level (CL) for $m_{H}=125$~\GeV, obtained from a combination of direct searches and indirect constraints from Higgs boson coupling measurements~\cite{HIGG-2015-03, CMS-HIG-16-016}. 

In this paper, a search for DM particles produced in association with a hadronically decaying $W$ or $Z$ boson (\monoWZ~search) is performed for specific DM models, including DM production via invisible Higgs boson decays. The analysis uses LHC $pp$ collision data at a centre-of-mass energy of 13~\TeV\ collected by the ATLAS experiment in 2015 and 2016, corresponding to a total integrated luminosity of \lumi. The results are also expressed in terms of upper limits on visible cross sections, allowing the reinterpretation of the search results in alternative models. In addition to the \monoWZ search, the as yet unexplored hypothesis of DM production in association with a potentially new vector boson $Z'$~\cite{Autran:2015mfa} is studied using the same collision data (\monoZprime~search). Compared to the analysis presented in Ref.~\cite{EXOT-2015-08}, the results are obtained from a larger data sample, and event selection and definition of the signal regions are further optimized, including new signal regions based on the tagging of jets from heavy-flavour hadrons and on jet topologies. Event topologies with two well separated jets from the vector boson decay are studied (referred to as the {\textit{resolved topology}}), as well as topologies with one large-radius jet from a highly boosted vector boson (referred to as the {\textit{merged topology}}).

The paper is organized as follows. A brief introduction to the ATLAS detector is given in Section~\ref{sec:exp}. The signal models are introduced in Section~\ref{sec:models}, while the samples of simulated signal and background processes are described in Section~\ref{sec:mc}. The  algorithms for the reconstruction and identification of final-state particles are summarized in Section~\ref{sec:objects}. Section~\ref{sec:selection} describes the criteria for the selection of candidate signal events. The background contributions are estimated with the help of dedicated control regions in data, as described in Section~\ref{sec:BG}. The experimental and theoretical
systematic uncertainties (Section~\ref{sec:sys}) are taken into account in the statistical interpretation of data, with the results presented in Section~\ref{sec:results}. Concluding remarks are given in Section~\ref{sec:summary}.

%% file: experiment.tex
The ATLAS detector~\cite{PERF-2007-01}  is a general-purpose detector with forward-backward symmetric cylindrical geometry.\footnote{The ATLAS experiment uses a right-handed coordinate system with its origin at the nominal interaction point (IP) in the centre of the detector and the $z$-axis along the beam pipe. The $x$-axis points from the IP to the centre of the LHC ring, and the $y$-axis points upward. Cylindrical coordinates $(r,\phi)$ are used in the transverse plane, $\phi$ being the azimuthal angle around the $z$-axis. The pseudorapidity is defined in terms of the polar angle $\theta$ as $\eta=-\ln\tan(\theta/2)$. Transverse momentum is computed from the three-momentum, $\vec{p}$, as $\pt = |\vec{p}| \sin\theta $.} It consists of an inner tracking detector (ID), electromagnetic (EM) and hadronic calorimeters and a muon spectrometer (MS) surrounding the interaction point.  A new innermost silicon pixel layer~\cite{IBL, CERN-LHCC-2012-009} was added to the ID before the start of data-taking in 2015. The inner tracking system, providing precision tracking in the pseudorapidity range $|\eta | < 2.5$, is immersed in a 2~T axial magnetic field, while toroidal magnets in the MS provide a field integral ranging from 2~Tm to 6~Tm across most of the MS. The electromagnetic calorimeter is a lead/liquid-argon (LAr) sampling calorimeter with an accordion geometry covering the pseudorapidity range $|\eta| < 3.2$.
The hadronic calorimetry is provided by a steel/scintillator-tile calorimeter in the range $|\eta | < 1.7$ and two copper/LAr calorimeters spanning $1.5 < |\eta | < 3.2$. The calorimeter coverage is extended to $|\eta|<4.9$ by copper/LAr and tungsten/LAr forward calorimeters providing both electromagnetic and hadronic energy measurements. The data are collected with a two-level trigger system~\cite{Aaboud:2016leb}. The first-level trigger selects events based on custom-made hardware and uses information from muon detectors and calorimeters with coarse granularity. The second-level trigger is based on software algorithms similar to those applied in the offline event reconstruction and uses the full detector granularity.

%% file: models.tex
\begin{figure}[!hbt]
\begin{center}
\subfigure[\label{fig:sigFeyn_a}]{\includegraphics[width=0.35\textwidth]{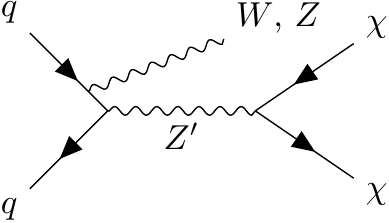}}
\subfigure[\label{fig:sigFeyn_b}]{\hskip1.3cm\includegraphics[width=0.35\textwidth]{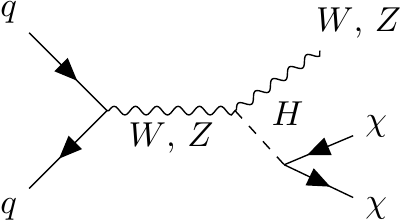}}
\subfigure[\label{fig:sigFeyn_c}]{\includegraphics[width=0.35\textwidth]{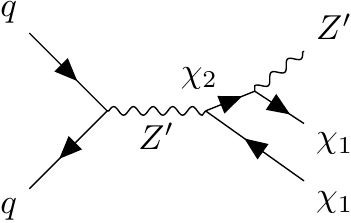}}
\subfigure[\label{fig:sigFeyn_d}]{\hskip1.3cm\includegraphics[width=0.35\textwidth]{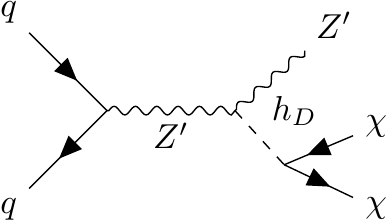}}
\caption{\label{fig:sigFeyn} Examples of dark matter particle ($\chi$) pair-production (a) in association with a $W$ or $Z$ boson in a simplified model with a vector mediator $Z'$ between the dark sector and the SM~\cite{vector}; (b) via decay of the Higgs boson $H$ produced in association with the vector boson~\cite{hinv_ref1, hinv_ref2, hinv_ref3, hinv_ref4, hinv_ref5}; (c) in association with a final-state $Z'$ boson via an additional heavy dark-sector fermion ($\chi_2$) ~\cite{Autran:2015mfa} or (d) via a dark-sector Higgs boson ($h_{\mathrm D}$)~\cite{Autran:2015mfa}.}
\end{center}
\end{figure}

Two signal models are used to describe DM production in the \monoWZ~final state. The first is a \emph{simplified vector-mediator model}, illustrated by the Feynman diagram in Figure~\ref{fig:sigFeyn_a}, in which a pair of Dirac DM particles is produced via an $s$-channel exchange of a vector mediator ($Z'$)~\cite{vector, dmf}.  There are four free parameters in this model: the DM and the mediator masses ($m_{\chi}$ and \mZprime, respectively), and the mediator couplings to the SM and DM particles (\gsm and \gdm, respectively).  The minimal total mediator decay width is assumed, allowing only vector mediator decays into DM or quarks. Its value is determined by the choice of the coupling values \gsm and \gdm~\cite{dmf} and it is much smaller than the mediator mass. The second is a model with \emph{invisible Higgs boson decays} in which a Higgs boson $H$ produced in SM Higgs boson production processes decays into a pair of DM particles which escape detection. The production process with a final state closest to the \monoWZ signature is associated production with a hadronically decaying $W$ or $Z$ boson ($VH$~production, see Figure~\ref{fig:sigFeyn_b}). The $WH$ and $ZH$ signals are predominantly produced via quark--antiquark annihilation ($\qqbar \to VH$), with an additional $ZH$ contribution from gluon--gluon fusion ($gg \to ZH$).
The production of a Higgs boson via gluon--gluon fusion ($ggH$) or vector boson fusion (VBF) followed by the Higgs boson decay into DM particles can also lead to events with large \met and two or more jets. Especially the $ggH$ signal has a contribution comparable to or even stronger than the $VH$ process, since its cross section is about 20 times larger and the jets originating from initial state radiation are more central than in the VBF process. The free parameter of this model is the branching ratio \BHinv. The cross sections for the different Higgs boson production modes are taken to be given by the SM predictions.

Two signal models describe DM production in the mono-$Z'$ final state~\cite{Autran:2015mfa}. Both models contain a $Z'$ boson in the final state; the $Z'$ boson is allowed to decay only hadronically. The $Z' \to \ttbar$ decay channel, kinematically allowed for very heavy $Z'$ resonances, is expected to contribute only negligibly to the selected signal events and therefore the branching ratio $\mathcal{B}_{Z' \to \ttbar}$ is set to zero. 
%, i.e.\ $Z'\to\qqbar$, into all possible quark flavours except for the $\ttbar$ decay channel. 
In the first model, the so-called \emph{dark-fermion model}, the intermediate $Z'$ boson couples to a heavier dark-sector fermion $\chi_2$ as well as the lighter DM candidate fermion $\chi_1$, see Figure~\ref{fig:sigFeyn_c}. 
The mass $m_{\chi_2}$ of the heavy fermion $\chi_2$ is a free parameter of the model, in addition to the DM candidate mass $m_{\chi_1}$, the mediator mass \mZprime, and the $Z'$ couplings  to $\chi_1\chi_2$ (\gdm) and to all SM particles (\gsm). The total $Z'$ and $\chi_2$ decay widths are determined by the choice of the mass and coupling parameter values, assuming that the only allowed decay modes are $\chi_2 \to Z'\chi_1$, $Z'\to\qqbar$ and $Z'\to\chi_2\chi_1$. Under these assumptions the decay widths are small compared to the experimental dijet and large-radius-jet mass resolutions. In the second, so-called \emph{dark-Higgs model}, a dark-sector Higgs boson $h_{\mathrm D}$ which decays to a $\chi\chi$ pair is radiated from the $Z'$ boson as illustrated in Figure~\ref{fig:sigFeyn_d}.  The masses $m_{h_{\mathrm D}}$, $m_\chi$, \mZprime and the constants \gsm and \gdm  are free parameters of the model. The latter is defined as the coupling of the dark Higgs boson $h_{\mathrm D}$ to the vector boson $Z'$. Similar to the dark-fermion model, the total decay widths of the $Z'$ and $h_{\mathrm D}$ bosons are determined by the values of the mass and coupling parameters, assuming that the $Z'$ boson can only decay into quarks or radiate an $h_{\mathrm D}$ boson. The dark Higgs boson is assumed to decay only into $\chi\chi$ or $Z'Z'^{(*)}$. The latter decay mode is suppressed for $m_{h_{\mathrm D}}<$~2$m_{Z'}$, which is the case for the parameter space considered in this paper.

%% file: mc.tex
All signal and background processes from hard-scatter $pp$ collisions were modelled by simulating the detector response to particles produced with Monte Carlo~(MC) event generators. The interaction of generated particles with the detector material was modelled with the {\textsc{Geant4}}~\cite{Agostinelli:2002hh, SOFT-2010-01} package and the same particle reconstruction algorithms were employed in simulation as in the data. Additional $pp$ interactions in the same and nearby bunch crossings (pile-up) were taken into account in simulation. The pile-up events were generated using~\textsc{Pythia~8.186}~\cite{Sjostrand:2007gs} with the A2 set of tuned parameters~\cite{ATL-PHYS-PUB-2012-003} and the  MSTW2008LO set of parton distribution functions (PDF)~\cite{Martin:2009iq}. The simulation samples were weighted to reproduce the observed distribution of the mean number of interactions per bunch crossing in the data.

The \monoWZ signal processes within the simplified $Z'$ vector-mediator model, as well as all \monoZprime signal processes, were modelled at leading-order (LO) accuracy with the {\textsc MadGraph5\textunderscore aMC@NLO}~v2.2.2 generator~\cite{Alwall:2014hca} interfaced to the {\textsc{Pythia~8.186}} and {\textsc{Pythia~8.210}} parton shower models, respectively. The A14 set of tuned parameters~\cite{ATL-PHYS-PUB-2014-021} was used together with the NNPDF23lo PDF set~\cite{Ball:2014uwa} for these signal samples. 
The \monoWZ signal samples within the simplified vector-mediator model were generated in a grid of mediator and DM particle masses, with coupling values set to $\gsm=0.25$ and $\gdm=1$ following the `V1' scenario from Ref.~\cite{Albert:2017onk}. The mediator mass \mZprime and the DM particle mass $m_\chi$ range from 10~\GeV\ to 10~\TeV\ and from 1~\GeV\ to 1~\TeV\ respectively. Two samples with $m_{\chi}=1$~\GeV\ were used to evaluate the impact of theory uncertainties on the signal, one with a mediator mass of 300~\GeV\ and the other with a mediator mass of 600~\GeV. The \monoZprime samples were simulated for mediator masses between 50~\GeV\ and 500~\GeV, with the \gdm coupling value set to $\gdm=1$. Following the current experimental constraints from dijet resonance searches~\cite{EXOT-2017-01, CMS-EXO-17-001, EXOT-2015-03, CMS-EXO-16-032}, in particular those for the mediator mass range below about 500~\GeV\ studied in this analysis, the \gsm coupling value was set to 0.1. For this choice of the couplings, the width of the $Z'$ boson is negligible compared to the experimental resolution, allowing limits to be set on the coupling product $\gsm \cdot \gdm$. For each choice of $m_{Z'}$, two signal samples were simulated in both \monoZprime models, each with a different choice of masses $m_{\chi_2}$ or $m_{h_{\mathrm D}}$ of intermediate dark-sector particles as summarized in Table~\ref{tab:mczprime}. Out of the two samples for a given $m_{Z'}$ value, the one with a lower (higher) mass of the intermediate dark-sector particle is referred to as the `light dark sector' (`heavy dark sector') scenario.
The mass $m_\chi$ in the dark-Higgs model was set to 5~\GeV, since it can be assumed that the kinematic properties are determined by the masses $m_{Z'}$ and $m_{h_{\mathrm D}}$ unless the mass $m_{\chi}$ is too large.
\begin{table}[!htbp]
\caption{Particle mass settings in the simulated \monoZprime samples for a given mediator mass \mZprime.}
\label{tab:mczprime}
\centering
\begin{tabular}{lll}
\hline \hline
Scenario & Dark-fermion model & Dark-Higgs model \\
\hline
\multirow{4}{*}{Light dark sector} & $m_{\chi_1}= 5$ \GeV &  $m_{\chi}= 5$ \GeV  
\\
    &        \multirow{2}{*}{$m_{\chi_2}= m_{\chi_1} + \mZprime + 25$ \GeV} &   \multirow{2}{*}{$m_{h_{\mathrm D}} = \begin{cases}
                \mZprime  & ,\ \mZprime < 125\ \textrm{\GeV} \\
                125\  \textrm{\GeV}      & ,\ \mZprime > 125\ \textrm{\GeV} \\
        \end{cases}$}\\
   & & \\
\   & & \\
\hline
\multirow{4}{*}{Heavy dark sector} & $m_{\chi_1}= \mZprime/2 $ &  $m_{\chi}= 5$ \GeV  \\
       &  \multirow{2}{*}{$m_{\chi_2} = 2\mZprime$} & \multirow{2}{*}{$m_{h_{\mathrm D}} = \begin{cases}
                125\  \textrm{\GeV}      & ,\ \mZprime < 125\ \textrm{\GeV} \\
                \mZprime  & ,\ \mZprime > 125\ \textrm{\GeV} \\
        \end{cases}$} \\
   & & \\
 \  & & \\
\hline \hline
\end{tabular}
\end{table}

Processes in the  \monoWZ final state involving invisible Higgs boson decays originate from the $VH$, $ggH$ and VBF SM Higgs boson production mechanisms and were all generated with the {\textsc{Powheg-Box~v2}}~\cite{Nason:2004rx,Frixione:2007vw,Alioli:2010xd} generator 
%~\cite{powheg5,Luisoni:2013kna} 
interfaced to {\textsc{Pythia~8.212}} for the parton shower, hadronization and the underlying event modelling. The detailed description of all generated production processes  together with the corresponding cross-section calculations can be found in Refs.~\cite{HIGG-2016-25, HIGG-2016-29}. The Higgs boson mass in these samples was set to $m_{H}=$~125~\GeV\ and the Higgs boson was decayed through the $H \rightarrow ZZ^* \rightarrow \nu\nu\nu\nu$ process to emulate the decay of the Higgs boson into invisible particles with a branching ratio of \BHinv = 100\%.

The major sources of background are the production of top-quark pairs ($t\bar{t}$) and the production of $W$ and $Z$ bosons in association with jets ($V$+jets, where $V\equiv W~\textrm{or}~Z$). The event rates and the shape of the final discriminant observables for these processes are constrained with data from dedicated control regions (see Section~\ref{sec:BG}). Other small background contributions include diboson ($WW,~WZ~\textrm{and}~ZZ$) and single top-quark production. Their contribution is estimated from simulation.

Events containing leptonically decaying $W$ or $Z$ bosons with associated jets were simulated using the {\textsc{Sherpa}}~2.2.1 generator~\cite{Gleisberg:2008ta}, with matrix elements calculated for up to two partons at next-to-leading order (NLO) and four partons at LO using  {\textsc{Comix}}~\cite{Gleisberg:2008fv} and {\textsc{OpenLoops}}~\cite{Cascioli:2011va} and merged with the {\textsc{Sherpa}} parton shower~\cite{Schumann:2007mg} using the ME+PS@NLO prescription~\cite{Hoeche:2012yf}. The NNPDF3.0 next-to-next-to-leading order (NNLO) PDF set~\cite{Ball:2014uwa} was used in conjunction with dedicated parton shower tuning developed by the {\textsc{Sherpa}} authors. The inclusive cross section was calculated up to NNLO in QCD~\cite{Melnikov:2006kv}.

For the generation of $t\bar{t}$ events, {\textsc{Powheg-Box v2}} was used with the CT10 PDF set~\cite{Lai:2010vv} in the NLO matrix element calculations. Electroweak $t$-channel, $s$-channel and $Wt$-channel single-top-quark events were generated with {\textsc{Powheg-Box}} v1. This event generator uses the four-flavour scheme for the NLO matrix element calculations together with the fixed four-flavour PDF set CT10f4~\cite{Lai:2010vv}. For all top-quark processes, top-quark spin correlations are preserved (for $t$-channel top-quark production, top quarks were decayed using {\textsc{MadSpin}}~\cite{Artoisenet:2012st}). The parton shower, hadronization, and the underlying event were simulated using {\textsc{Pythia~6.428}}~\cite{Sjostrand:2006za} with the CTEQ6L1 PDF set~\cite{Pumplin:2002vw} and the corresponding Perugia 2012 set of tuned parameters~\cite{Skands:2010ak}. The top-quark mass was set to 172.5~\GeV. The {\textsc{EvtGen}} 1.2.0 program~\cite{EvtGen} was used for the properties of $b$- and $c$-hadron decays. The inclusive $t\bar{t}$ cross section was calculated up to NNLO with soft gluon resummation at next-to-next-to-leading-logarithm (NNLL) accuracy~\cite{Czakon:2013goa}. Single top-quark production cross sections were calculated at NLO accuracy~\cite{PhysRevD.56.5919, PhysRevD.56.5919, Stelzer:1998ni, Smith:1996ij, Kidonakis:2013zqa}. 

Diboson events with one of the bosons decaying hadronically and the other leptonically were generated with the {\textsc{Sherpa 2.1.1}} event generator. Matrix elements were calculated for up to one ($ZZ$) or zero ($WW$, $WZ$) additional partons at NLO and up to three additional partons at LO using {\textsc{Comix}}  and {\textsc{OpenLoops}}, and merged with the {\textsc{Sherpa}} parton shower according to the ME+PS@NLO prescription. The CT10 PDF set was used in conjunction with dedicated parton shower tuning developed by the {\textsc{Sherpa}} authors. The event generator cross sections at NLO were used in this case. 
In addition, the Sherpa diboson sample cross section is scaled to
account for the cross section change when switching to the $G_\mu$ scheme for
the electroweak parameters, resulting in an effective value of $\alpha \approx
1/132$.

%In addition, the {\textsc{Sherpa}} diboson sample cross section was scaled down to account for its use of $\alpha$=1/129 rather than 1/132 corresponding to the use of current PDG parameters as input to the G$_\mu$ scheme~\cite{PDG_review}.

%% file: objects.tex
%Two experimental signatures of DM production are exploited in the search, depending on the Lorentz-boost of the vector boson. For bosons with smaller boosts, two distinct particle jets from the vector boson decay can be observed in the detector. If the boson is strongly boosted, angular separation of the two jets is reduced, leading to an event topology with a single large-radius jet instead. Both signal topologies are further characterized by a large missing transverse momentum. Since no high-$\pt$ isolated leptons are expected from the signal, a veto is imposed on leptons (electrons or muons) in the final state to define the signal region. Complementary to this signal region, the presence of exactly one muon or exactly two electrons or two muons is required in dedicated data control regions for $V$+jets and $t\bar{t}$ backgrounds. To improve the sensitivity to mono-$Z$ signatures with $Z\rightarrow bb$ decays and better constrain different background components, events in signal and control regions are classified according to the $b$-hadron content using jet $b$-tagging algorithms.

The selection of \monoWZ and \monoZprime candidate signal events and events in dedicated one-muon and two-lepton (electron or muon) control regions relies on the reconstruction and identification of jets, electrons and muons, as well as on the reconstruction of the missing transverse momentum. These are described in the following.

Three types of jets are employed in the search. They are reconstructed from noise-suppressed topological calorimeter energy clusters~\cite{PERF-2014-07}  ({\em{``small-$R$''}} and {\em{``large-$R$''}}~jets) or inner detector tracks ({\em{``track''}}~jets) using the anti-$k_{t}$ jet clustering algorithm~\cite{Cacciari:2008gp, Cacciari:2011ma} with different values of the radius parameter~$R$.

Small-$R$ jets ($j$) with radius parameter $R=0.4$ are used to identify vector bosons with a relatively low boost. Central jets (forward jets) within $|\eta|<2.5$ ($2.5\leq|\eta|<4.5$) are required to satisfy $\pt>20$~\GeV\ ($\pt>30$~\GeV). The small-$R$ jets satisfying $\pt<60$~\GeV\ and $|\eta|<2.4$ are required to be associated with the primary vertex using the jet-vertex-tagger discriminant~\cite{ATLAS-CONF-2014-018} in order to reject jets originating from pile-up vertices.  The vertex with the highest $\sum{\pt^2}$ of reconstructed tracks is selected as the primary vertex. Jet energy scale and resolution, as well as the corresponding systematic uncertainties, are determined with simulation and data at $\sqrt{s}=13$~\TeV~\cite{PERF-2016-04, ATLAS-CONF-2015-017}. Jets within $|\eta|<$~2.5 containing $b$-hadrons are identified using the MV2c10 $b$-tagging algorithm~\cite{PERF-2012-04, ATL-PHYS-PUB-2016-012, ATL-PHYS-PUB-2015-039} at an operating point with a 70\% $b$-tagging efficiency measured in simulated $t\bar{t}$ events.

Large-$R$ jets ($J$)~\cite{PERF-2015-03,ATL-PHYS-PUB-2015-033} are reconstructed with a radius parameter of $R=1.0$ to allow the detection of merged particle jets from a boosted vector boson decay. The trimming algorithm~\cite{Krohn:2009th} is applied to remove the energy deposits from pile-up, the underlying event and soft radiation, by reclustering the large-$R$ jet constituents into sub-jets with radius parameter $R=0.2$. The sub-jets with transverse momenta below 5\% of the original jet transverse momentum are removed from the large-$R$ jet. The jet mass is calculated as the resolution-weighted mean of the mass measured using only calorimeter information and the track-assisted mass measurement~\cite{ATLAS-CONF-2016-035}. Large-$R$ jets are required to satisfy $\pt>200$~\GeV\ and $|\eta|<2.0$. In the \monoWZ search, these jets are tagged as originating from a hadronic $W$- or $Z$-boson decay using $\pt$-dependent requirements on the jet mass and substructure variable \DTwoBetaOne~\cite{d2variable,c2variable}. The latter is used to select jets with two distinct concentrations of energy within the large-$R$ jet~\cite{PERF-2012-02,ATLAS-CONF-2015-035}.  The jet mass and \DTwoBetaOne selection criteria are adjusted as a function of jet $\pt$ to select $W$ or $Z$ bosons with a constant efficiency of 50\% measured in simulated events. In the \monoZprime search, large-$R$ jets are tagged as originating from the hadronic decay of a  $Z'$ boson using a jet-mass requirement and requiring \DTwoBetaOne$<$1.2, chosen to optimize the search sensitivity. The momenta of both the large-$R$ and small-$R$ jets are corrected for energy losses in passive material and for the non-compensating response of the calorimeter. Small-$R$ jets are also corrected for the average
additional energy due to pile-up interactions.

Track jets with radius parameter $R=0.2$~\cite{ATL-PHYS-PUB-2014-013} are used to identify large-$R$ jets containing $b$-hadrons~\cite{ATLAS-CONF-2016-039}. Inner detector tracks originating from the primary vertex, selected by impact parameter requirements, are used in the track jet reconstruction. Track jets are required to satisfy $\pt>10$~\GeV\ and $|\eta|<2.5$, and are matched to the large-$R$ jets via ghost-association~\cite{Cacciari:2008gn}. As for the small-$R$ jets, the track jets containing $b$-hadrons are identified using the MV2c10 algorithm at a working point with 70\% efficiency.

Simulated jets are labelled according to the flavour of the hadrons with $\pt>$~5~\GeV\ which are found within a cone of size $\Delta R \equiv \sqrt{(\Delta\phi)^2 + (\Delta\eta)^2}=$~0.3 around the jet axis. If a $b$-hadron is found, the jet is labelled as a $b$-jet. If no $b$-hadron, but a $c$-hadron is found, the jet is labelled as a $c$-jet. Otherwise the jet is labelled as a light jet ($l$) originating from $u$-, $d$-, or $s$-quarks or gluons. Simulated $V$+jets events are categorized according to this particle-level labelling into three separate categories:  $V$~+~heavy~flavour ($V$+HF) events, $V+cl$ events and  $V$~+~light flavour ($V$+LF) events. The first category consists of $V+bb$, $V+bc$, $V+cc$ and $V+bl$ components, while the last one is given by the $V+ll$ component alone. In the very rare case that after the final selection only one jet is present in addition to the $V$ boson, the missing jet is labelled as a light jet.

Electron candidates are reconstructed from energy clusters in the electromagnetic calorimeter that are associated to an inner detector track. The electron candidates  are identified using a likelihood-based procedure~\cite{Aaboud:2016vfy,ATLAS-CONF-2016-024} in combination with additional track hit requirements. All electrons, including those employed for the electron veto in the signal and in the one-muon and two-muon control regions, must satisfy the `loose' likelihood criteria. An additional, more stringent criterion is applied in the two-electron control region, requiring that at least one of the electrons passes the `medium' likelihood criteria. Each electron is required to have $\pt>7$~\GeV, and $|\eta|<2.47$, with their energy calibrated as described in Ref.~\cite{PERF-2013-05, ATL-PHYS-PUB-2016-015}. To suppress the jets misidentified as electrons, electron isolation is required, defined as an upper limit on the scalar sum of the $\pt^i$ of the tracks $i$ (excluding the track associated to the electron candidate) within a cone of size $\Delta R=0.2$ around the electron, $(\sum{\pt^i})^{\Delta R = 0.2}$, relative to electron $\pt$. The $\pt$- and $\eta$-dependent limits corresponding to an isolation efficiency of 99\% are applied. In addition, to suppress electrons not originating from the primary vertex, requirements are set on the longitudinal impact parameter, $|z_0\sin{\theta}|<$~0.5~mm, and the transverse impact parameter significance, $|d_0|/\sigma(d_0)<$~5.

Muon candidates are primarily reconstructed from a combined fit to inner detector hits and muon spectrometer segments~\cite{PERF-2015-10}. In the central detector region ($|\eta|<$~0.1) lacking muon spectrometer coverage, muons are also identified by matching a reconstructed inner detector track to calorimeter energy deposits consistent with a minimum ionizing particle. Two identification working points with different purity are used. All muons, including those employed for the muon veto in the signal and in the two-electron control regions, must satisfy the `loose' criteria. In addition, the muon in the one-muon control region and at least one of the two muons in the two-muon control region must pass the `medium' selection criteria. Each muon is required to have $\pt>7$~\GeV\ and $|\eta|<2.7$ and satisfy the impact parameter criteria $|z_0\sin{\theta}|<$~0.5~mm and $|d_0|/\sigma(d_0)<$~3. All muons are required to be isolated by requiring an upper threshold on the scalar sum  $(\sum{\pt^i})^{\Delta R = 0.3}$ relative to the muon $\pt$ that corresponds to a 99\% isolation efficiency, similarly to the electrons. In the one-muon control region,  tighter isolation criteria with $(\sum{\pt^i})^{\Delta R = 0.3}/\pt<$~0.06 are applied. In both cases, the muon $\pt$ is subtracted from the scalar sum.

The vector missing transverse momentum $\vec{\met}$~is calculated as the negative vector sum of the transverse momenta of calibrated small-$R$ jets and leptons, together with the tracks which are associated to the primary interaction vertex but not associated to any of these physics objects~\cite{PERF-2014-04}. A closely related quantity, $\vec{\metnolep}$, is calculated in the same way but excluding the reconstructed muons or electrons. The missing transverse momentum is given by the magnitude of these vectors, $\met=|\vec{\met}|$ and $\metnolep=|\vec{\metnolep}|$. In addition, the track-based missing transverse momentum vector, $\vec{\mpt}$, and similarly $\vec{\mptnolep}$, is calculated as the negative vector sum of the transverse momenta of tracks with $\pt>0.5$~\GeV\ and $|\eta|<$~2.5 originating from the primary vertex.

%% file: selection.tex
Events studied in this analysis are accepted by a combination of  $\met$ triggers with thresholds between 70~\GeV\ and 110~\GeV, depending on the data-taking periods. The trigger efficiency is measured in data using events with large \met accepted by muon triggers. The triggers are found to be fully efficient for $\met>200$~\GeV\ and the inefficiency at lower \met values and the corresponding uncertainty are taken into account. At least one collision vertex with at least two associated tracks is required in each event, and for the signal region selection a veto is imposed on all events with loose electrons or muons in the final state. Depending on the Lorentz boost of the vector boson, 
two distinct event topologies are considered: a $\emph{merged~topology}$ where the decay products of the vector boson are reconstructed as a single large-$R$ jet, and a $\emph{resolved~topology}$ where they are reconstructed as individual small-$R$ jets. Each event is first passed through the merged-topology selection and, if it fails, it is passed through the resolved-topology selection. Thus, there is no overlap of events between the two final-state topologies. For the \monoZprime search, the categorization into merged and resolved event topologies is only performed for the mediator mass hypothesis of $m_{Z'}$ below 100~\GeV. For heavier mediator masses, the angular separation of jets from the $Z'$ boson decay is expected to be larger than the size of a large-$R$ jet. Thus, only the resolved-topology selection criteria are applied in this case. 

The \monoWZ and \monoZprime event selection criteria applied for each of the two topologies are summarized in Table~\ref{tab:selection}. The criteria have been optimized to obtain the maximum expected signal significance. In the merged (resolved) event topology, at least one large-$R$ jet (at least two small-$R$ jets) and $\met$ values above 250~\GeV\ (above 150~\GeV) are required in the final state. In order to suppress the $t\bar{t}$ and $V$+jets background with heavy-flavour jets, all events with merged topology containing $b$-tagged track jets not associated to the large-$R$ jet  via ghost-association are rejected. In the resolved topology, all events with more than two $b$-tagged small-$R$ jets are rejected. The highest-$\pt$ large-$R$ jet in an event is considered as the candidate for a hadronically decaying vector boson in the merged topology. Similarly, in the resolved topology the two highest-\pT (leading) $b$-tagged small-$R$ jets are selected as the candidate for a hadronically decaying $W$ or $Z$ boson and, if there are fewer than two $b$-jets in the final state, the highest-$\pt$ remaining jets are used to form the hadronic $W$ or $Z$ boson decay candidate. Additional criteria are applied in both merged and resolved topologies to suppress the contribution from multijet events. Since the vector bosons in signal events are recoiling against the dark matter particles, a threshold is applied on the azimuthal separation between the $\vec{\met}$ vector and the highest-$\pt$ large-$R$ jet (system of the two highest-$\pt$ jets) in the merged (resolved) topology, $\Delta\phi(\vec{\met}, J~\mathrm{or}~jj)>$~120$^{\textrm{o}}$. Also, the angles between $\vec{\met}$ and each of the up to three highest-$p_{\mathrm{T}}$ small-$R$ jets should be sufficiently large, $\min\left[\Delta\phi(\vec{\met}, j)\right]>$~20$^{\textrm{o}}$, in order to suppress events with a significant \met contribution from mismeasured jets. Events with a large \met value originating from calorimeter mismeasurements are additionally suppressed by the requirement of a non-vanishing track-based missing transverse momentum, $\mpt>$~30~\GeV, and a requirement on the azimuthal separation between the calorimeter-based and track-based missing transverse momenta, $\Delta\phi(\vec{\met},\vec{\mpt})<$~90$^{\textrm{o}}$. The \mpt~requirements also reduce non-collision background from beam halo or beam--gas interactions that produce signal in time with the colliding proton bunches. Such events are characterized mainly by energy deposits in the calorimeters in the absence of track activity. In the categories with two $b$-tagged jets the non-collision background is 
negligible and the expected discovery significance is higher without the \mpt~requirement, which is not applied. Further criteria are imposed on events with the resolved topology. The leading jet is required to have $\pt^{j_1}>$~45~\GeV. To improve the modelling of the trigger efficiency with MC events, the scalar sum of the transverse momenta of all jets is required to be $\sum{\pt^{j_i}}>$~120~(150)~\GeV\ in events with two (at least three) jets.  

After these general requirements, the events are classified according to the number of $b$-tagged jets into events with exactly zero ($0b$), one ($1b$) and two ($2b$) $b$-tagged jets to improve the signal-to-background ratio and the sensitivity to $Z\to bb$ decays. Small-$R$ jets (track jets) are used for the $b$-tagging in the resolved (merged) category. Further selection criteria defining the final signal regions are introduced separately for the \monoWZ and \monoZprime searches.

For the \monoWZ search, the events in the $0b$ and $1b$ categories with merged topology are further classified into high-purity (HP) and low-purity (LP) regions; the former category consists of events satisfying the $\pt$-dependent requirements on the jet substructure variable \DTwoBetaOne, allowing an improved discrimination for jets containing $V\to\qqbar$ decays, while the latter one selects all the remaining signal events. In the signal region with resolved topology, the angular separation $\Delta R_{jj}$ between the two leading jets  is required to be smaller than 1.4 (1.25) in the $0b$ and $1b$ ($2b$) categories. 
%The azimuthal distance between the two leading jets is also required to be small, $\Delta\phi_{jj}<$~140$^\textrm{o}$.  
Finally, a mass window requirement is imposed on the vector boson candidate in each of the eight resulting signal categories. In the $0b$ and $1b$ merged-topology categories, a mass requirement depending on the large-$R$ jet \pt~is applied. The large-$R$ jet mass and \DTwoBetaOne requirements have been optimized within a dedicated study of the $W/Z$ tagger performance~\cite{ATL-PHYS-PUB-2015-033, PERF-2015-03, ATLAS-CONF-2017-064}. In the $2b$ merged-topology category, in which the signal is expected to come predominantly from $Z\to bb$ decays, a mass window requirement of 75~\GeV~$<m_J<$~100~\GeV\ is applied. The large-$R$ jet substructure variable \DTwoBetaOne is not considered in this channel in order to obtain a higher signal efficiency and higher expected discovery significance. In the resolved $0b$ and $1b$ ($2b$) categories, the mass of the  dijet system composed of the two leading jets is required to be 65~\GeV~$<m_{jj}<$~105~\GeV\ (65~\GeV~$<m_{jj}<$~100~\GeV). For the \monoZprime search, a similar classification by the $b$-tagging multiplicity, and by the substructure variable \DTwoBetaOne into high- and low-purity regions in the merged-topology category, is performed, using slightly different requirements on the substructure of the large-$R$ jet. A \pt-independent requirement on the substructure variable \DTwoBetaOne $<$~1.2 is used in signal regions with merged topology, as this is found to provide the maximum expected signal significance. Additional criteria also differ from the criteria applied in the \monoWZ search. No criteria are applied on the $\Delta R_{jj}$ variable in events with the resolved topology, since the high-mass $Z'$ bosons in dark-fermion or dark-Higgs models are less boosted than $W$ or $Z$ bosons in the simplified vector-mediator model, leading to a larger angular separation of jets from the $Z'$ boson decays. The requirements on the mass of the $Z'$ candidate are optimized for each event category as summarized in Table~\ref{tab:selection}. 
\begin{table}[!htbp]
\caption{Event selection criteria in the \monoWZ and \monoZprime signal regions with merged and resolved event topologies. The symbols ``$j$'' and ``$J$'' denote the reconstructed small-$R$ and large-$R$ jets, respectively. The abbreviations HP and LP denote respectively the high- and low-purity signal regions with merged topology, as defined by the cut on the large-$R$ jet substructure variable \DTwoBetaOne . }
\centering
\resizebox{\linewidth}{!}{
\begin{tabular}{l| c|c|c|c|c| |c|c|c}
\hline
\hline

 &
\multicolumn{5}{c||}{{\textbf{Merged topology}}}&  
\multicolumn{3}{c}{{\textbf{Resolved topology}}}\\
\hline
\hline
\multicolumn{9}{l}{{~}}\\
\multicolumn{9}{l}{{\textbf{General requirements}}}\\

\hline

\met &
\multicolumn{5}{c||}{$>$~250~\GeV}&         
\multicolumn{3}{c}{$>$~150~\GeV}\\

Jets, leptons &
\multicolumn{5}{c||}{$\ge$1$J$, 0$\ell$}&    
\multicolumn{3}{c}{$\ge$2$j$, 0$\ell$}\\

$b$-jets &
\multicolumn{5}{c||}{no $b$-tagged track jets outside of $J$}&    
\multicolumn{3}{c}{$\leq$~2 $b$-tagged small-$R$ jets}\\

\hline

&
\multicolumn{8}{c}{$\Delta\phi(\vec{\met}, J~\mathrm{or}~jj)>$~120$^\textrm{o}$}\\

Multijet&
\multicolumn{8}{c}{$\min_{i\in \{1,2,3\}}\left[\Delta\phi(\vec{\met},j_i)\right]>$~20$^\textrm{o}$}\\

suppression&
\multicolumn{8}{c}{$\mpt>$~30~\GeV\ or $\geq$2~$b$-jets}\\

&
\multicolumn{8}{c}{$\Delta\phi(\vec{\met},\vec{\mpt})<$~90$^\textrm{o}$}\\

\hline

Signal&
\multicolumn{5}{c||}{}& 
\multicolumn{3}{c}{$\pt^{j_1}>$~45~\GeV}\\

properties&
\multicolumn{5}{c||}{}& 
\multicolumn{3}{c}{$\sum{\pt^{j_i}}>$~120~(150)~\GeV\ for 2 ($\geq$~3) jets}\\

\hline
\hline

\multicolumn{9}{l}{{~}}\\
\multicolumn{9}{l}{{\textbf{Mono-$W/Z$ signal regions}}}\\
\hline

& 
\multicolumn{1}{>{\centering}p{0.7cm}|}{{\textbf{~~$0b$}}}& 
\multicolumn{1}{>{\centering}p{0.7cm}|}{{\textbf{~~$0b$}}}& 
\multicolumn{1}{>{\centering}p{0.7cm}|}{{\textbf{~~$1b$}}}& 
\multicolumn{1}{>{\centering}p{0.7cm}|}{{\textbf{~~$1b$}}}& 
\multicolumn{1}{>{\centering}p{0.7cm}||}{{\textbf{$2b$}}}& 
\multicolumn{1}{>{\centering}p{1.7cm}|}{{\textbf{$0b$}}}& 
\multicolumn{1}{>{\centering}p{1.7cm}|}{{\textbf{$1b$}}}& 
\multicolumn{1}{>{\centering}p{1.7cm}}{{\textbf{$2b$}}}\\ 

& 
\multicolumn{1}{>{\centering}p{0.7cm}|}{{\textbf{HP}}}& 
\multicolumn{1}{>{\centering}p{0.7cm}|}{{\textbf{LP}}}& 
\multicolumn{1}{>{\centering}p{0.7cm}|}{{\textbf{HP}}}& 
\multicolumn{1}{>{\centering}p{0.7cm}|}{{\textbf{LP}}}& 
\multicolumn{1}{>{\centering}p{0.7cm}||}{}& 
\multicolumn{1}{>{\centering}p{1.7cm}|}{}& 
\multicolumn{1}{>{\centering}p{1.7cm}|}{}& 
\multicolumn{1}{>{\centering}p{1.7cm}}{}\\ 
\hline

%$\Delta\phi_{jj}$&
%-& -& -& -& -&
%\multicolumn{1}{c|}{$<~140^{\textrm{o}}$}&
%\multicolumn{1}{c|}{$<~140^{\textrm{o}}$}&
%\multicolumn{1}{c}{$<~140^{\textrm{o}}$}\\

$\Delta R_{jj}$&
--& --& --& --& --& 
$<$~1.4& 
$<$~1.4&
$<$~1.25\\

\DTwoBetaOne \ptJ-dep.&
pass& fail& pass& fail& --& 
--& 
--&
--\\

\hline
Mass requirement&
\multicolumn{4}{c|}{$m_J$}& $m_J$&
\multicolumn{2}{c|}{$m_{jj}$}& $m_{jj}$\\
\lbrack\GeV] &
\multicolumn{4}{c|}{$W/Z$ tagger requirement}& 
[75, 100]&
\multicolumn{2}{c|}{[65, 105]}& 
[65, 100]
\\

\hline
\hline 

\multicolumn{9}{l}{{~}}\\
\multicolumn{9}{l}{{\textbf{Mono-$Z'$ signal regions}}}\\
\hline

& 
\multicolumn{1}{>{\centering}p{0.7cm}|}{{\textbf{~~$0b$}}}& 
\multicolumn{1}{>{\centering}p{0.7cm}|}{{\textbf{~~$0b$}}}& 
\multicolumn{1}{>{\centering}p{0.7cm}|}{{\textbf{~~$1b$}}}& 
\multicolumn{1}{>{\centering}p{0.7cm}|}{{\textbf{~~$1b$}}}& 
\multicolumn{1}{>{\centering}p{0.7cm}||}{{\textbf{$2b$}}}& 
\multicolumn{1}{>{\centering}p{1.7cm}|}{{\textbf{$0b$}}}& 
\multicolumn{1}{>{\centering}p{1.7cm}|}{{\textbf{$1b$}}}& 
\multicolumn{1}{>{\centering}p{1.7cm}}{{\textbf{$2b$}}}\\ 

& 
\multicolumn{1}{>{\centering}p{0.7cm}|}{{\textbf{HP}}}& 
\multicolumn{1}{>{\centering}p{0.7cm}|}{{\textbf{LP}}}& 
\multicolumn{1}{>{\centering}p{0.7cm}|}{{\textbf{HP}}}& 
\multicolumn{1}{>{\centering}p{0.7cm}|}{{\textbf{LP}}}& 
\multicolumn{1}{>{\centering}p{0.7cm}||}{}& 
\multicolumn{1}{>{\centering}p{1.7cm}|}{}& 
\multicolumn{1}{>{\centering}p{1.7cm}|}{}& 
\multicolumn{1}{>{\centering}p{1.7cm}}{}\\ 
\hline

\DTwoBetaOne$<$1.2& 
pass& fail& pass& fail& --& 
--& 
--&
--\\

\hline
& 
\multicolumn{5}{l||}{\underline{For \mZprime $<$~100~\GeV:}}& 
\multicolumn{3}{l}{\underline{For \mZprime $<$~200~\GeV:}}\\ 
& 
\multicolumn{2}{>{\centering}p{1.5cm}|}{[0.85\mZprime,}& 
\multicolumn{3}{>{\centering}p{3.4cm}||}{[0.75\mZprime,}& 
\multicolumn{1}{>{\centering}p{1.7cm}|}{[0.85\mZprime,}& 
\multicolumn{2}{>{\centering}p{3.4cm}}{[0.75\mZprime,}\\ 
Mass requirement& 
\multicolumn{2}{>{\centering}p{1.5cm}|}{\mZprime + 10]}& 
\multicolumn{3}{>{\centering}p{3.4cm}||}{\mZprime+  10]}& 
\multicolumn{1}{>{\centering}p{1.7cm}|}{\mZprime + 10]}& 
\multicolumn{2}{>{\centering}p{3.4cm}}{\mZprime + 10]}\\ 

[\GeV]&   
\multicolumn{5}{l||}{~}& 
\multicolumn{3}{l}{~}\\ 

&
\multicolumn{5}{l||}{\underline{For \mZprime $\ge$~100~\GeV:}}& 
\multicolumn{3}{l}{\underline{For \mZprime $\ge$~200~\GeV:}}\\ 

&
\multicolumn{5}{>{\centering}p{5.5cm}||}{no merged-topology}& 
\multicolumn{1}{>{\centering}p{1.7cm}|}{[0.85\mZprime,}& 
\multicolumn{2}{>{\centering}p{3.4cm}}{[0.80\mZprime,}\\ 
& 
\multicolumn{5}{>{\centering}p{5.5cm}||}{selection applied}& 
\multicolumn{1}{>{\centering}p{1.7cm}|}{\mZprime + 20]}& 
\multicolumn{2}{>{\centering}p{3.4cm}}{\mZprime + 20]}\\ 

\hline
\hline
\end{tabular}
}
\label{tab:selection}
\end{table}

\begin{figure}[!htbp]
\centering
\subfigure[]{\includegraphics[width=0.49\textwidth]{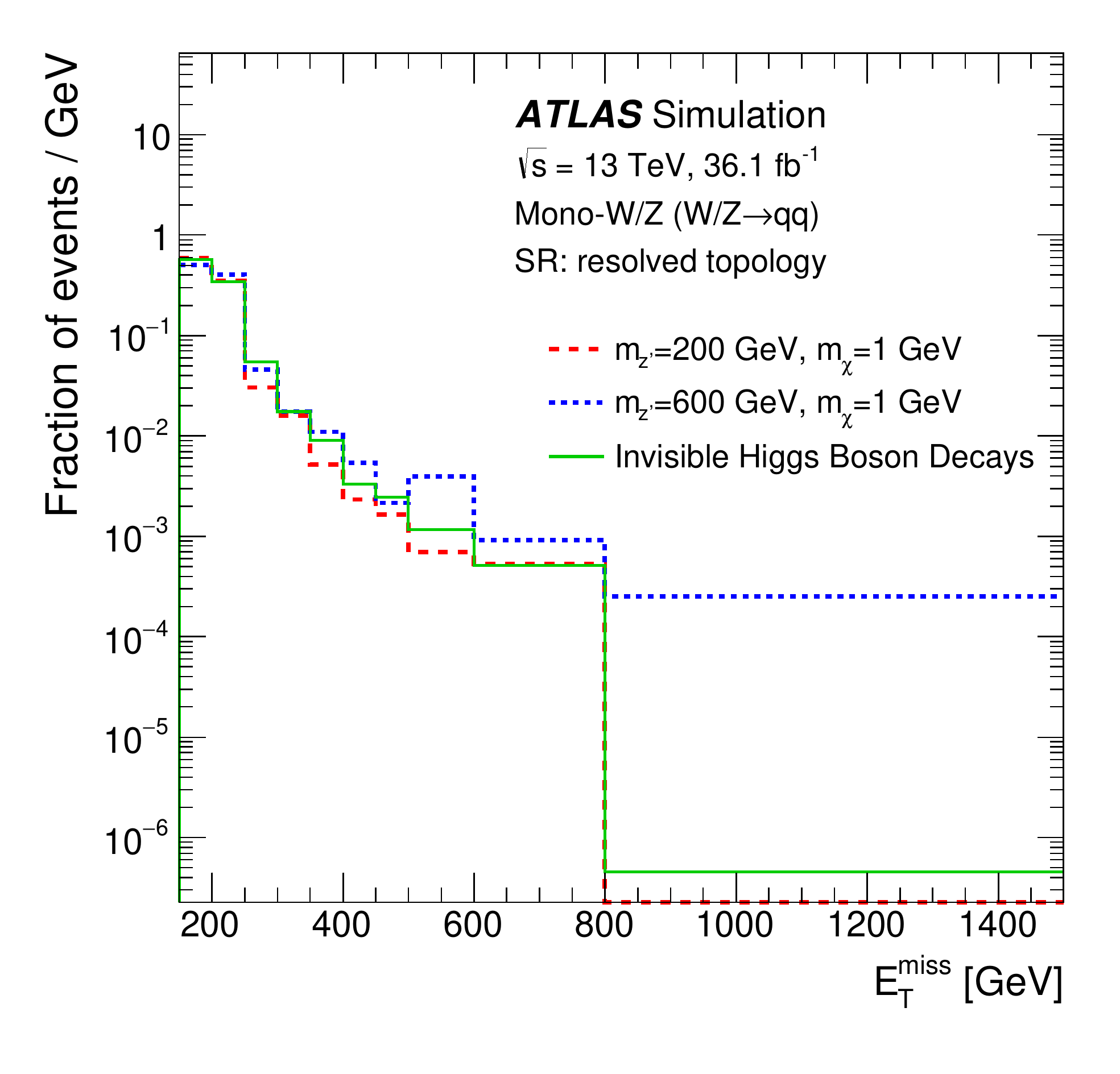}}
\subfigure[]{\includegraphics[width=0.49\textwidth]{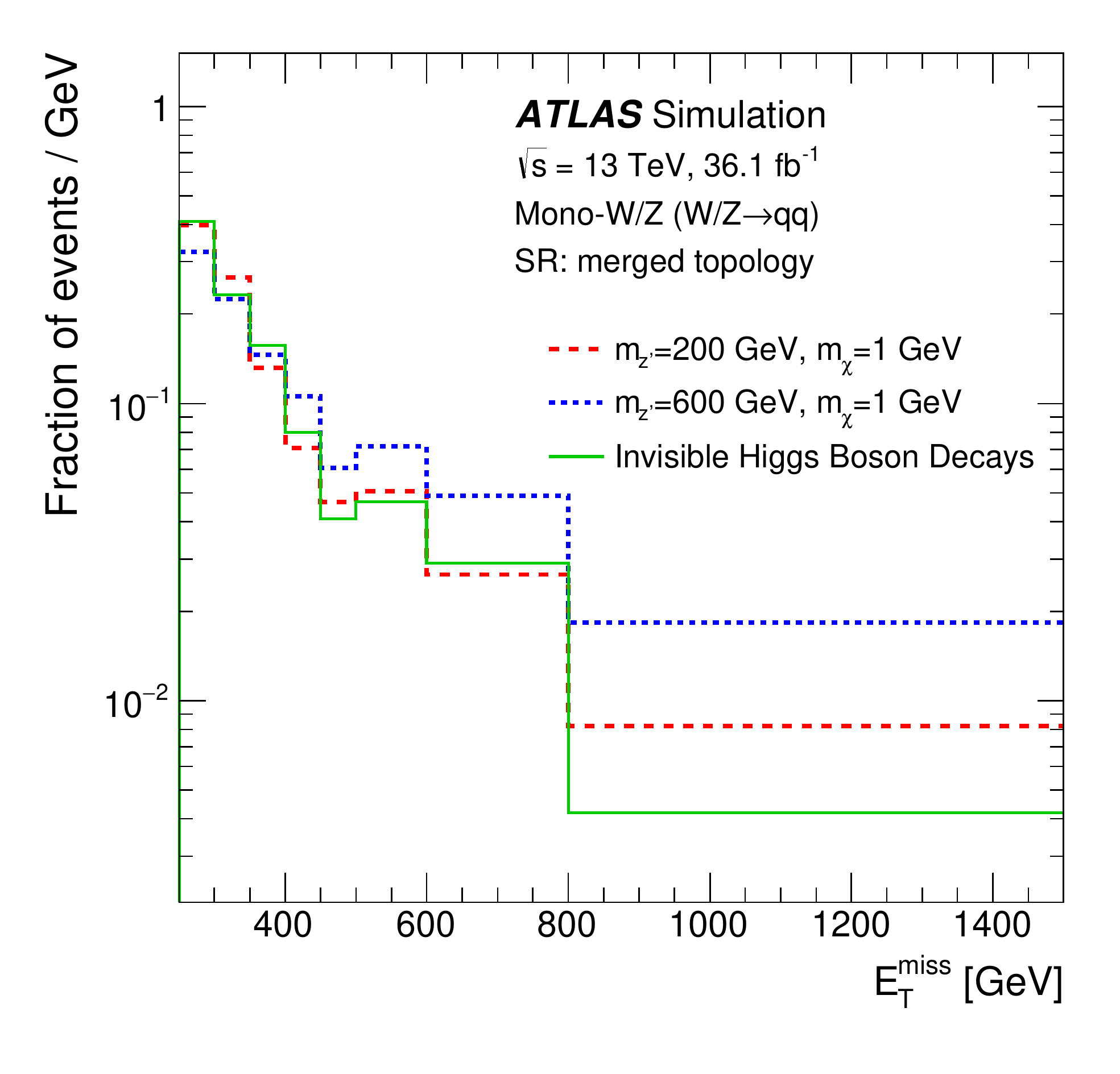}}
\subfigure[]{\includegraphics[width=0.49\textwidth]{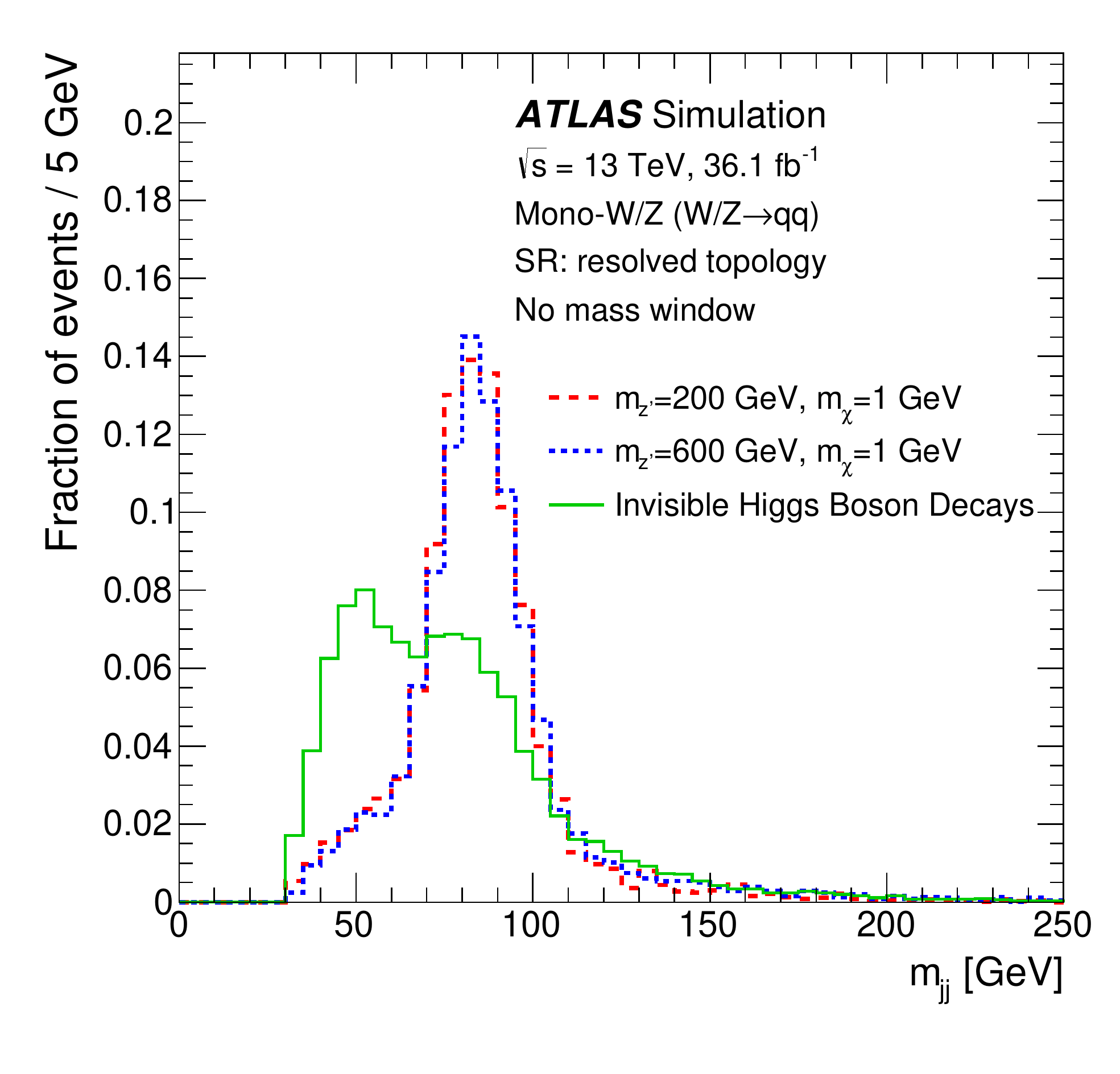}}
\subfigure[]{\includegraphics[width=0.49\textwidth]{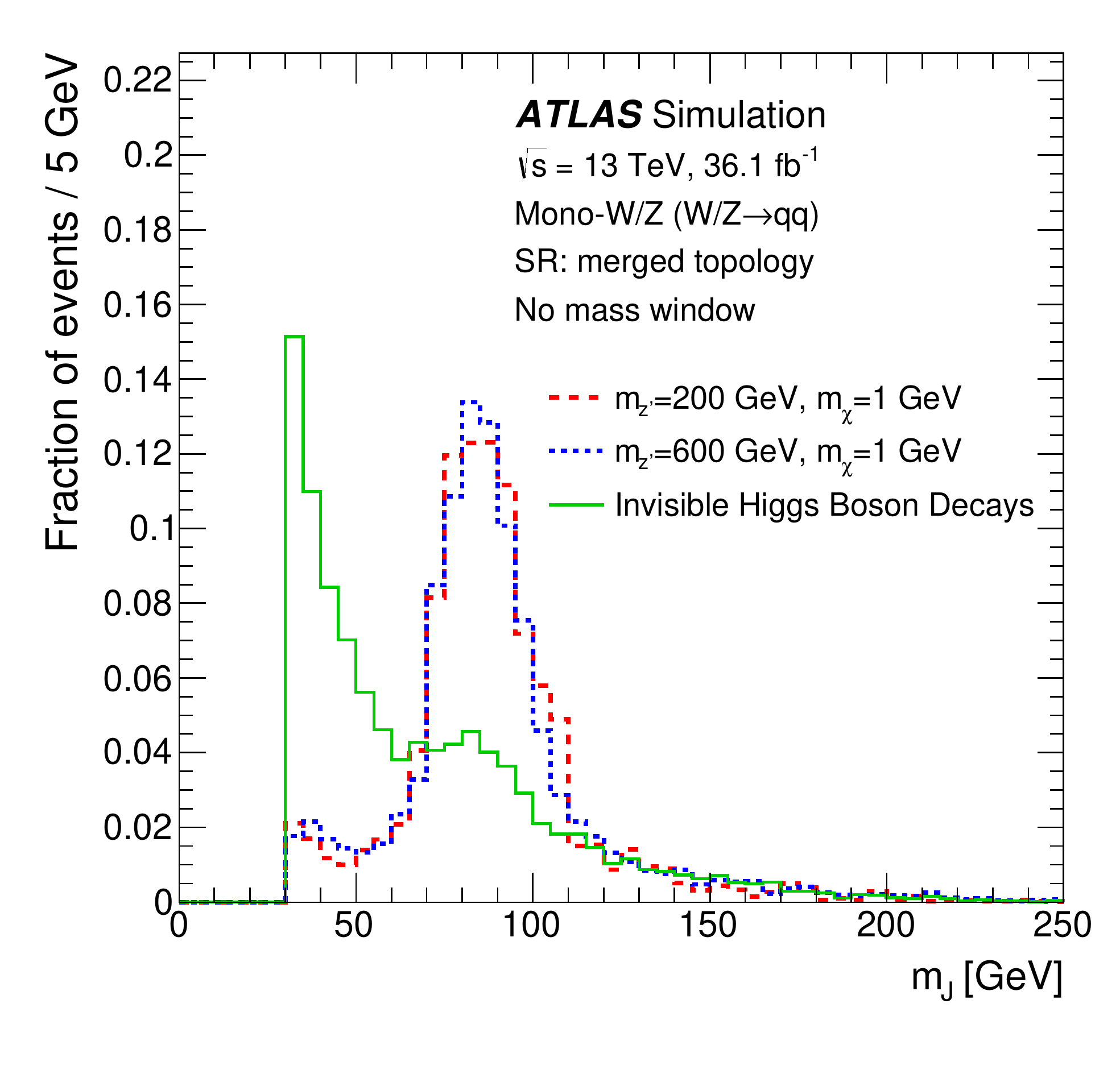}}
\caption{Expected distributions of missing transverse momentum, \met, normalized to unit area, for the simplified vector-mediator model and invisible Higgs boson decays after the full selection in the (a) resolved and (b) merged event topologies, and the expected invariant mass distributions (c) $m_{jj}$ in the resolved and (d) $m_{J}$ in the merged event topologies, before the mass window requirement. The signal contributions from each resolved (merged) category are summed together. The invisible Higgs boson decays include a large contribution from $ggH$ events, which results in the observed mass distribution.
}
\label{fig:signal_distributions_monoWZ}
\end{figure}

\begin{figure}[!htbp]
\centering
\subfigure[]{\includegraphics[width=0.41\textwidth]{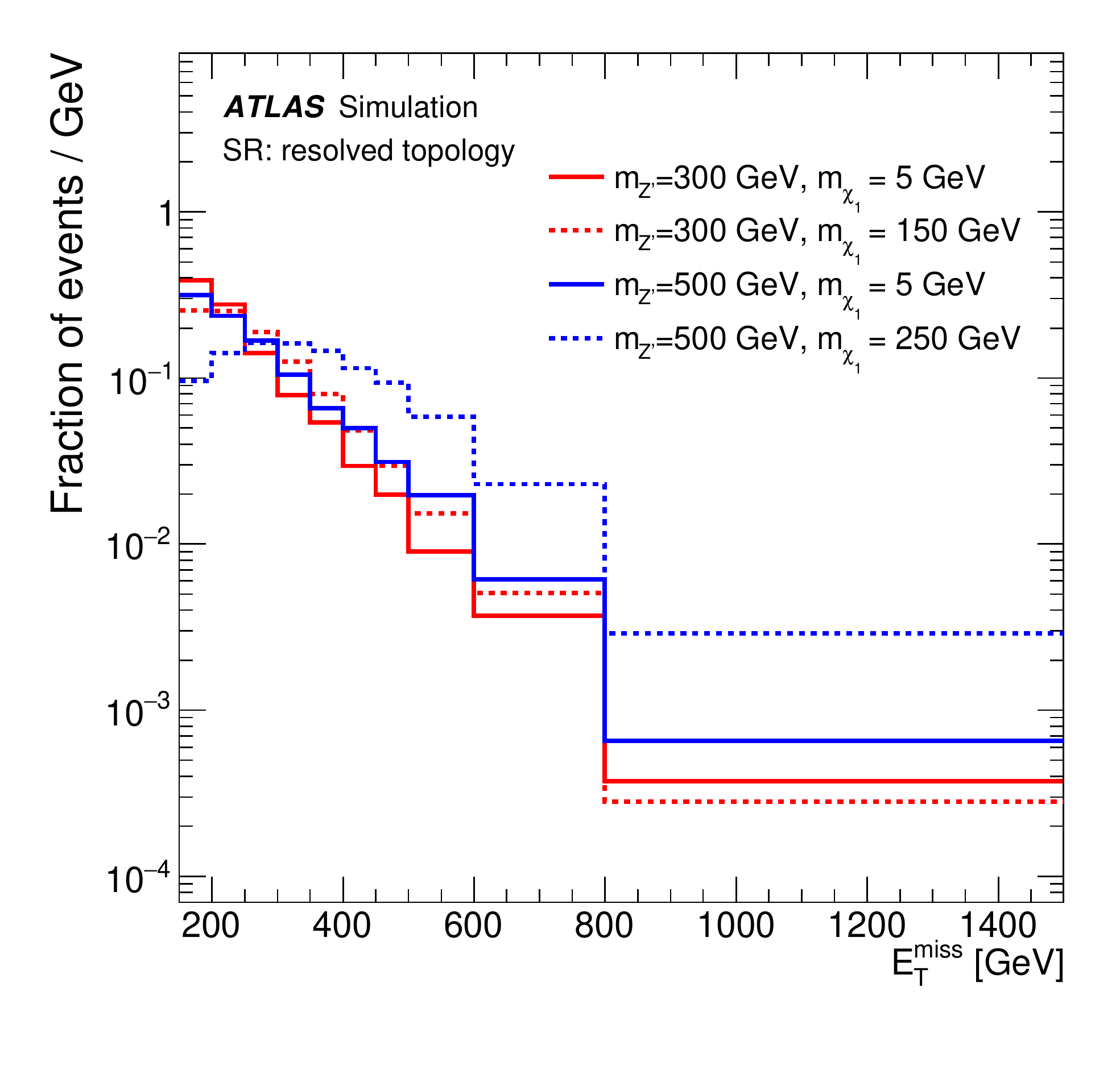}}
\subfigure[]{\includegraphics[width=0.41\textwidth]{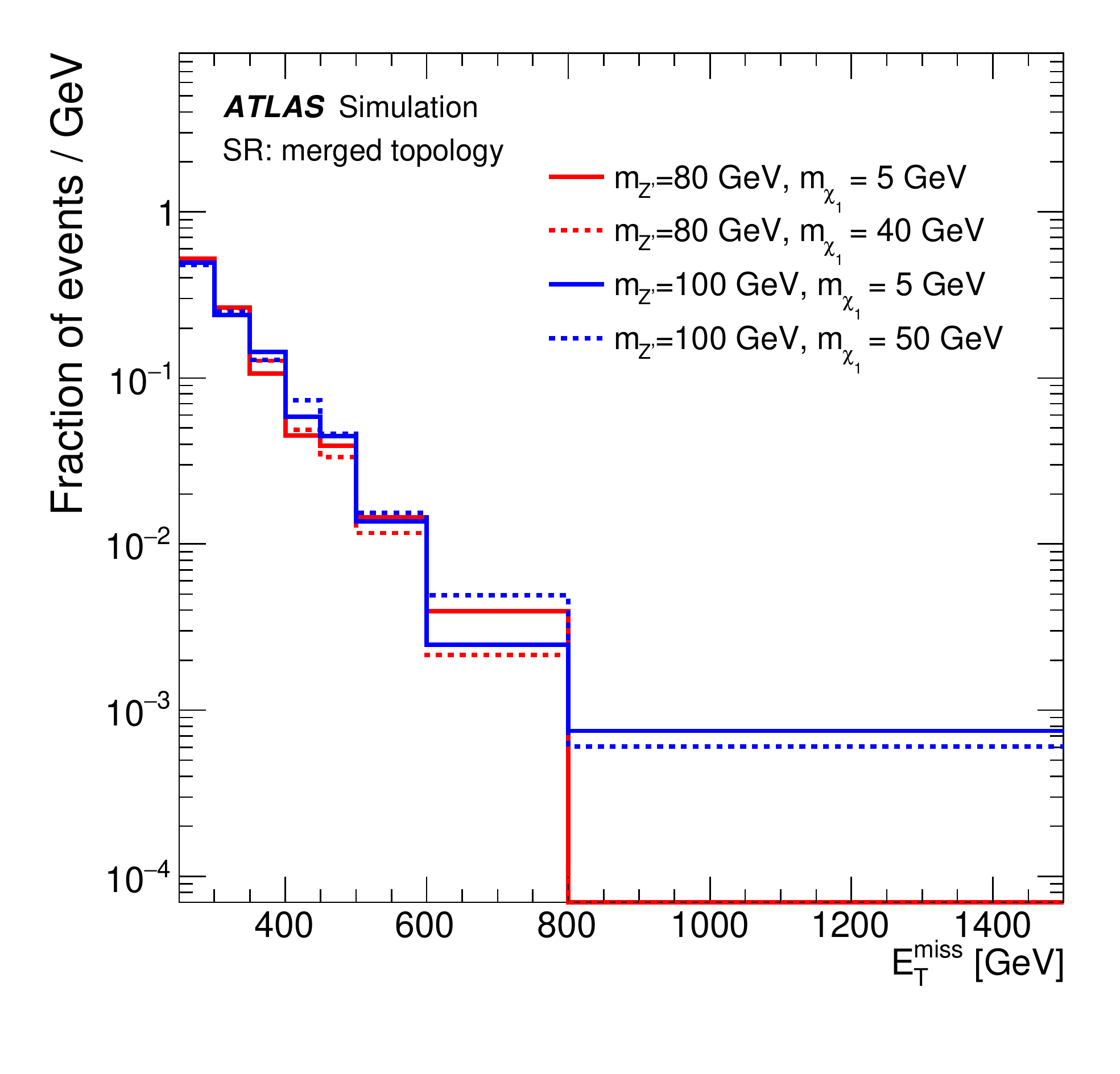}}
\vskip-0.2cm
\subfigure[]{\includegraphics[width=0.41\textwidth]{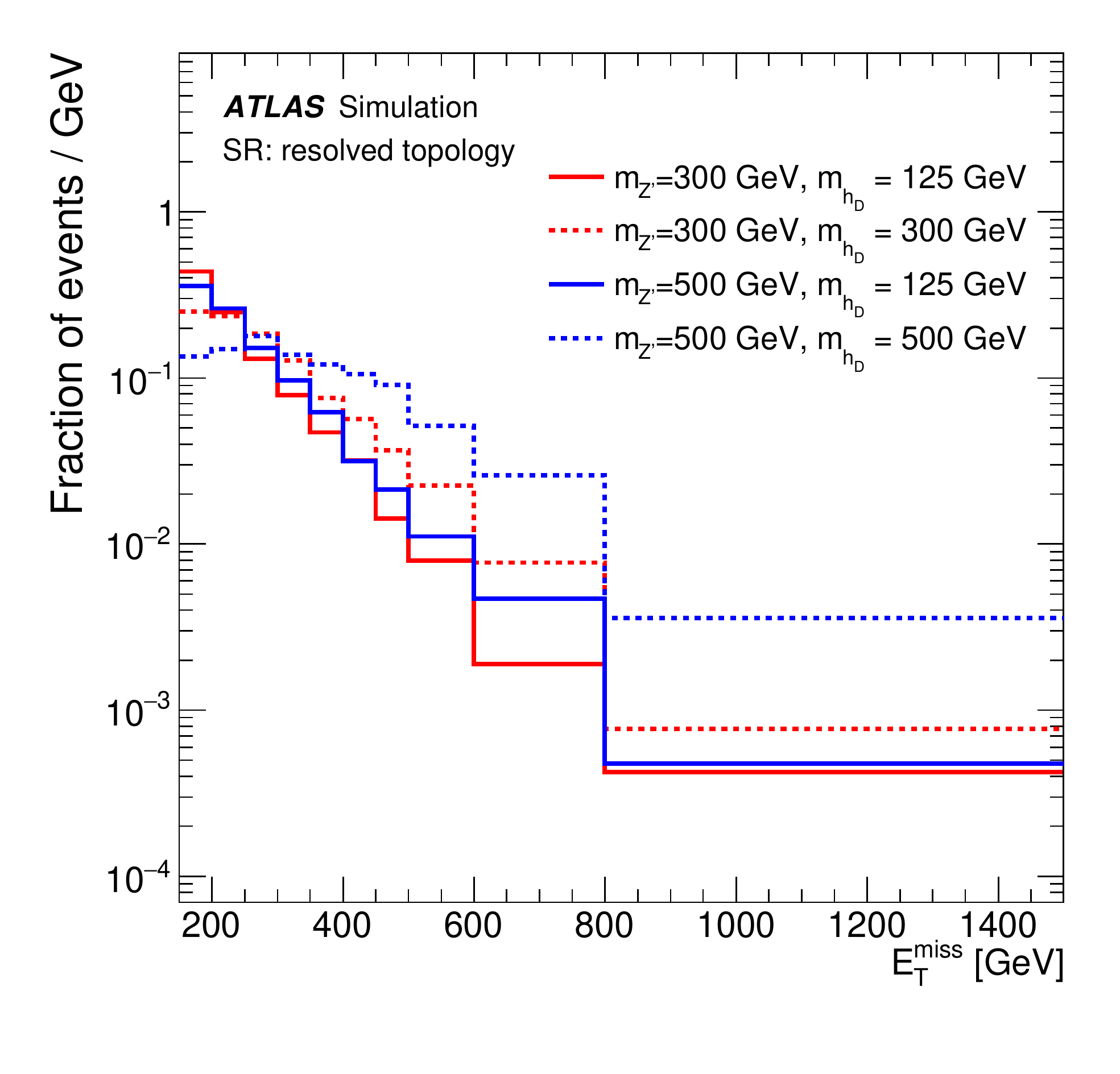}}
\subfigure[]{\includegraphics[width=0.41\textwidth]{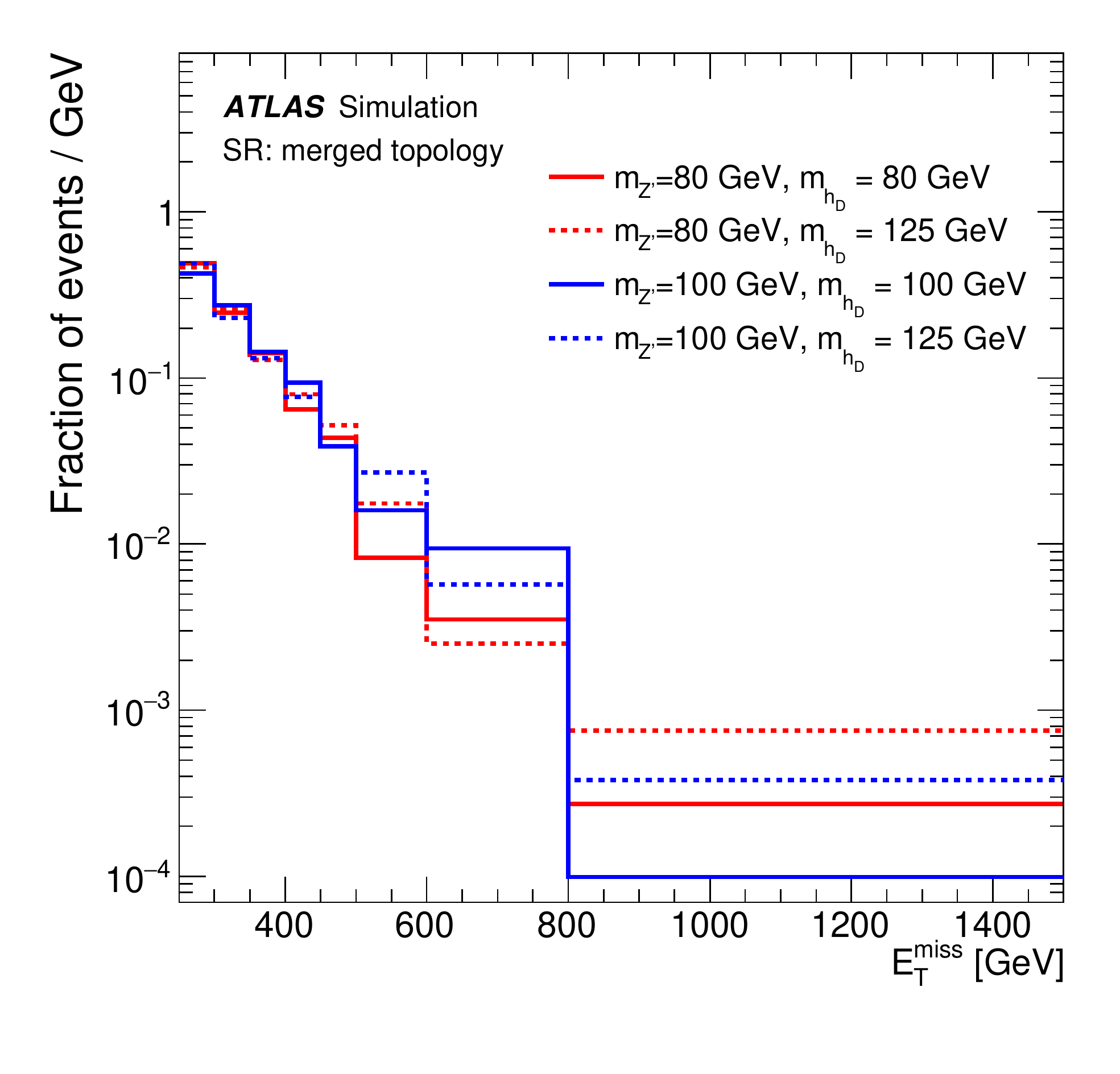}}
\vskip-0.2cm
\subfigure[]{\includegraphics[width=0.41\textwidth]{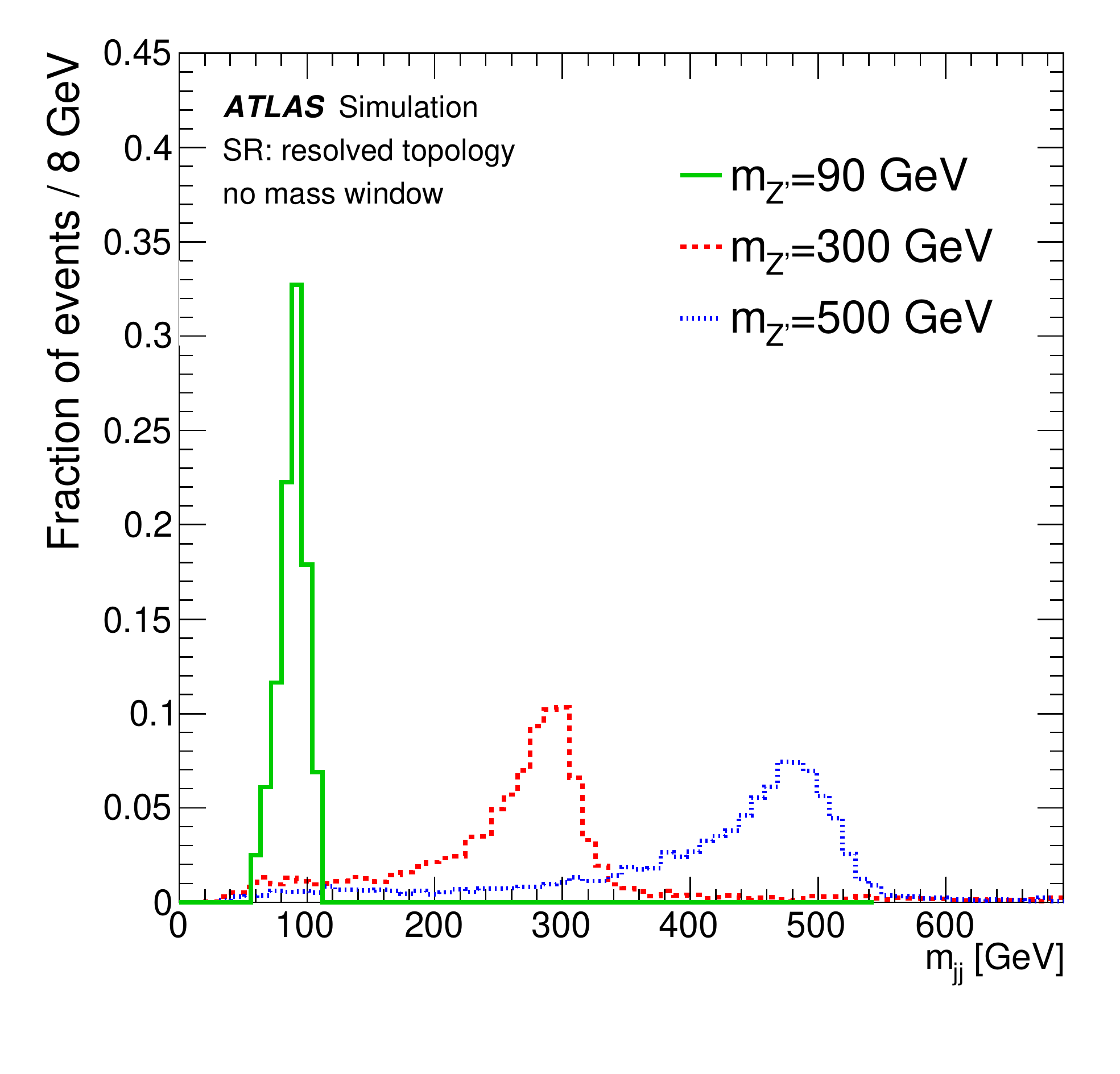}}
\subfigure[]{\includegraphics[width=0.41\textwidth]{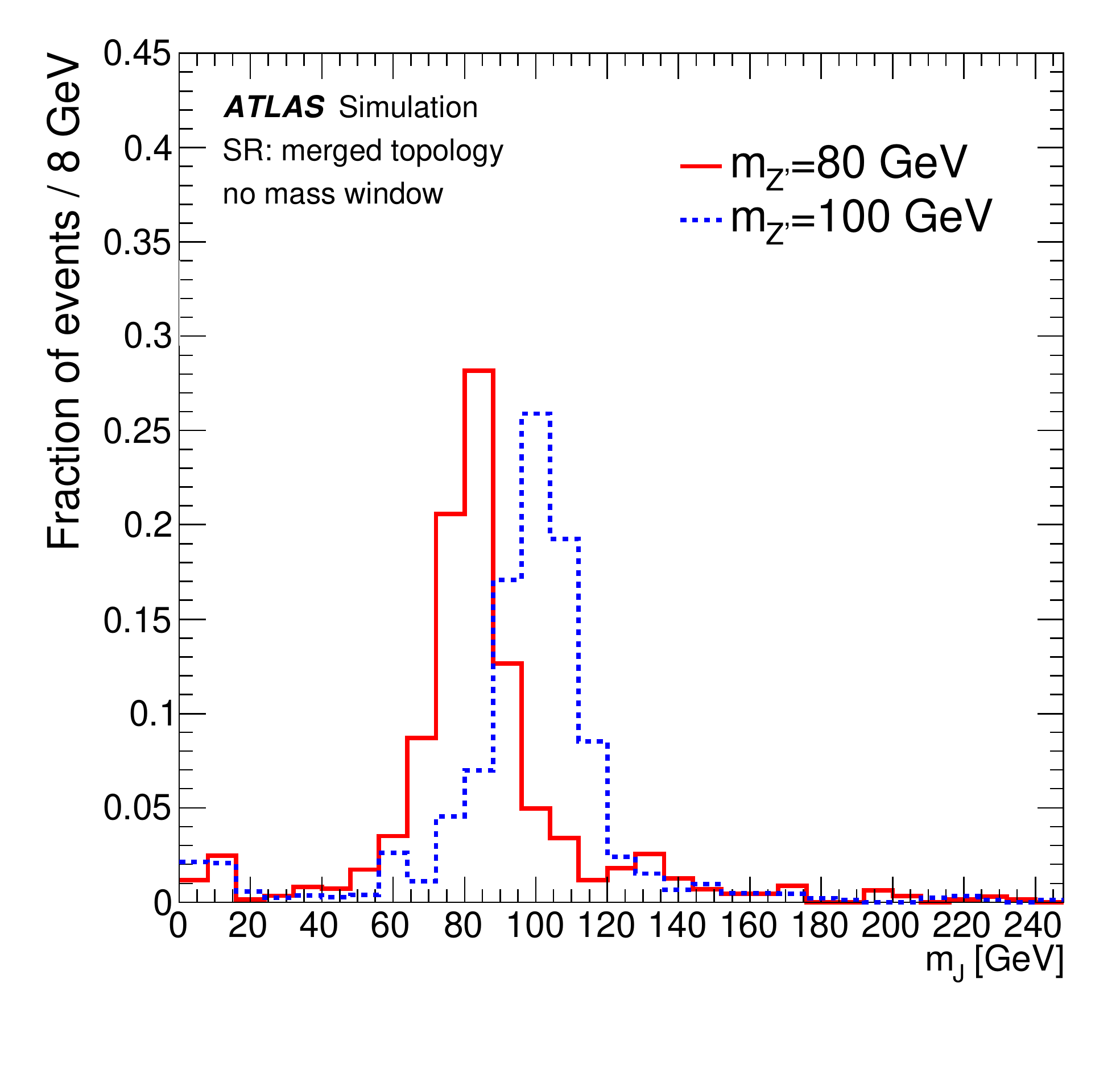}}
\vskip-0.3cm
\caption{Expected distributions of missing transverse momentum, \met, normalized to unit area,  after the full selection for the dark-fermion \monoZprime model in the (a) resolved and (b) merged event topologies, the dark-Higgs \monoZprime model in the (c) resolved and (d) merged event topologies, as well as the expected invariant mass distribution (e) $m_{jj}$ in the resolved and (f) $m_{J}$ in the merged event topologies for the dark-fermion \monoZprime model in the light dark-sector scenario before the mass window requirement. Similar mass distributions are also observed in the simulation of the other \monoZprime models.}
\label{fig:signal_distributions_mzp}
\end{figure}

For both the \monoWZ and the \monoZprime search, the \met~distribution in each event category is used as the final discriminant in the statistical interpretation of the data, since for the models with very large \met values a better sensitivity can be achieved compared to the $V$-candidate mass discriminant. The \met distributions after the full selection, as well as the $m_{J}$ and $m_{jj}$ distributions before the mass window requirement, are shown for various signal models in Figures~\ref{fig:signal_distributions_monoWZ} and~\ref{fig:signal_distributions_mzp}.

\begin{figure}[!htbp]
\centering
\includegraphics[width=0.7\textwidth]{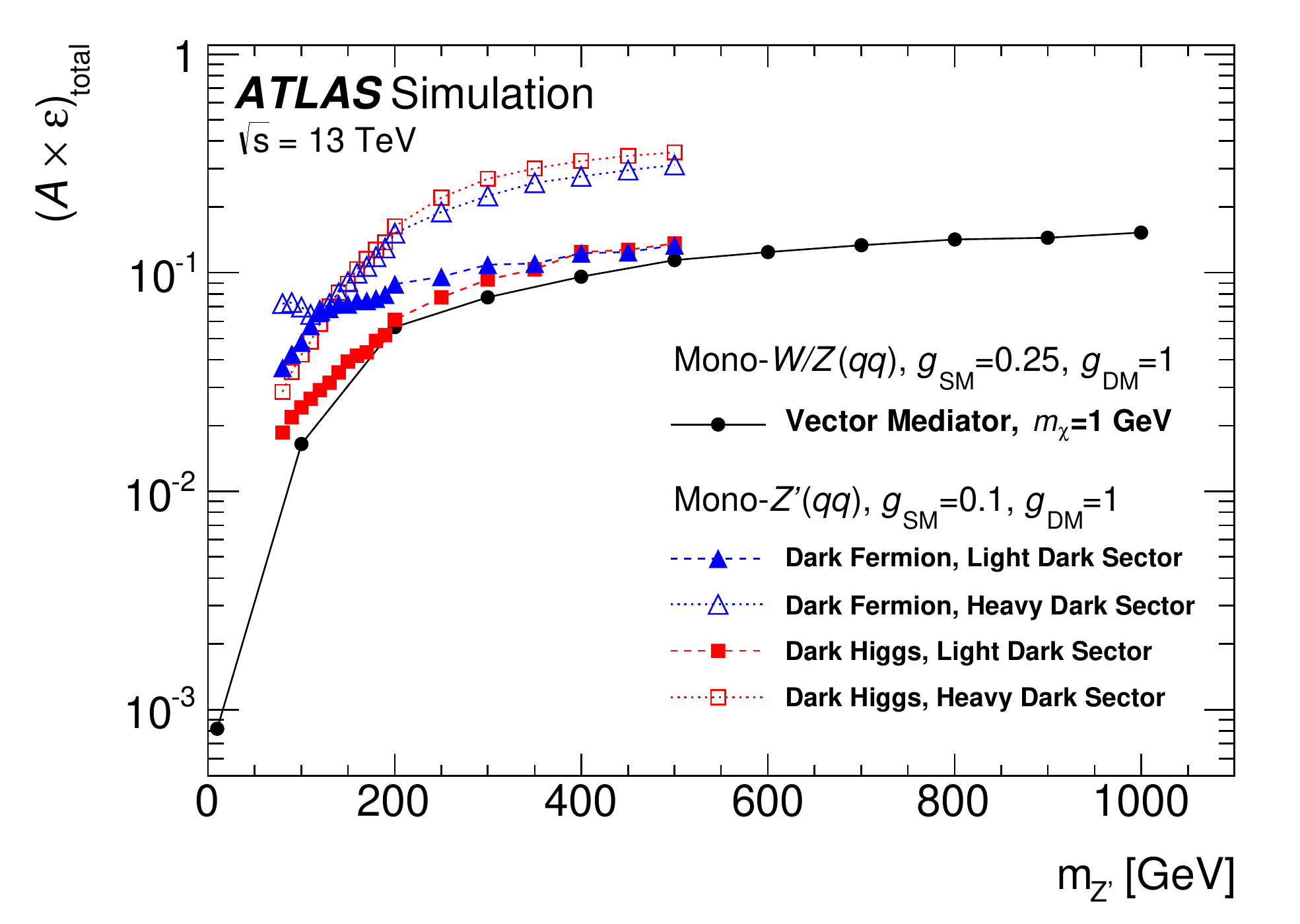}
\caption{ The product of acceptance and efficiency $(\mathcal{A} \times \varepsilon)_{\mathrm{total}}$, defined as the number of signal events satisfying the full set of selection criteria, divided by the total number of generated signal events, for the combined mono-$W$ and mono-$Z$ signal of the simplified vector-mediator model and for the \monoZprime dark-fermion and dark-Higgs signal models, shown in dependence on the mediator mass $m_{Z'}$. For a given model, the signal contributions from each category are summed together. The lines  are drawn to guide the eye.}
\label{fig:ae_total}
\end{figure}

Figure~\ref{fig:ae_total} shows the product $(\mathcal{A} \times \varepsilon)_{\mathrm{total}}$ of the signal acceptance $\mathcal{A}$ and selection efficiency $\varepsilon$ for the simplified vector-mediator model and for the dark-fermion and dark-Higgs \monoZprime signal models after the full event selection. This product is defined as the number of signal events satisfying the full set of selection criteria, divided by the total number of generated signal events.  For all signal models, the main efficiency loss is caused by the minimum \met requirement.

In the simplified vector-mediator model, the $(\mathcal{A} \times \varepsilon)_{\mathrm{total}}$, obtained by summing up signal contributions from all event categories, increases from 1\% for low to 15\% for high mediator mass due to the increase of the missing transverse momentum in the final state.

Similarly, for the  \monoZprime signal models, the  $(\mathcal{A} \times \varepsilon)_{\mathrm{total}}$  increases  with increasing mediator mass from 2\% to 15\% (from a few \% to up to 40\%) in scenarios with a light (heavy) dark sector. The $(\mathcal{A} \times \varepsilon)_{\mathrm{total}}$ for invisible Higgs boson decays is 0.5\% when summing over all signal regions. About 58\% of that signal originates from $ggH$, 35\% from $VH$ and 7\% from VBF production processes, with $(\mathcal{A} \times \varepsilon)_{\mathrm{total}}$ values of 0.3\%, 5.7\% and 0.5\%, respectively. 

The number of signal events in a given signal-region category, relative to the total number of signal events selected in all signal categories, depends on the signal model and mediator mass.
The largest fraction is expected in the $0b$ category with resolved topology, where it ranges from 40\% to 80\%. This is followed by the $0b$-HP and $0b$-LP merged-topology categories with 10\% to 20\% of signal events in each of the two. In the \monoZprime signal models, the $1b$ and $2b$ categories with resolved topology contain about 7\% to 10\% of the total signal contribution. The signal contributions in every other category are below 5\%.

%% file: background.tex
The dominant background contribution in the signal region originates from $t\bar{t}$ and $V$+jets production. In the latter case, the biggest contributions are from decays of $Z$ bosons into neutrinos ($Z\to\nu\nu$) and $W\to\tau\nu$, together with $W\to (e\nu, \mu\nu)$  with non-identified electrons and muons. The normalization of the $t\bar{t}$ and $V$+jets background processes and the corresponding shapes of the final \met~discriminant are constrained using two dedicated background-enriched data control regions with leptons in the final state. The multijet background contribution is estimated by employing additional multijet-enriched control regions. Events in each control region are selected using criteria similar to, while at the same time disjoint from, those in the signal region. Events are also categorized into merged and resolved topologies, each divided into three categories with different $b$-tagged jet multiplicities. No requirement is imposed on the large-$R$ jet substructure or $\Delta R_{jj}$ and therefore there is no further classification of the merged-topology events into low- and high-purity control regions, as is the case for the signal regions. The remaining small contributions from diboson and single-top-quark production are determined from simulation.

The two control regions with one and two leptons in the final state are defined to constrain the $W$+jets and $Z$+jets background respectively, together with the $t\bar{t}$ contribution in the one lepton control region. The latter process is dominant in $2b$ control-region categories. The {\emph{one-lepton control region}} is defined by requiring no `loose' electrons and exactly one muon with `medium' identification, $\pt>$~25~\GeV\ and satisfying `tight'  isolation criteria. Events are collected by \met triggers, as these triggers enhance most efficiently contributions from events with a signal-like topology. The {\emph{two-lepton control region}} uses events passing a single-lepton trigger. One of the two reconstructed leptons has to be matched to the corresponding trigger lepton. A pair of `loose' muons or electrons with invariant dilepton mass $66~\GeV <m_{\ell\ell}<116$~\GeV\ is required in the final state. At least one of the two leptons is required to have $\pt>$~25~\GeV\ and to satisfy the stricter `medium' identification criteria. To emulate the missing transverse momentum from non-reconstructed leptons (neutrinos) in $W$ ($Z$) boson decays, the $\metnolep$ and $\mptnolep$ variables are used instead of $\met$ and $\mpt$, respectively, for the event selection in the one-lepton and two-lepton control regions. The $\metnolep$ distribution is employed in the statistical interpretation as the final discriminant in these control regions. The control-region data are also used to confirm the good modelling of other discriminant variables such as the invariant mass of the vector boson candidate and the large-$R$ jet substructure variable \DTwoBetaOne in events with signal-like topology.

The multijet background contribution is estimated separately for each signal region category from a {\emph{multijet control region}} selected by inverting the most effective requirement used to discriminate against multijet events in the signal region, i.e.\ by requiring $\min [\Delta\phi(\metvec, j)] \equiv \min [\Delta\phi] < 20^{\textrm{o}}$. The $\met$ distribution observed in this region is used as an expected multijet background shape after a simulation-based subtraction of a small contribution from non-multijet background. To account for the inversion of the $\min [\Delta\phi]$ requirement, the distribution is scaled by the corresponding normalization scale factor. This normalization scale factor is determined in an equivalent control region, but with both the $\min [\Delta\phi]$  and $\Delta\phi(\metvec,\mptvec)$ requiremens removed and the mass window criterion inverted to select only events in the mass sidebands. In this new control region, the $\met$ distribution from events with $\min [\Delta\phi]<20^{\textrm{o}}$ is fitted to the data with $\min [\Delta\phi]>20^{\textrm{o}}$, together with other background contributions, and the resulting normalization factor is applied to the $\met$ distribution from the multijet control region. For the \monoWZ search, the high-mass sideband is used, ranging from the upper mass window bound to 250~\GeV. Since $\Delta R_{jj}$ and $\Delta\phi_{jj}$ criteria are not applied in the \monoZprime search, the event topology in the high-mass sideband is in general not close enough to the topology of the signal region. Therefore, the low-mass sideband is used for the estimate of the multijet contribution in the \monoZprime search. The sideband mass range depends on the mass of the $Z'$ boson: the upper sideband bound is set to the lower bound of the signal region mass window and the size of the sideband is the same as the size of the mass window in the signal region. The multijet contribution is estimated to contribute up to a few percent of the total background yield depending on the signal category. The contribution from the multijet background in the one-lepton and two-lepton control regions is negligible.

For the \monoWZ searches, all background contributions are additionally constrained by the {\emph{mass sideband regions}} in the zero-lepton final state. These regions are defined by the same selection criteria as introduced in Section~\ref{sec:selection}, except for the requirements on the large-$R$ jet and dijet mass values, which are required to be above the signal mass window and below 250~\GeV. Events in this region are topologically and kinematically very similar to those in the full signal region, with a similar background composition. The corresponding sideband regions are also introduced for the one-lepton and the two-lepton control regions. While there is no signal contamination expected in the one-lepton and two-lepton control regions, the signal contribution in the zero-lepton mass sideband region is not negligible. Compared to the total signal contribution in the signal region described in the previous section, about 20\% of additional signal events are expected in the sidebands in the case of the simplified vector-mediator model. For the invisible Higgs boson decays, the original signal contribution is increased by about 35\% after including the sideband region, dominated by the $ggH$ production process. No sideband regions are employed for the \monoZprime searches. Since the hypothesized mass of the $Z'$ boson is a free parameter, the zero-lepton sideband regions cannot be considered free from signal contamination.

The final estimate of background contributions is obtained from a simultaneous fit of the expected final discriminants to data in all signal, sideband and control regions (see Section~\ref{sec:results}). The signal contributions in the mass sideband regions are taken into account in the fit.

%% file: systematics.tex
Several experimental and theoretical systematic uncertainties affect the results of the analysis. Their impact is evaluated in each bin of an \met distribution. In this section, the impact of different sources of uncertainty on the expected signal and background yields is summarized, while the overall impact on the final results is discussed in the next section. 

Theoretical uncertainties in the signal yield due to variations of the QCD renormalization and factorization scale, uncertainties in the parton distribution functions, and the underlying event and parton shower description, are estimated to be about 10--15\% for the simplified vector-mediator model. For the invisible decays of the Higgs boson produced via $VH$ and $ggH$ processes, the theory uncertainties affect the signal yields by 5\% and 10\% respectively for the resolved event topology and are about two times larger for the merged topology. 
%The MC description for light-flavour jets is confirmed to be consistent with observed data within this uncertainty using $Z$+jets events, which have a similar topology to the $ggH$ signal with gluon jets.
%The fake rate of tagging hadronic $W/Z$ decays in MC was confirmed within this uncertainty using $Z$+jets events, which have a similar topology to the $ggH$ signal with gluon jets. 
No systematic uncertainty in the VBF signal is considered, since it has a negligible impact on the final results. No theoretical uncertainty is considered for the \monoZprime signals, since it is negligible compared to the experimental uncertainties.

A number of theoretical modelling systematic uncertainties are considered for the background processes, affecting mostly the expected shape of the \met distribution. These uncertainties are estimated following the studies of Ref.~\cite{HIGG-2016-29} and are briefly summarized here. The uncertainties in the $V$+jets background contribution come mainly from limited knowledge of the jet flavour composition in terms of the $V$+HF categorization introduced in Section~\ref{sec:objects}, as well as the modelling of the vector boson transverse momentum (\ptv) and dijet mass (\mjj) distributions. The former are evaluated by means of scale variations in the generated {\textsc{Sherpa}} samples. In addition, the difference between the {\textsc{Sherpa}} nominal sample and an alternative {\textsc MadGraph5\textunderscore aMC@NLO}~v2.2.2 sample produced with a different matrix-element generator is added in quadrature to
yield the total uncertainty. The uncertainty in the modelling of the \ptv~and \mjj~distributions is obtained from the comparison of simulated events with dedicated control-region data, as well as comparisons with alternative generator predictions. For $t\bar{t}$ production, uncertainties in the shapes of the top-quark transverse momentum distribution, and the \mjj~ and \ptv~distributions of the $V$ boson candidate, are considered by comparing the nominal simulated sample to alternative samples with different parton shower, matrix element generation and tuning parameters. A similar procedure is applied for the diboson and single-top-quark backgrounds. While the overall $V$+jets and  $t\bar{t}$ normalization is determined from the fit to data, the comparison between different generators is also employed to assign a normalization uncertainty to single-top-quark and diboson production since their contributions are estimated from simulation.

An uncertainty of 100\% is assigned to the multijet normalization in both the \monoWZ and \monoZprime searches due to the statistical uncertainty in the control data, the impact of non-multijet background and the extrapolation from multijet control regions to signal regions. The shapes of the multijet background distributions are subject to an uncertainty of the order of 10\%, depending on the amount of non-multijet background in each signal region.

In both the \monoWZ and \monoZprime searches, the largest source of experimental systematic uncertainty in the merged topology is the modelling of the large-$R$ jet properties. The large-$R$ jet mass scale and resolution uncertainty~\cite{PERF-2012-02, ATLAS-CONF-2015-035, ATLAS-CONF-2017-064} has an impact of up to 5\% on the expected background yields, and up to 5\%, 10\% and 15\% on the signal yields from invisible Higgs boson decays, the simplified vector-mediator model and \monoZprime models respectively. The uncertainty in the large-$R$ jet energy resolution affects the simplified vector-mediator signal by 3\% and background by 1\%. The impact on the \monoZprime signal and the signal from invisible Higgs boson decays is at the sub-percent level. The uncertainty in the scale of the \DTwoBetaOne substructure parameter affects the migration between the high-purity and low-purity regions, with a 5--10\%  (2--5\%) impact on the background (\monoWZ and \monoZprime signal) yields. The combined impact of all other large-$R$ jet uncertainties is below a few percent. The combined impact of large-$R$ jet uncertainties on events within the resolved-topology categories is negligible for the \monoWZ search and below 2\% for the \monoZprime searches. The small-$R$ jet uncertainties are dominated by the energy scale and resolution uncertainties. The small-$R$ jet energy scale uncertainty has an up to 10\% (up to 6\%) impact on the background (signal) yields. The uncertainty in the small-$R$ jet energy resolution has a 2--5\% impact on the signal yields. The corresponding impact of this uncertainty on the background yield is at a sub-percent level in the mass window around the $W$- and $Z$-boson mass, growing to around 1.5\% for the \monoZprime search in the mass window around $m_{Z'}=500$~\GeV. The $b$-tagging calibration uncertainty affects the migration of signal and background events between categories with different $b$-tag multiplicities by up to 10\%. The uncertainty in the missing transverse momentum component which is not associated with any of the selected objects with high transverse momentum affects the background (signal) yields by about 1--3\% (2--10\%). The uncertainties in the trigger efficiency, lepton reconstruction and identification efficiency, as well as the lepton energy scale and resolution, affect the signal and background contributions only at a sub-percent level. 

The uncertainty in the combined 2015+2016 integrated luminosity is 2.1\%. It is derived, following a methodology similar to that detailed in Ref.~\cite{DAPR-2013-01}, from a calibration of the luminosity scale using $x$--$y$ beam-separation scans performed in August 2015 and May 2016.

%% file: results.tex
\subsection{Statistical interpretation}

A profile likelihood fit~\cite{Cowan:2010js} is used in the interpretation of the
data to search for dark matter production. The likelihood function used to fit the data is defined as the product of conditional probabilities $P$ over binned distributions of discriminating observables in each event category $j$,
\begin{equation*}
\begin{split}
& \mathcal{L}(\mu, \vec{\theta}) =  \displaystyle  
\prod_{j}^{N_{\mathrm{categories}}} \prod_{i}^{N_{\mathrm{bins}}} 
 P\left(N_{ij} ~|\mu S_{ij}(\vec{\theta}) ~+ B_{ij}(\vec{\theta}) \right)   
\prod_{k}^{N_{\mathrm{nuisance}}} \mathcal{G}({\theta_{k}}) \;.
\end{split}
\label{eq:likelihood_simplified}
\end{equation*}
The likelihood function depends on the signal strength $\mu$, defined as the signal yield relative to the prediction from simulation, and on the vector of nuisance parameters $\vec{\theta}$ accounting for the background normalization and systematic uncertainties introduced in Section~\ref{sec:sys}. The Poisson distributions $P$ correspond to the observation of $N_{ij}$ events in each bin $i$ of the discriminating observable given the expectations for the background, $B_{ij}(\vec{\theta})$, and for the signal, $S_{ij}(\vec{\theta})$. A constraint on a nuisance parameter $\theta_k$ is represented by the Gaussian function $\mathcal{G}({\theta_{k}})$. The correlations between nuisance parameters across signal and background processes and categories are taken into account.

For the \monoWZ search, the event categories include all eight zero-lepton signal regions (see Section~\ref{sec:selection}), six one-lepton and six two-lepton control regions, as well as the corresponding sideband regions for each of these twenty categories (see Section~\ref{sec:BG}). In comparison, no sideband regions are employed for the \monoZprime search and only categories with the resolved topology are considered for $m_{Z'}>$~100~\GeV.  In the zero-lepton signal and sideband regions, the $\met$ distribution is used as the discriminating variable since the signal process results in relatively large $\met$ values compared to the backgrounds. In order to constrain the backgrounds and the $\met$ shape in the signal region, the $\metnolep$ variables are used in the fit in the one- and two-lepton control regions. The normalizations of the $W$+HF, $W$+LF, $Z$+HF, $Z$+LF and $t\bar{t}$ background components are treated as unconstrained parameters in the fit, independent from each other and correlated across all event categories. The uncertainties in the flavour composition of the $V$+HF processes are taken into account following the studies outlined in Section~\ref{sec:sys}. The normalization of other background components is constrained according to their theory uncertainty. A possible difference between the normalization factors in events with resolved and merged topologies for the $W$+jets, $Z$+jets and $t\bar{t}$ processes due to systematic modelling effects is taken into account by means of two additional constrained nuisance parameters. The multijet contribution is only considered in the signal regions and the corresponding mass sidebands, with uncorrelated normalization factors in each category. 

%
%%%%%%%%%%%%%%%%%%%%%%%%%%%%%%%%%%
%  SIGNAL REGION TABLES AND PLOTS
%%%%%%%%%%%%%%%%%%%%%%%%%%%%%%%%%%
%
%
%%%%%%%%%%%%%%%%%%
% mono W/Z tables
%%%%%%%%%%%%%%%%%%
%
%

\subsection{Measurement results}

\begin{figure}[!htbp]
\centering
\subfigure[]{\includegraphics[width=0.4\textwidth]{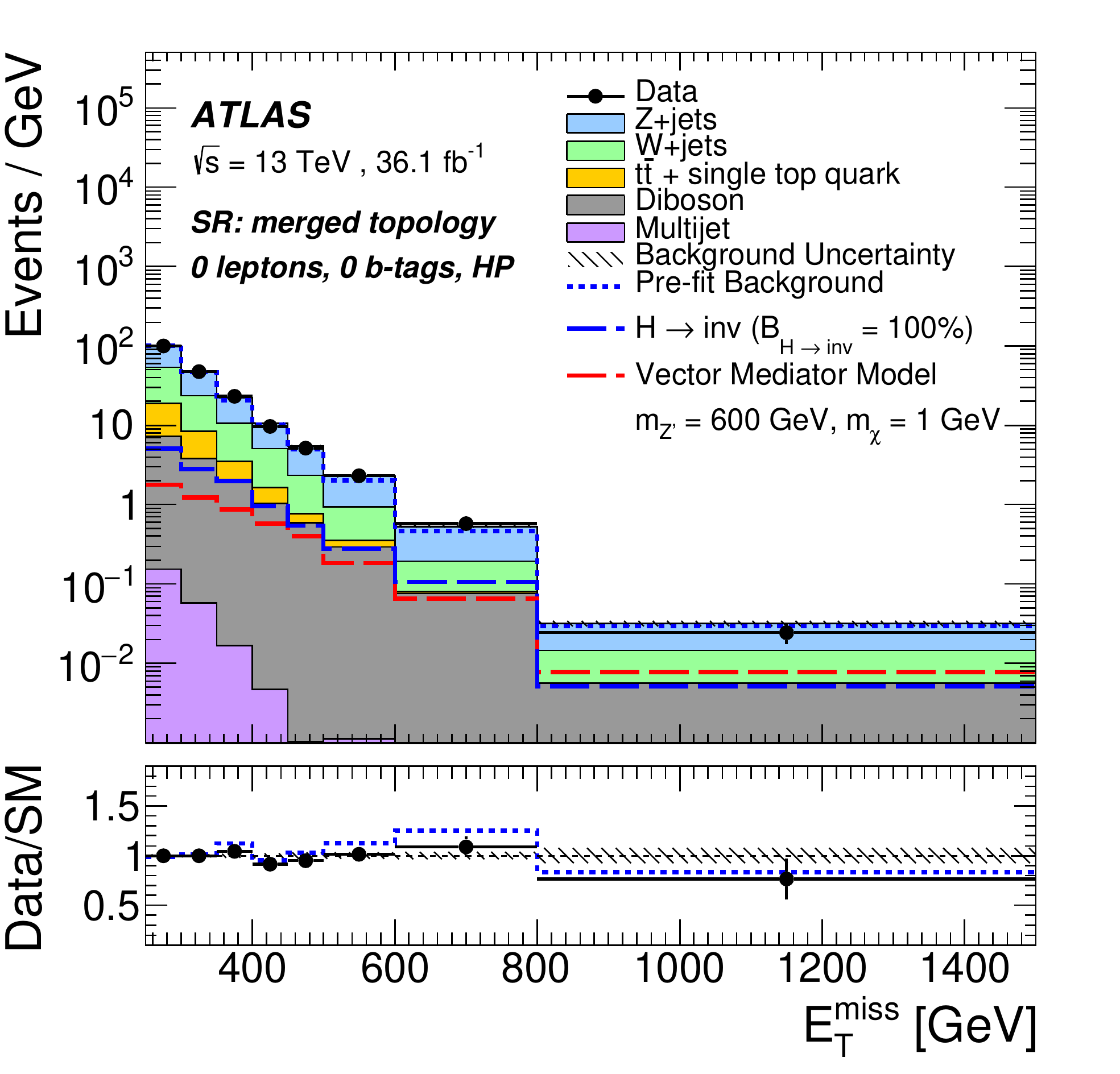}}   \subfigure[]{\includegraphics[width=0.4\textwidth]{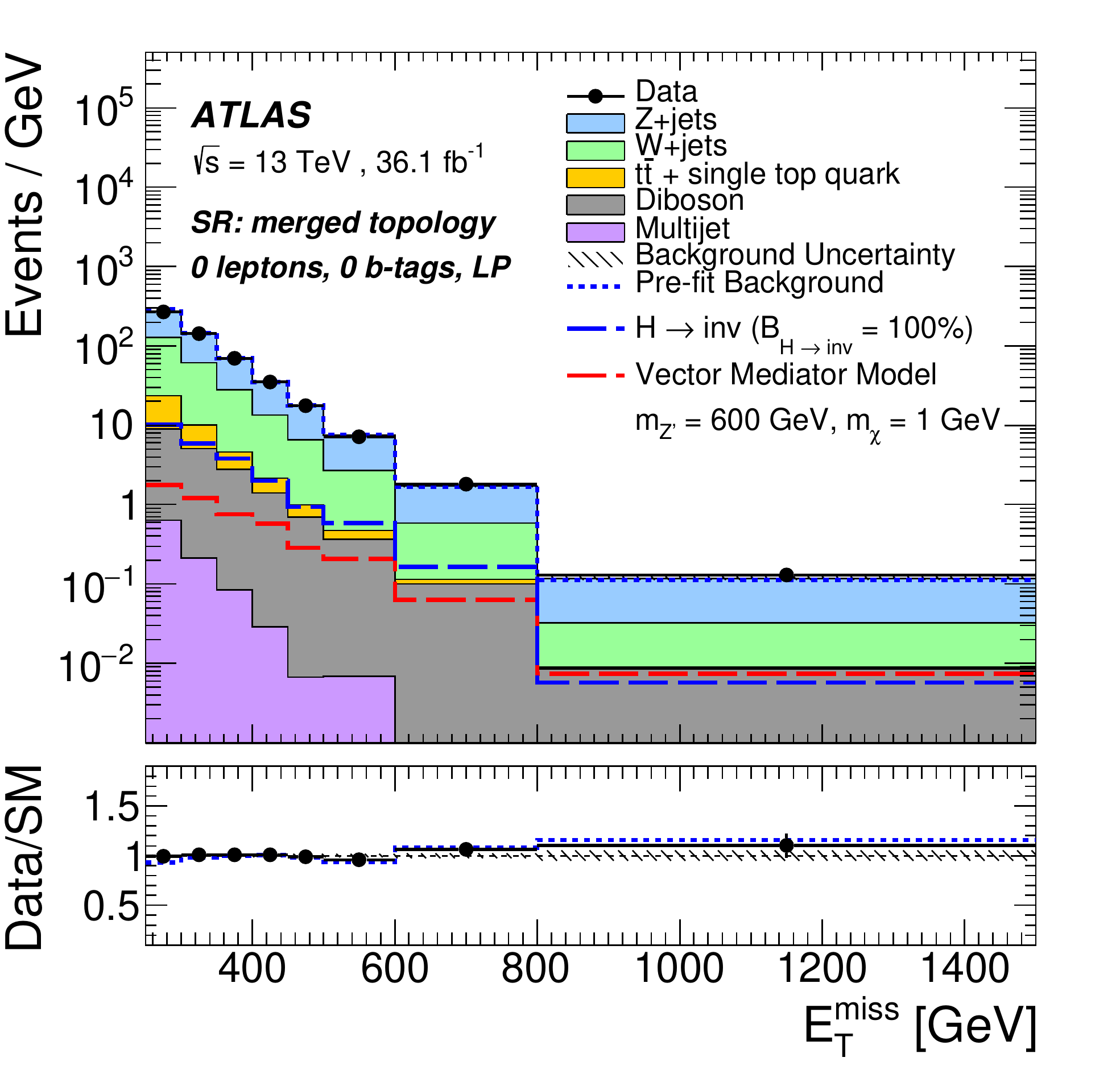}}  
\vskip-0.4cm
\subfigure[]{\includegraphics[width=0.4\textwidth]{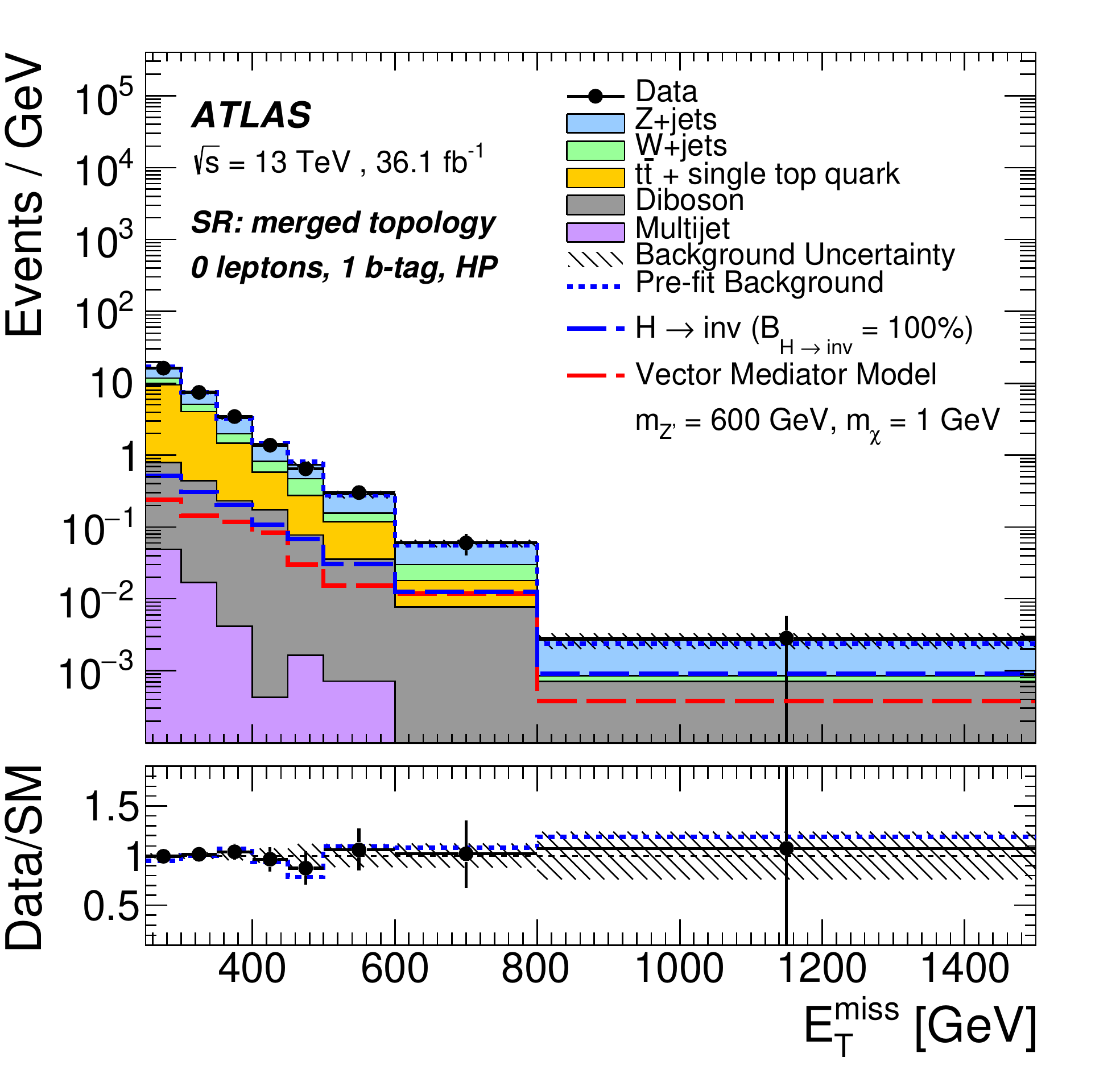}}   \subfigure[]{\includegraphics[width=0.4\textwidth]{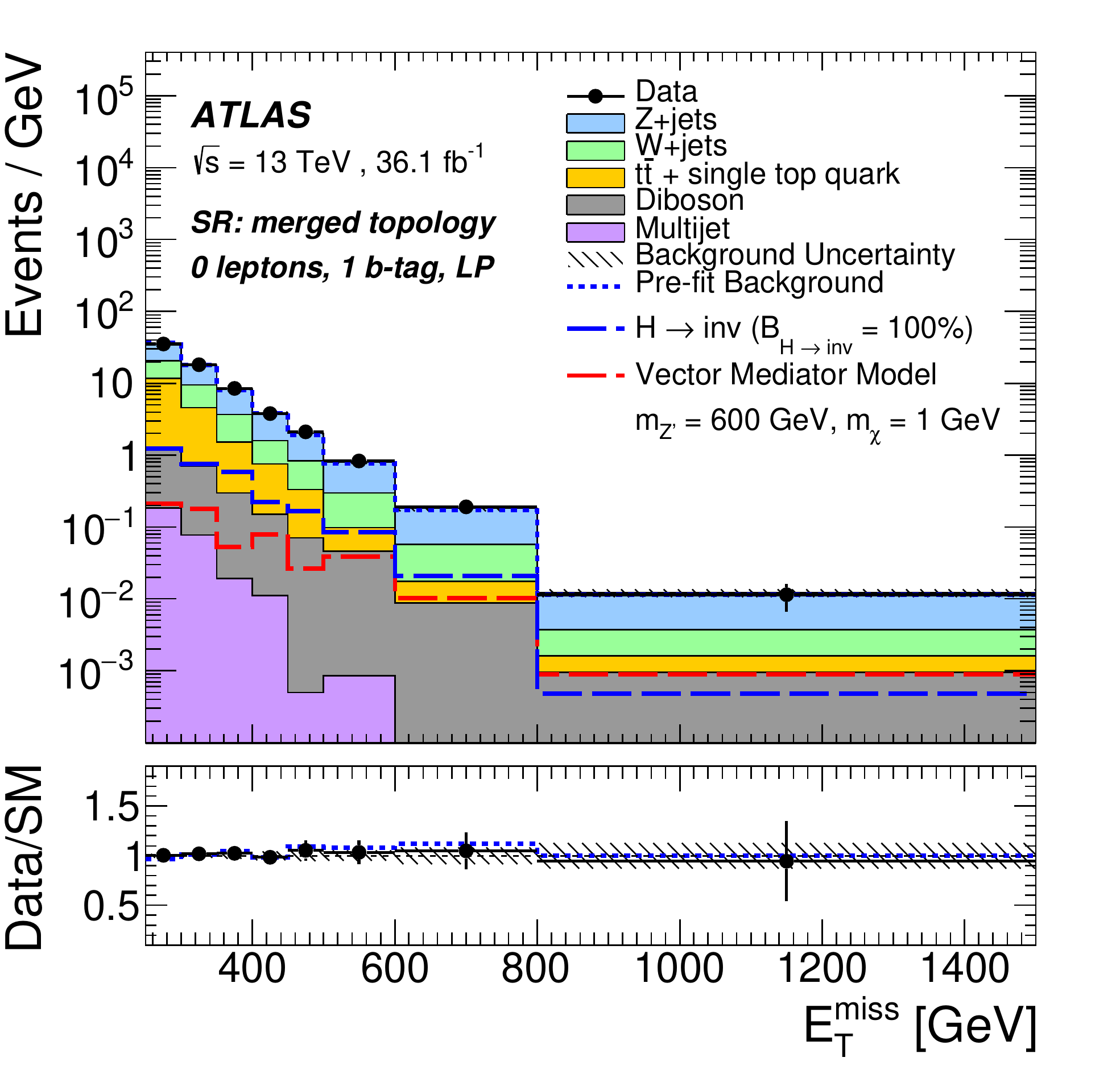}}   
\vskip-0.25cm
\subfigure[]{\includegraphics[width=0.4\textwidth]{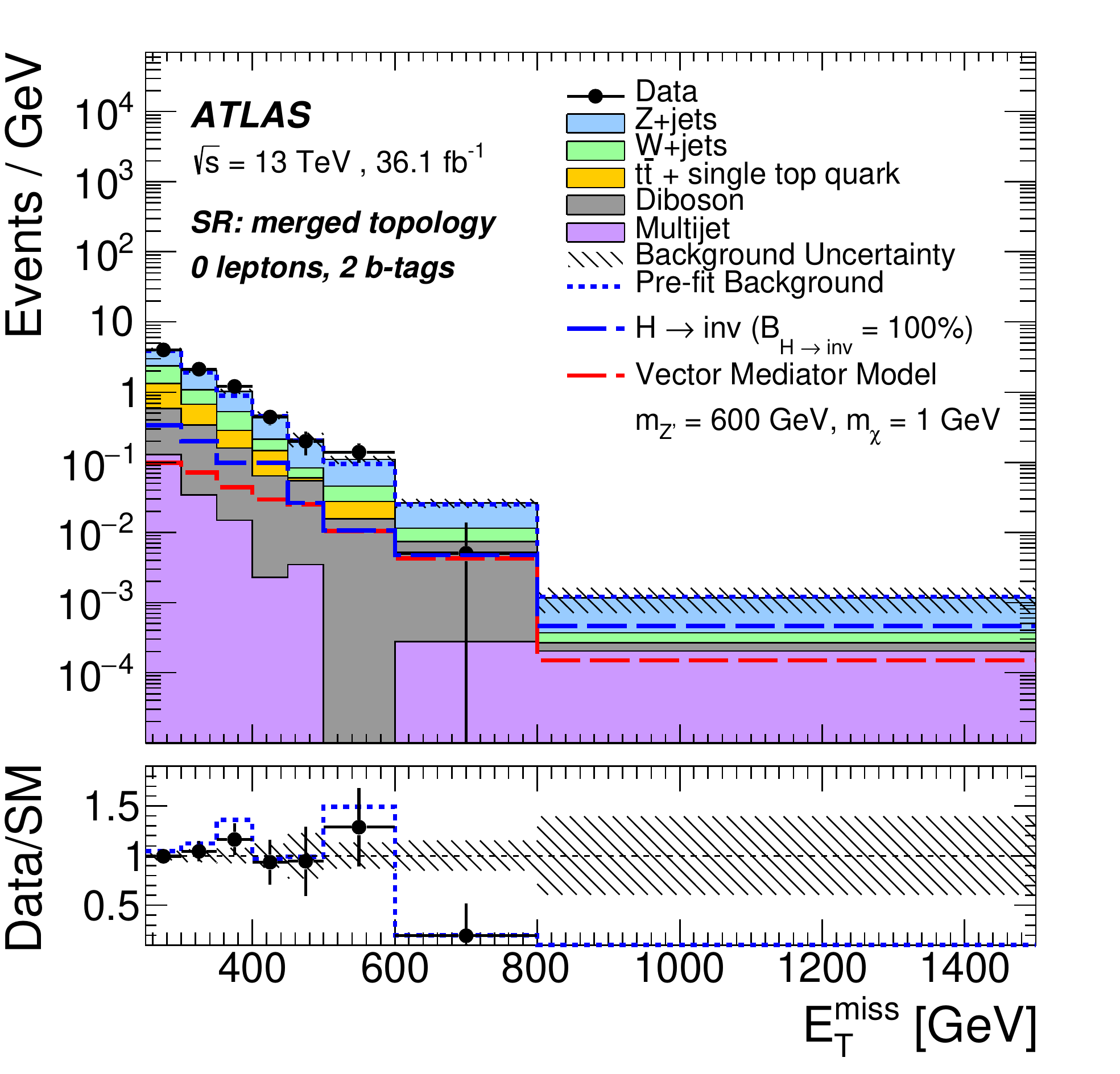}}      
\vskip-0.5cm
\caption{
\label{fig:postfit_signal_merged}
The observed (dots) and expected (histograms) distributions of missing transverse momentum, $\met$, obtained with 36.1~\ifb of data at $\sqrt{s}=$~13~\TeV\ in the \monoWZ signal region with the merged event topology after the profile likelihood fit (with $\mu = 0$), shown separately for the (a) $0b$-HP, (b) $0b$-LP, (c) $1b$-HP, (d) $1b$-LP, and (e) $2b$-tag event categories. The total background contribution before the fit to data is shown as a dotted blue line. The hatched area represents the total background uncertainty. The signal expectations for the simplified vector-mediator model with $m_{\chi}=1$~\GeV\ and $m_{Z'}=600$~\GeV\ (dashed red line) and for the invisible Higgs boson decays (dashed blue line) are shown for comparison. The inset at the bottom of each plot shows the ratio of the data to the total post-fit (dots) and pre-fit (dotted blue line) background expectation. 
}
\end{figure}

\begin{figure}[!htbp]
\centering
\subfigure[]{\includegraphics[width=0.48\textwidth]{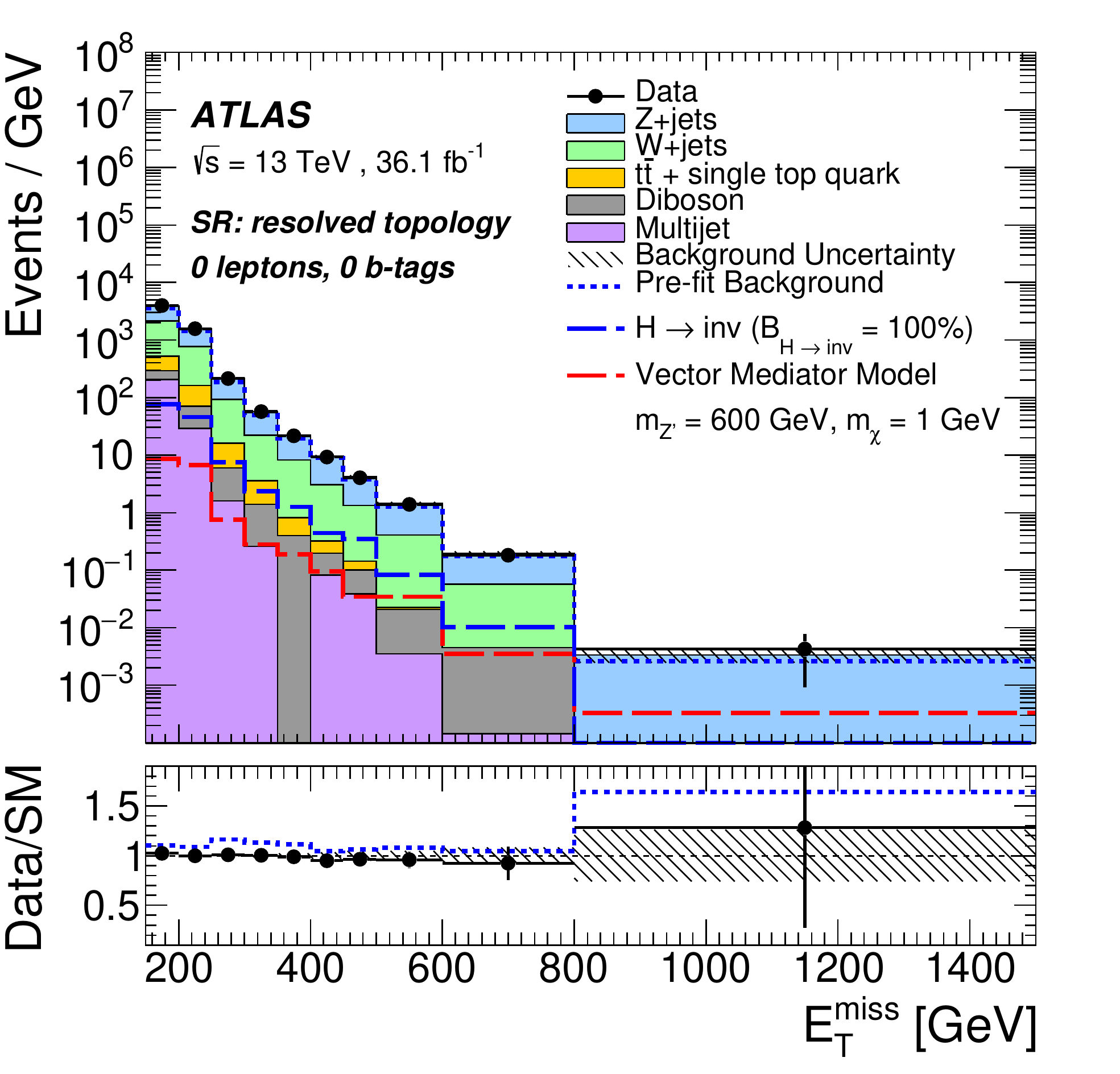}}      
\subfigure[]{\includegraphics[width=0.48\textwidth]{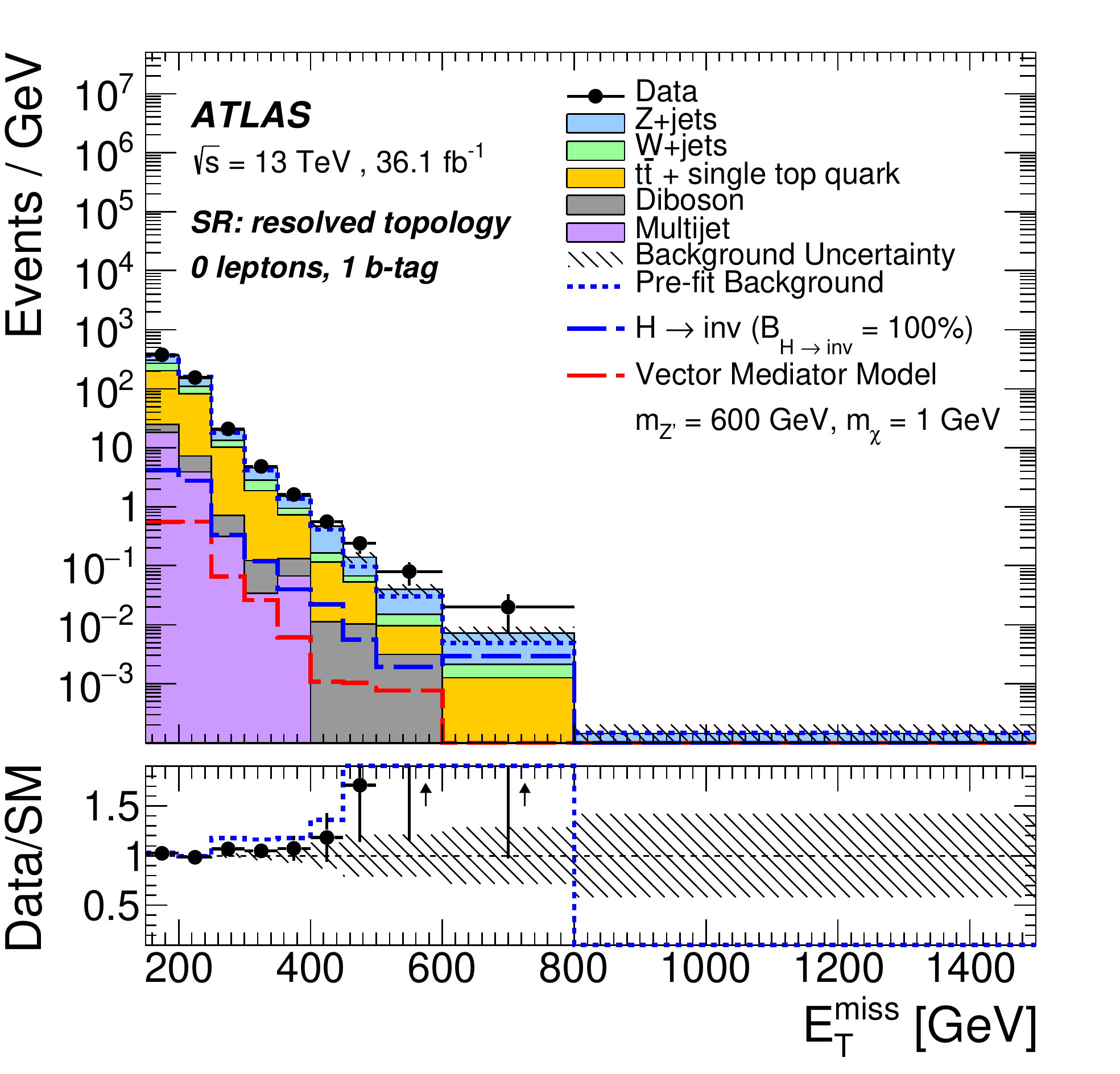}}      
\subfigure[]{\includegraphics[width=0.48\textwidth]{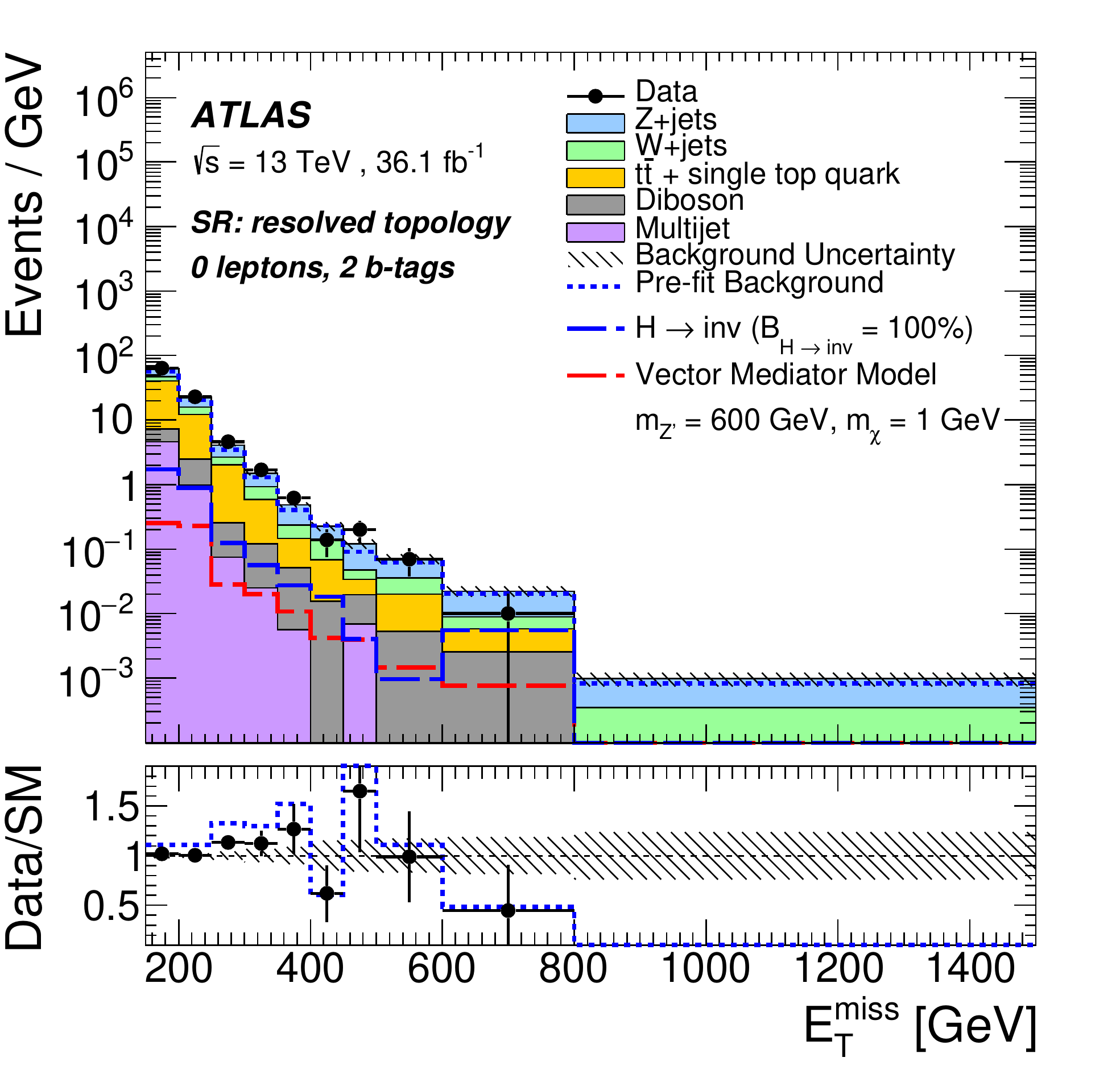}}      
\caption{
\label{fig:postfit_signal_resolved}
The observed (dots) and expected (histograms) distributions of missing transverse momentum, $\met$, obtained with 36.1~\ifb of data at $\sqrt{s}=$~13~\TeV\ in the \monoWZ signal region with the resolved event topology after the profile likelihood fit (with $\mu = 0$), shown separately for the (a) $0b$-, (b) $1b$- and (c) $2b$-tag categories. The total background contribution before the fit to data is shown as a dotted blue line. The hatched area represents the total background uncertainty. The signal expectations for the simplified vector-mediator model with $m_{\chi}=1$~\GeV\ and $m_{Z'}=600$~\GeV\ (dashed red line) and for the invisible Higgs boson decays (dashed blue line) are shown for comparison. The inset at the bottom of each plot shows the ratio of the data to the total post-fit (dots) and pre-fit (dotted blue line) background expectation. 
}
\end{figure}

The normalization of the $W$+HF, $W$+LF and $Z$+LF background components obtained from a fit to the data under the background-only hypothesis is in a good agreement with the SM expectation, while the $Z$+HF ($t\bar{t}$) normalization is 30\% higher (20\% lower) than the expected SM value. In addition to the normalization factors, the final background event yields in each event category are also affected by the systematic uncertainties discussed in Section~\ref{sec:sys}. For all backgrounds other than $Z$+HF and $t\bar{t}$, the number of background events obtained from the fit agrees well with the prediction from simulation in each event category individually. The observed number of events passing the final \monoWZ signal selection is shown for each event category in Table~\ref{tab:event_yields_monoWZ} together with the expected background contributions obtained from the fit under the background-only hypothesis. The expectations for several signal points within the simplified vector-mediator model and for the invisible Higgs boson decays are shown in addition for comparison. Figures~\ref{fig:postfit_signal_merged} and~\ref{fig:postfit_signal_resolved} show the corresponding distributions of the missing transverse momentum in the merged and resolved \monoWZ signal regions, respectively.
The background contributions which are illustrated here are obtained from a simultaneous fit of the expected final discriminants to data with a background-only hypothesis in all signal, sideband and control regions. In this scenario the signal regions lead to a strong constraint of the total background estimate, which is relaxed with a floating signal contribution in the final fit.  

\begin{table}[!htbp]
\begin{center}
\caption{
\label{tab:event_yields_monoWZ}
The expected and observed numbers of events for an integrated luminosity of 36.1~fb$^{-1}$ and $\sqrt{s}=$~13~\TeV, shown separately in each \monoWZ signal region category. The background yields and uncertainties are shown after the profile likelihood fit to the data (with $\mu = 0$). The quoted background uncertainties include both the statistical and systematic contributions, while the uncertainty in the signal is statistical only.  The uncertainties in the total background can be smaller than those in individual components due to anti-correlations of nuisance parameters. 
}
\resizebox{\linewidth}{!}{
\begin{tabular}{l cc cc cc}
\hline\hline

& \multicolumn{5}{c}{Merged topology}\\
Process &         
  \phantom{00}$0b$-HP&   \phantom{00}$0b$-LP&  \phantom{00}$1b$-HP&   \phantom{00}$1b$-LP&   \phantom{00}$2b$ \\
  \hline
Vector-mediator model, \\
$m_{\chi}=$1~\GeV, $m_{Z'}=$200~\GeV & 
\phantom{00}814 $\pm$ 48\phantom{0} & 
\phantom{00}759 $\pm$ 45\phantom{0} & 
\phantom{000}96 $\pm$ 18\phantom{0} & 
\phantom{000}99 $\pm$  16\phantom{0} & 
\phantom{000}49.5 $\pm$  4.3\phantom{00} \\
$m_{\chi}=$1~\GeV, $m_{Z'}=$600~\GeV & 
\phantom{00}280.9 $\pm$ 9.0\phantom{00} & 
\phantom{00}268.5 $\pm$ 8.8\phantom{00} & 
\phantom{000}34.7 $\pm$ 3.6\phantom{00} & 
\phantom{000}33.8 $\pm$ 3.1\phantom{00} & 
\phantom{000}15.38 $\pm$ 0.84\phantom{00} \\
\\
\multicolumn{6}{l}{Invisible Higgs boson decays ($m_H=$~125~\GeV, \BHinv~=~100\%)} \\
$VH$& 
\phantom{00}408.4 $\pm$ 2.1\phantom{00}& 
\phantom{00}299.3 $\pm$ 2.0\phantom{00}& 
\phantom{000}52.06 $\pm$ 0.85\phantom{00}& 
\phantom{000}44.06 $\pm$ 0.82\phantom{00}& 
\phantom{000}27.35 $\pm$ 0.52\phantom{00}\\

$ggH$& 
\phantom{00}184 $\pm$ 19\phantom{0}& 
\phantom{00}837 $\pm$ 35\phantom{0}& 
\phantom{000}11.7 $\pm$ 3.8\phantom{00}& 
\phantom{00}111 $\pm$ 30\phantom{0}& 
\phantom{000}12.3 $\pm$ 4.2\phantom{00}\\

VBF& 
\phantom{000}29.1  $\pm$ 2.5\phantom{00}& 
\phantom{000}96.0  $\pm$ 4.6\phantom{00}& 
\phantom{0000}2.43  $\pm$ 0.36\phantom{00}& 
\phantom{0000}5.83  $\pm$ 0.43\phantom{00}& 
\phantom{0000}0.50  $\pm$ 0.07\phantom{00}\\

\\
$W$+jets & 
\phantom{0}3170 $\pm$ 140\phantom{} & 
\phantom{}10120 $\pm$ 380\phantom{} & 
\phantom{00}218 $\pm$  28\phantom{0} & 
\phantom{00}890 $\pm$ 110\phantom{} & 
\phantom{000}91 $\pm$  12\phantom{0} \\

$Z$+jets & 
\phantom{0}4750 $\pm$ 200\phantom{} & 
\phantom{}15590 $\pm$ 590\phantom{} & 
\phantom{00}475 $\pm$ 52\phantom{0} & 
\phantom{0}1640 $\pm$ 180\phantom{} & 
\phantom{00}186 $\pm$ 12\phantom{0} \\

$t\bar{t}$ & 
\phantom{00}775 $\pm$ 48\phantom{0} & 
\phantom{00}937 $\pm$ 60\phantom{0} & 
\phantom{00}629 $\pm$ 27\phantom{0} & 
\phantom{00}702 $\pm$ 34\phantom{0} & 
\phantom{000}50 $\pm$ 11\phantom{0}\\

Single top-quark & 
\phantom{00}159 $\pm$  12\phantom{0} & 
\phantom{00}197 $\pm$  13\phantom{0} & 
\phantom{000}89.7 $\pm$  6.7\phantom{00} & 
\phantom{00}125.5 $\pm$  8.7\phantom{00} & 
\phantom{000}16.1 $\pm$  1.7\phantom{00}\\

Diboson & 
\phantom{00}770 $\pm$ 110\phantom{} & 
\phantom{00}960 $\pm$ 140\phantom{} & 
\phantom{000}88 $\pm$  14\phantom{0} & 
\phantom{00}115 $\pm$  18\phantom{0} & 
\phantom{000}54 $\pm$  10\phantom{0}\\

Multijet & 
\phantom{000}12 $\pm$ 35\phantom{0} & 
\phantom{000}49 $\pm$ 140\phantom{} & 
\phantom{0000}3.7 $\pm$ 3.3\phantom{00} & 
\phantom{000}15 $\pm$ 13\phantom{0} & 
\phantom{0000}9.3 $\pm$ 9.4\phantom{00}\\
\hline

Total background & 
\phantom{0}9642 $\pm$  87\phantom{0} & 
\phantom{}27850 $\pm$ 150\phantom{} & 
\phantom{0}1502 $\pm$  31\phantom{0} & 
\phantom{0}3490 $\pm$  52\phantom{0} & 
\phantom{00}407 $\pm$  15\phantom{0}\\

Data & 
\phantom{0}9627 \phantom{$\pm$ 000}& 
\phantom{}27856 \phantom{$\pm$ 000}& 
\phantom{0}1502 \phantom{$\pm$ 000}& 
\phantom{0}3525 \phantom{$\pm$ 000}& 
\phantom{00}414 \phantom{$\pm$ 000}\\

\hline\hline
 & \multicolumn{5}{c}{Resolved topology}\\
Process & 
\multicolumn{2}{c}{\phantom{00}$0b$}& \multicolumn{2}{c}{\phantom{00}$1b$}& \phantom{00}$2b$\\
\hline
Vector-mediator model,\\
$m_{\chi}=$1~\GeV, $m_{Z'}=$200~\GeV  & 
\multicolumn{2}{c}{\phantom{0}5050 $\pm$ 130\phantom{}} & 
\multicolumn{2}{c}{\phantom{00}342 $\pm$ 29\phantom{0}} & 
\phantom{00}136.7 $\pm$ 6.0\phantom{00}\\
$m_{\chi}=$1~\GeV, $m_{Z'}=$600~\GeV  & 
\multicolumn{2}{c}{\phantom{00}840 $\pm$ 16\phantom{0}} & 
\multicolumn{2}{c}{\phantom{000}59.9 $\pm$ 4.6\phantom{00}} & 
\phantom{000}27.86 $\pm$ 0.94\phantom{00}\\
\\

\multicolumn{4}{l}{Invisible Higgs boson decays ($m_H=$~125~\GeV, \BHinv~=~100\%)} \\
$VH$& 
\multicolumn{2}{c}{\phantom{0}2129.6 $\pm$ 6.4\phantom{00}}& 
\multicolumn{2}{c}{\phantom{00}171.7 $\pm$ 2.2\phantom{00}}& 
\phantom{00}104.7 $\pm$ 1.2\phantom{00}\\
$ggH$& 
\multicolumn{2}{c}{\phantom{0}4111 $\pm$ 78\phantom{0}}& 
\multicolumn{2}{c}{\phantom{00}178 $\pm$ 16\phantom{0}}& 
\phantom{000}37 $\pm$ 11\phantom{0}\\
VBF& 
\multicolumn{2}{c}{\phantom{00}514 $\pm$ 12\phantom{0}}& 
\multicolumn{2}{c}{\phantom{000}19.8 $\pm$ 2.3\phantom{00}}& 
\phantom{0000}2.33 $\pm$ 0.72\phantom{00}\\

\\
$W$+jets & 
\multicolumn{2}{c}{\phantom{}117500 $\pm$ 4600\phantom{}} & 
\multicolumn{2}{c}{\phantom{0}5000 $\pm$ 680\phantom{}} & 
\phantom{00}598 $\pm$ 98\phantom{0}\\

$Z$+jets & 
\multicolumn{2}{c}{\phantom{}135400 $\pm$ 5600\phantom{}} & 
\multicolumn{2}{c}{\phantom{0}7710 $\pm$ 780\phantom{}} & 
\phantom{0}1219 $\pm$ 67\phantom{0}\\

$t\bar{t}$ & 
\multicolumn{2}{c}{\phantom{}13800 $\pm$ 780\phantom{}} & 
\multicolumn{2}{c}{\phantom{}12070 $\pm$ 420\phantom{}} & 
\phantom{0}2046 $\pm$ 70\phantom{0}\\

Single top-quark & 
\multicolumn{2}{c}{\phantom{0}2360 $\pm$ 140\phantom{}} & 
\multicolumn{2}{c}{\phantom{0}1148 $\pm$ 71\phantom{0}} & 
\phantom{00}222 $\pm$ 14\phantom{0}\\

Diboson & 
\multicolumn{2}{c}{\phantom{0}6880 $\pm$ 950\phantom{}} & 
\multicolumn{2}{c}{\phantom{00}514 $\pm$ 71\phantom{0}} & 
\phantom{00}228 $\pm$ 34\phantom{0}\\

Multijet & 
\multicolumn{2}{c}{\phantom{0}11900 $\pm$ 2300\phantom{}} & 
\multicolumn{2}{c}{\phantom{0}1130 $\pm$ 370\phantom{}} & 
\phantom{00}290 $\pm$ \phantom{}150\phantom{}\\
\hline

Total background & 
\multicolumn{2}{c}{\phantom{}287770 $\pm$ 570\phantom{0}} & 
\multicolumn{2}{c}{\phantom{}27580 $\pm$ 170\phantom{}} & 
\phantom{0}4601 $\pm$ 90\phantom{0}\\

Data & 
\multicolumn{2}{c}{\phantom{}287722} \phantom{$\pm$ 0000} & 
\multicolumn{2}{c}{\phantom{}27586}  \phantom{$\pm$ 000} & 
\phantom{0}4642 \phantom{$\pm$ 000}\\

\hline\hline
\end{tabular}
}
\end{center}
\end{table}

%
%%%%%%%%%%%%%%%%%
% mono W/Z plots
%%%%%%%%%%%%%%%%%
%
%

\begin{figure}[!htbp]
\centering
\subfigure[]{\includegraphics[width=0.4\textwidth]{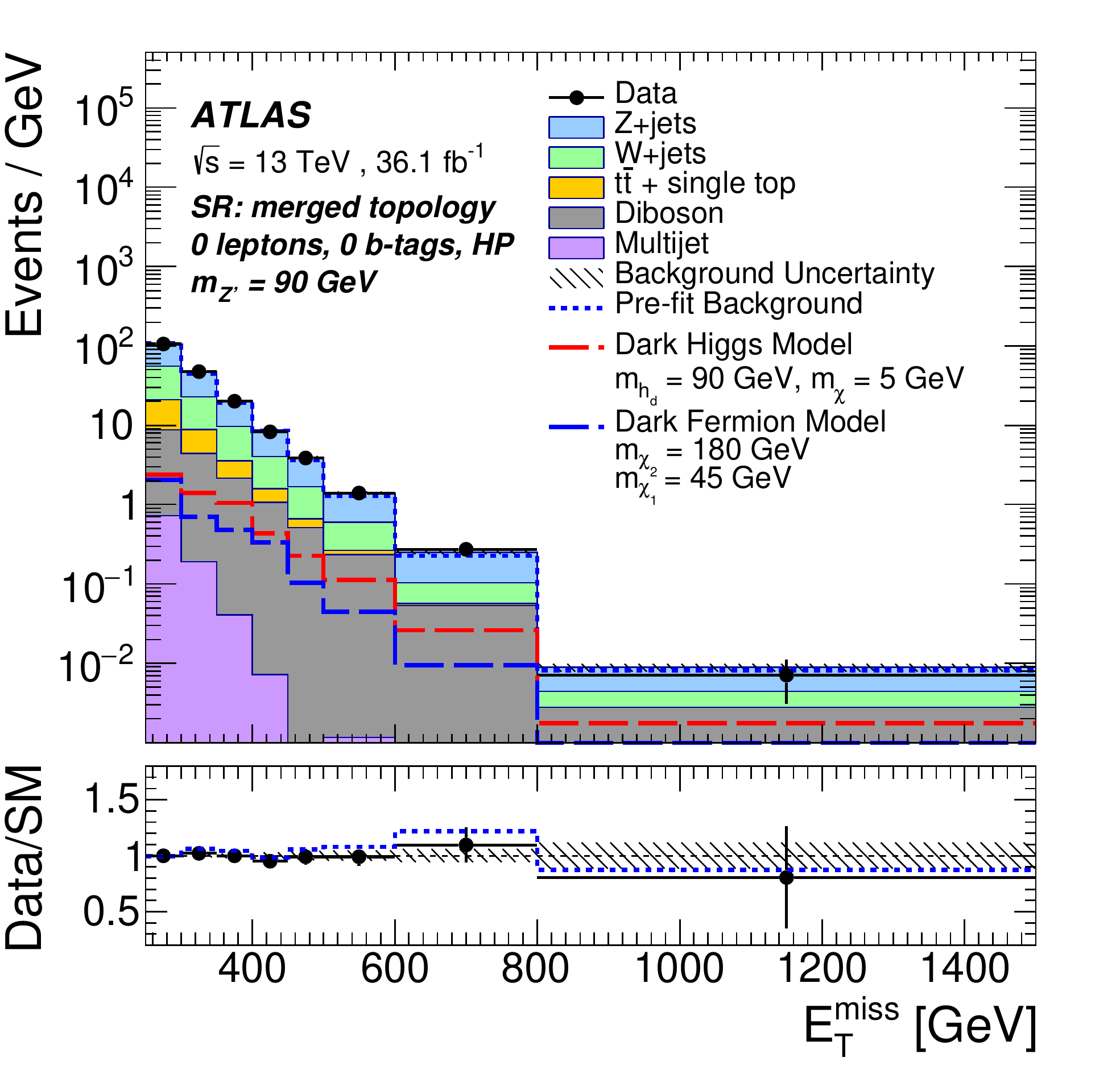}}
\subfigure[]{\includegraphics[width=0.4\textwidth]{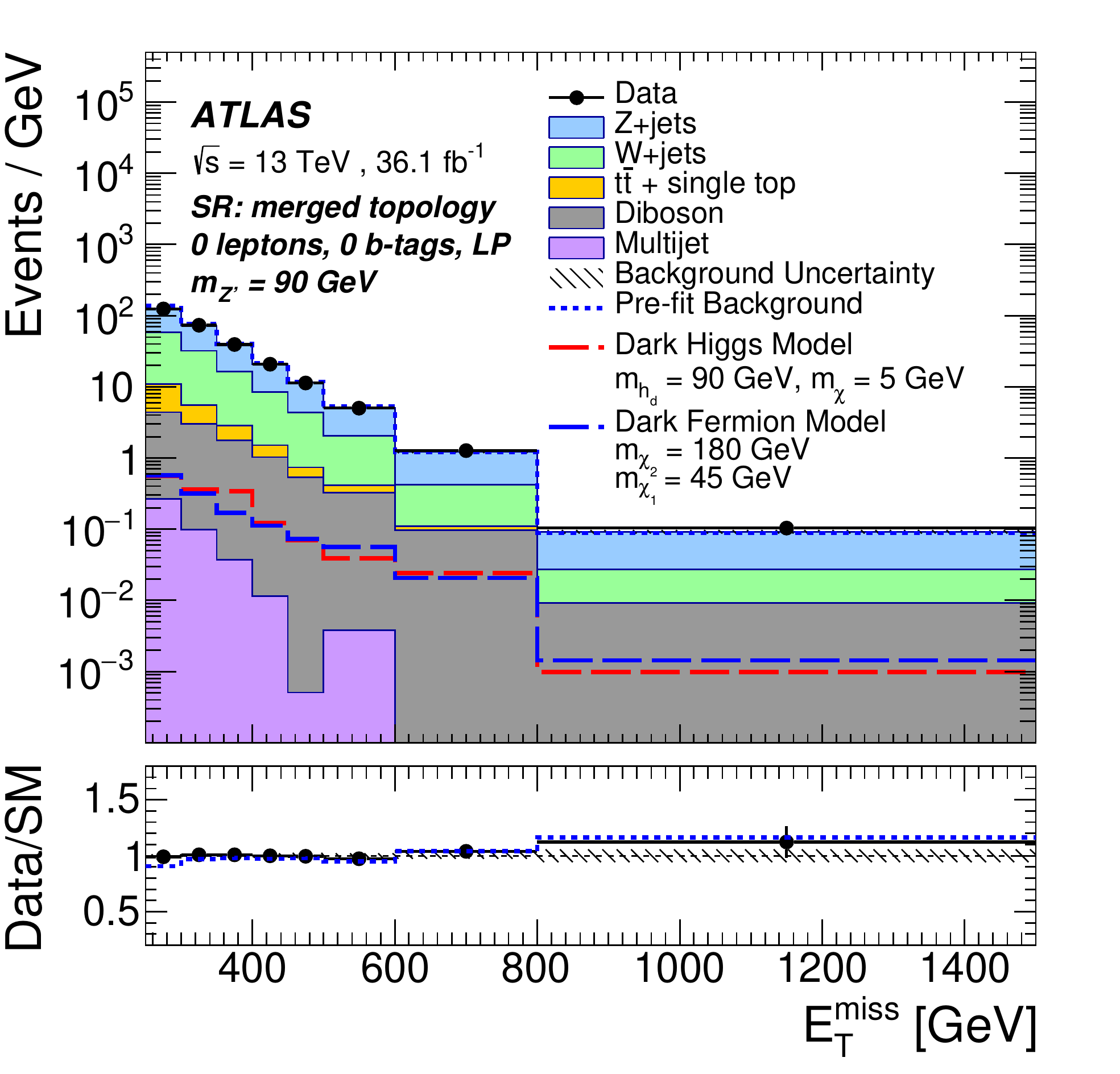}}
\vskip-0.4cm
\subfigure[]{\includegraphics[width=0.4\textwidth]{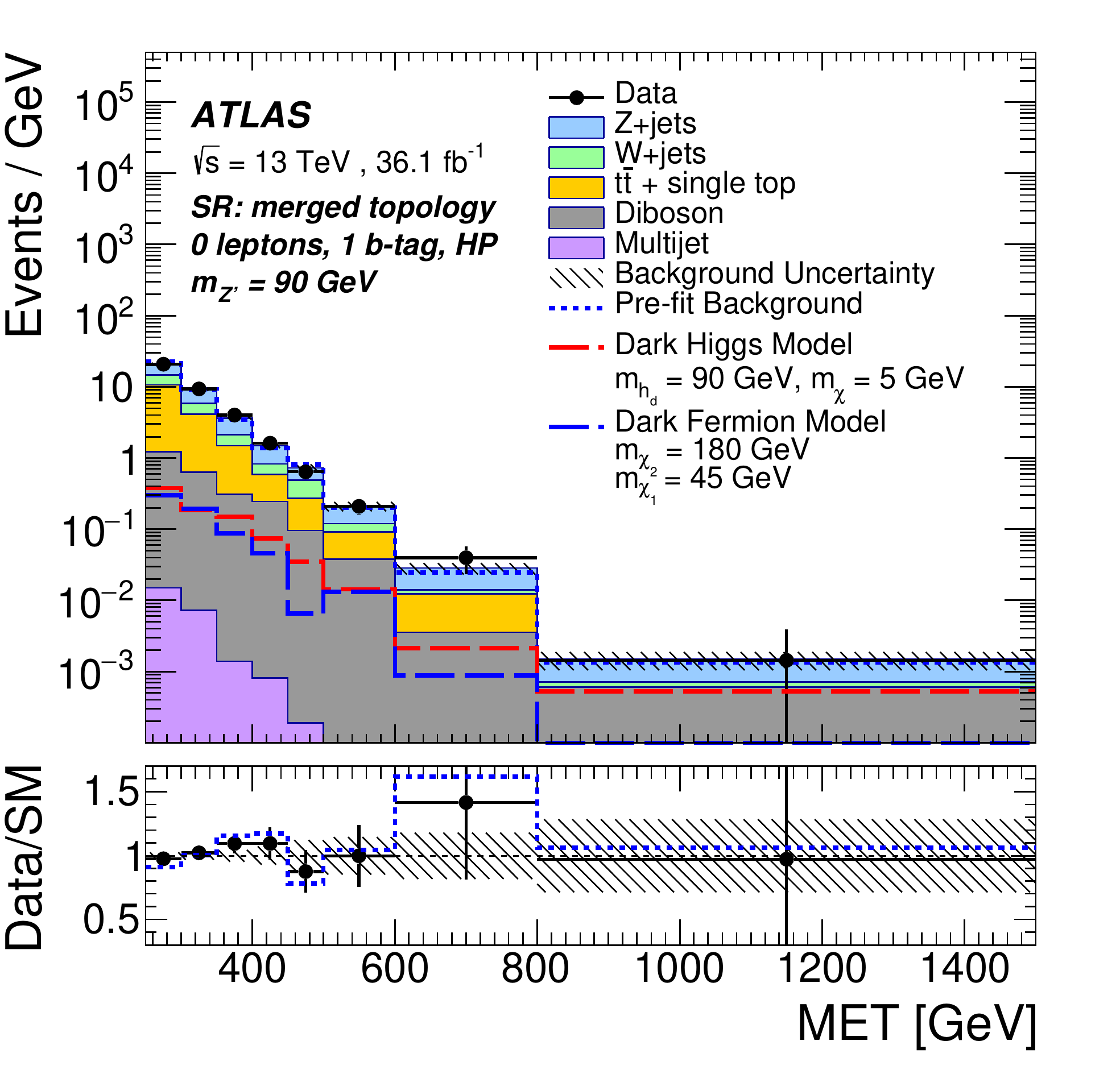}}
\subfigure[]{\includegraphics[width=0.4\textwidth]{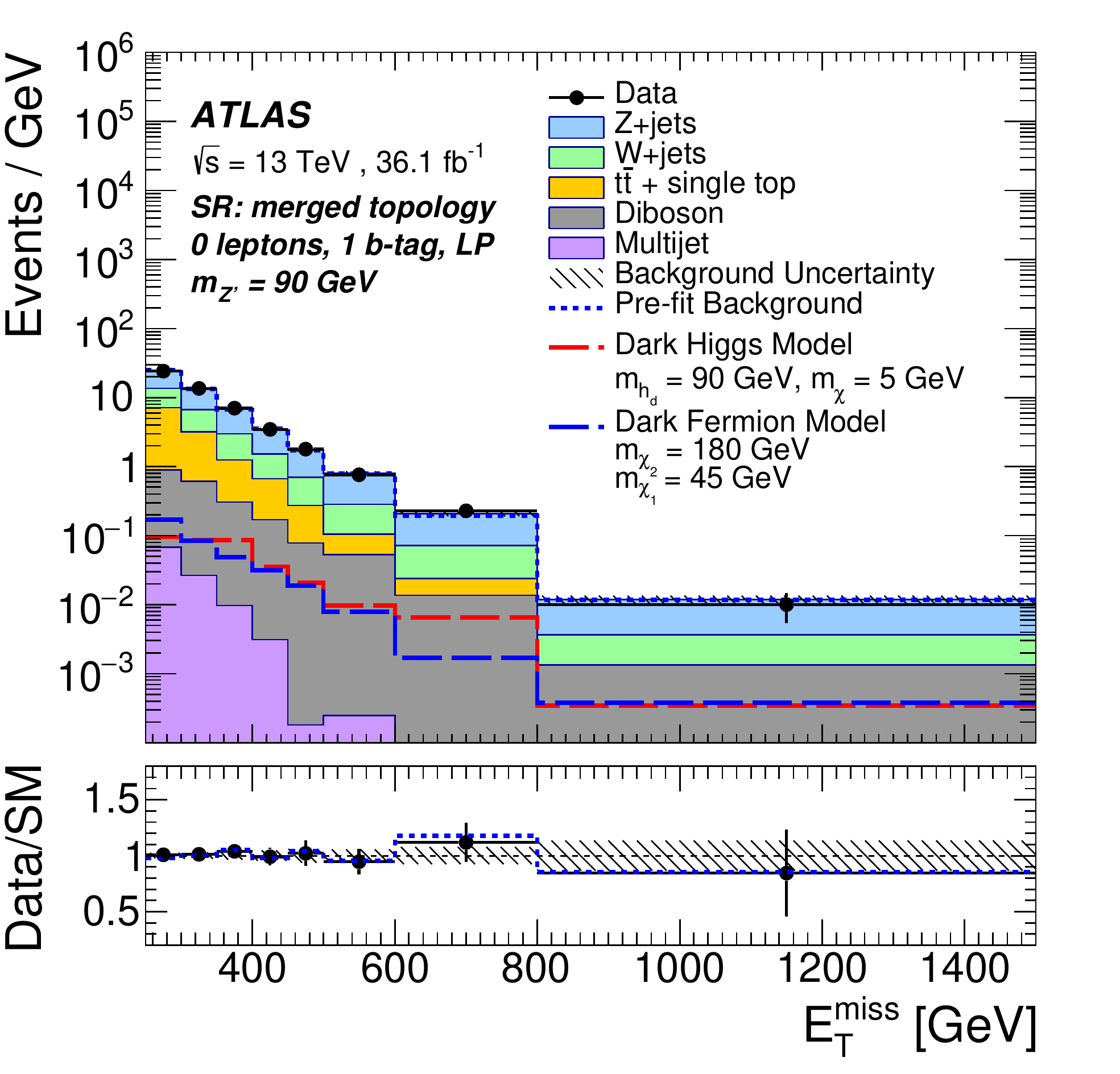}}
\vskip-0.2cm
\subfigure[]{\includegraphics[width=0.4\textwidth]{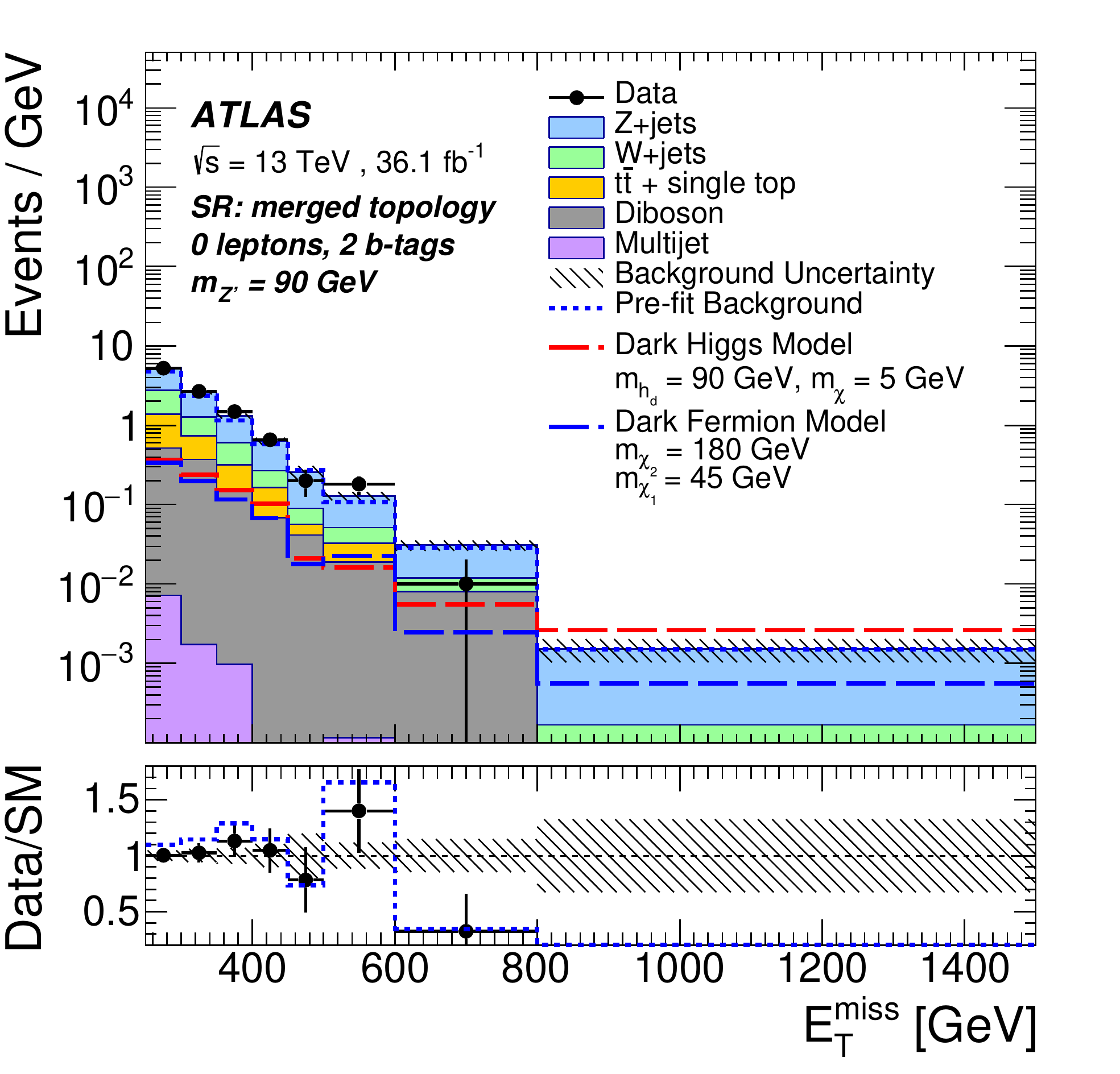}}
\vskip-0.4cm
\caption{
\label{fig:postfit_monozp_90_merged}
The observed (dots) and expected (histograms) distributions of missing transverse momentum, $\met$, obtained with 36.1~\ifb of data at $\sqrt{s}=$~13~\TeV\ in the \monoZprime signal region with $m_{Z'}=$~90~\GeV\ and the merged event topology after the profile likelihood fit (with $\mu = 0$), shown separately for the (a) $0b$-HP, (b) $0b$-LP, (c) $1b$-HP, (d) $1b$-LP, and (e) $2b$-tag event categories. The total background contribution before the fit to data is shown as a dotted blue line. The hatched area represents the total background uncertainty. The expectations for the selected dark-Higgs (dashed red line) and dark-fermion (dashed blue line) signal points are shown for comparison. The inset at the bottom of each plot shows the ratio of the data to the total post-fit (dots) and pre-fit (dotted blue line) background expectation. 
}
\end{figure}

\begin{figure}[!htbp]
\centering
\subfigure[]{\includegraphics[width=0.39\textwidth]{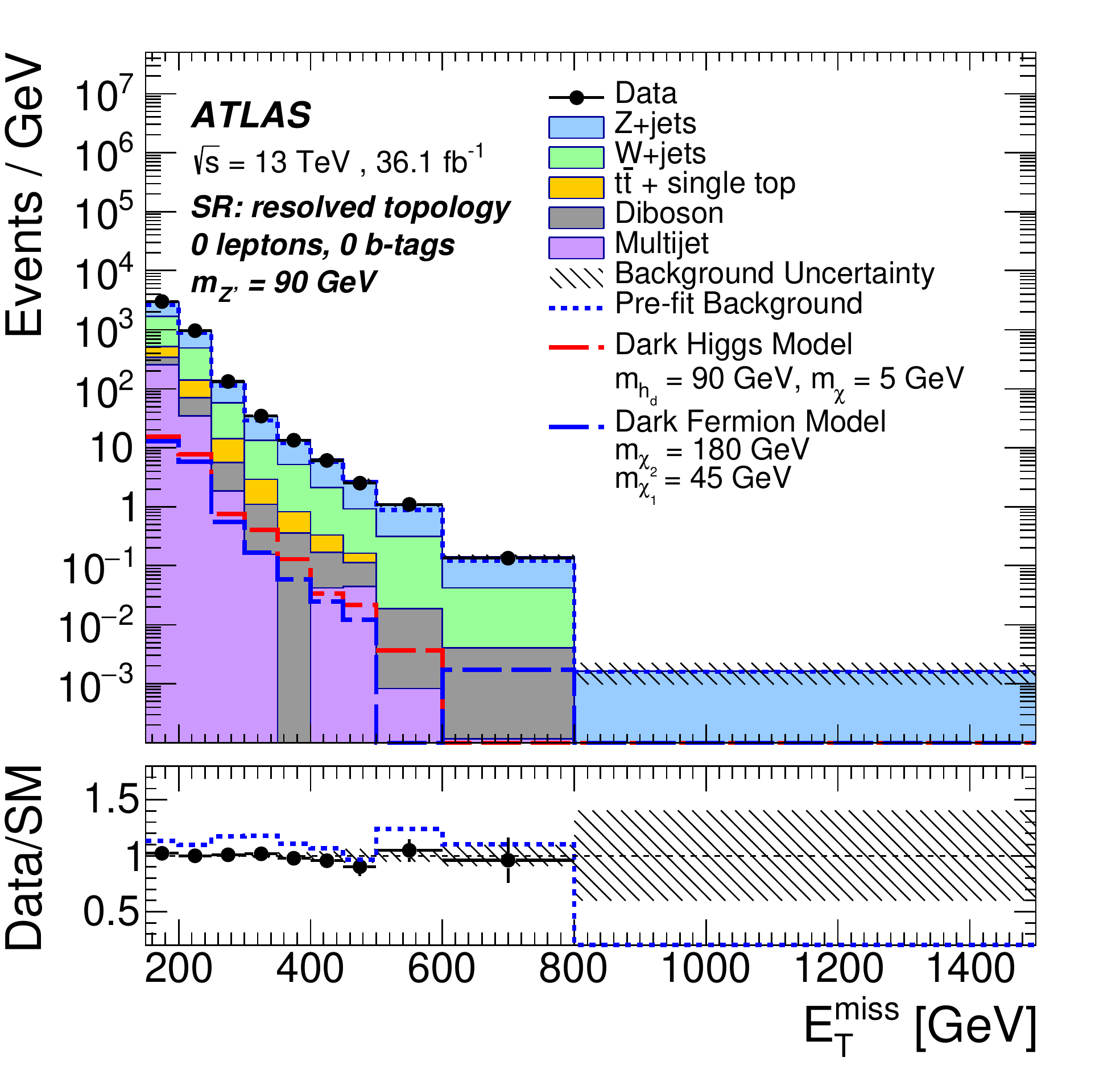}}
\subfigure[]{\includegraphics[width=0.39\textwidth]{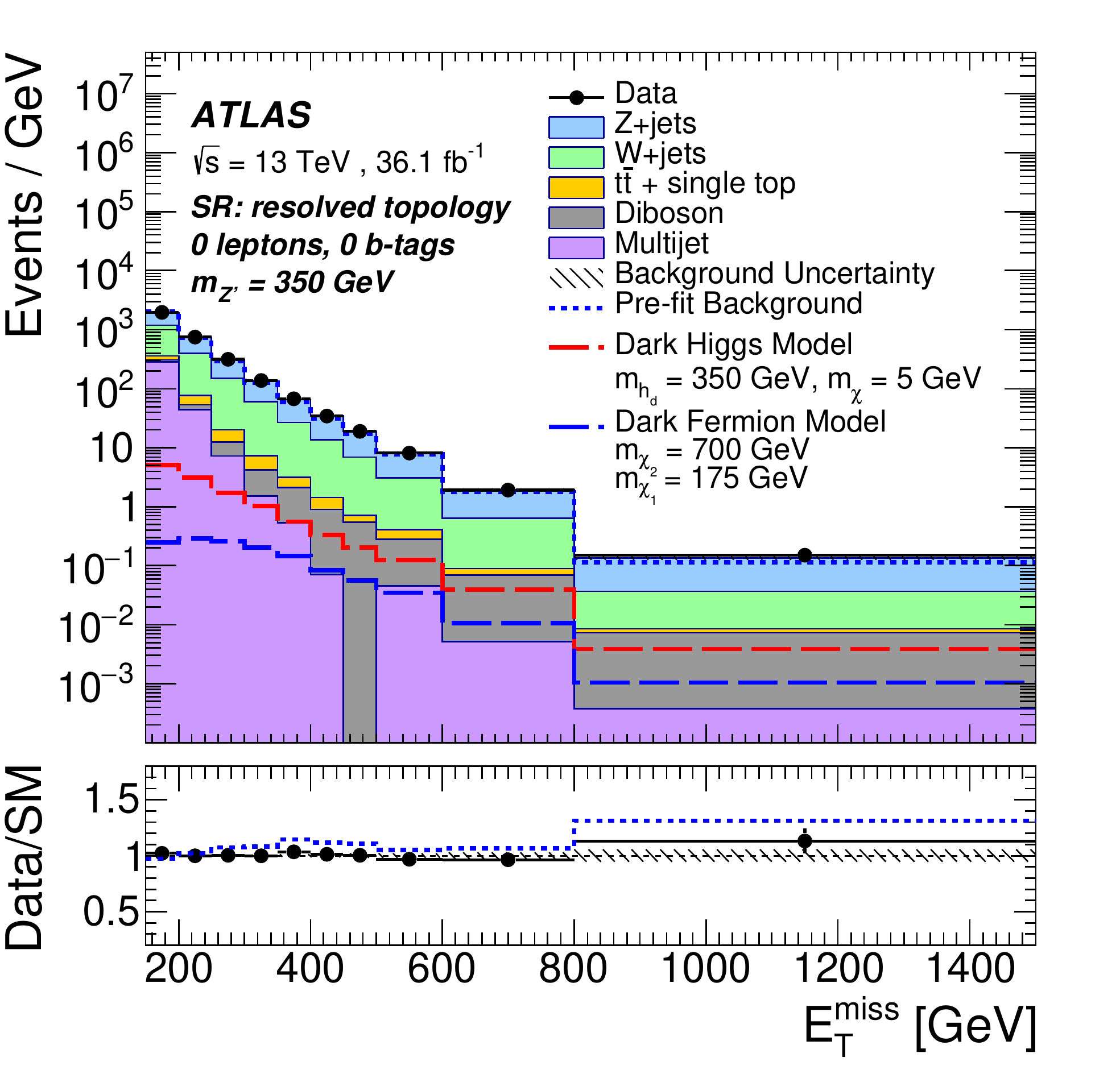}}
\vskip-0.4cm
\subfigure[]{\includegraphics[width=0.39\textwidth]{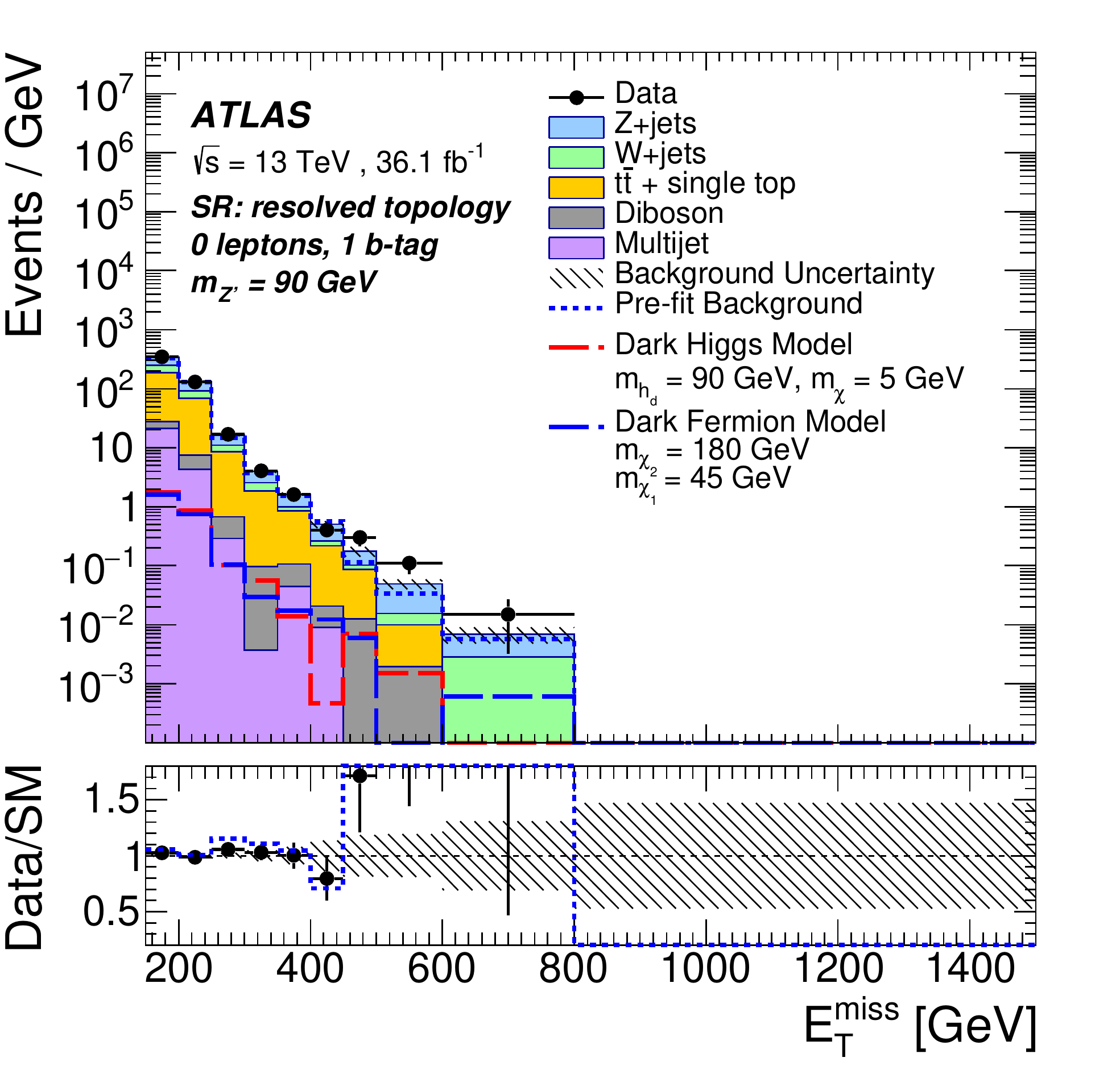}}
\subfigure[]{\includegraphics[width=0.39\textwidth]{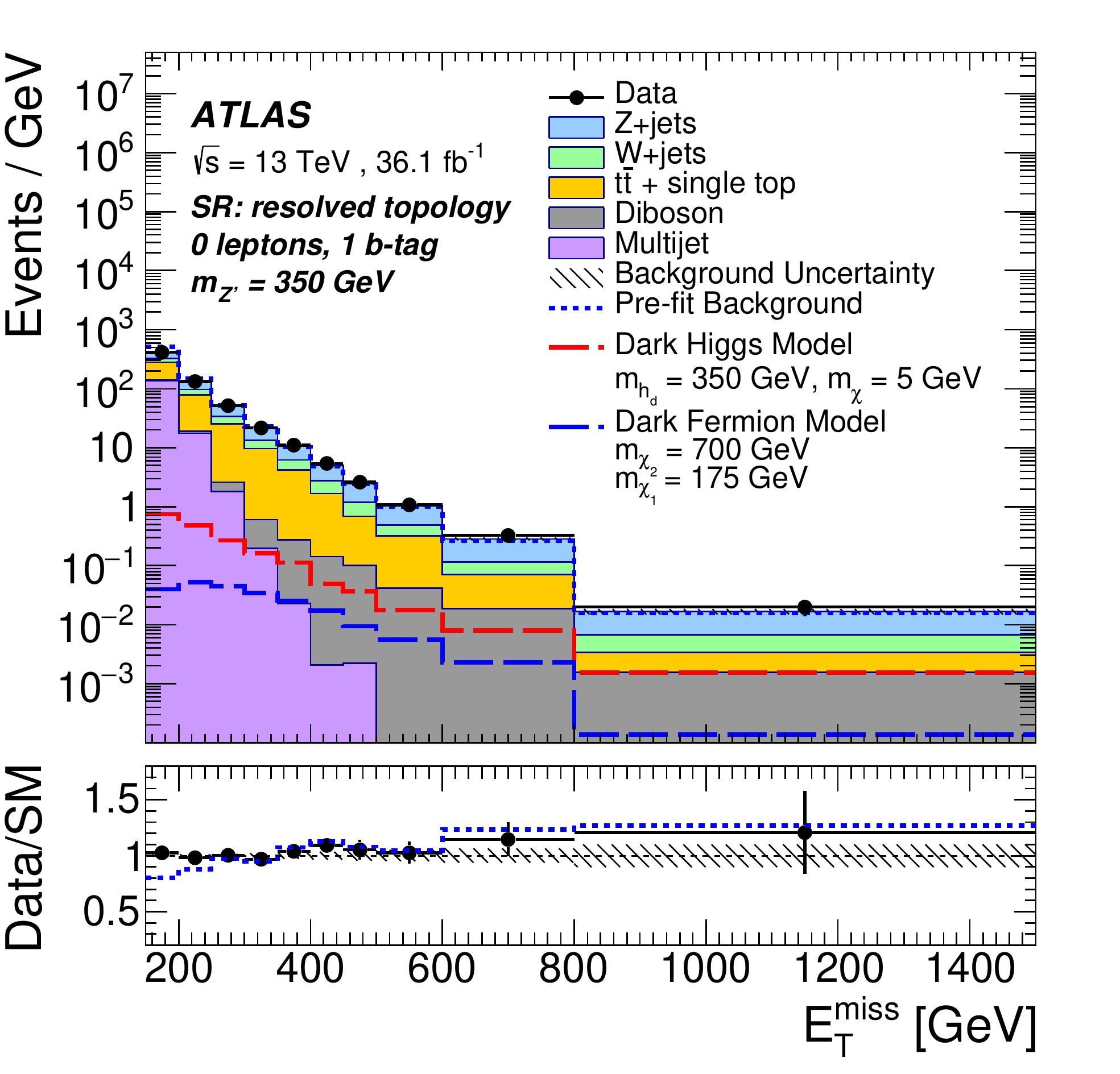}}
\vskip-0.4cm
\subfigure[]{\includegraphics[width=0.39\textwidth]{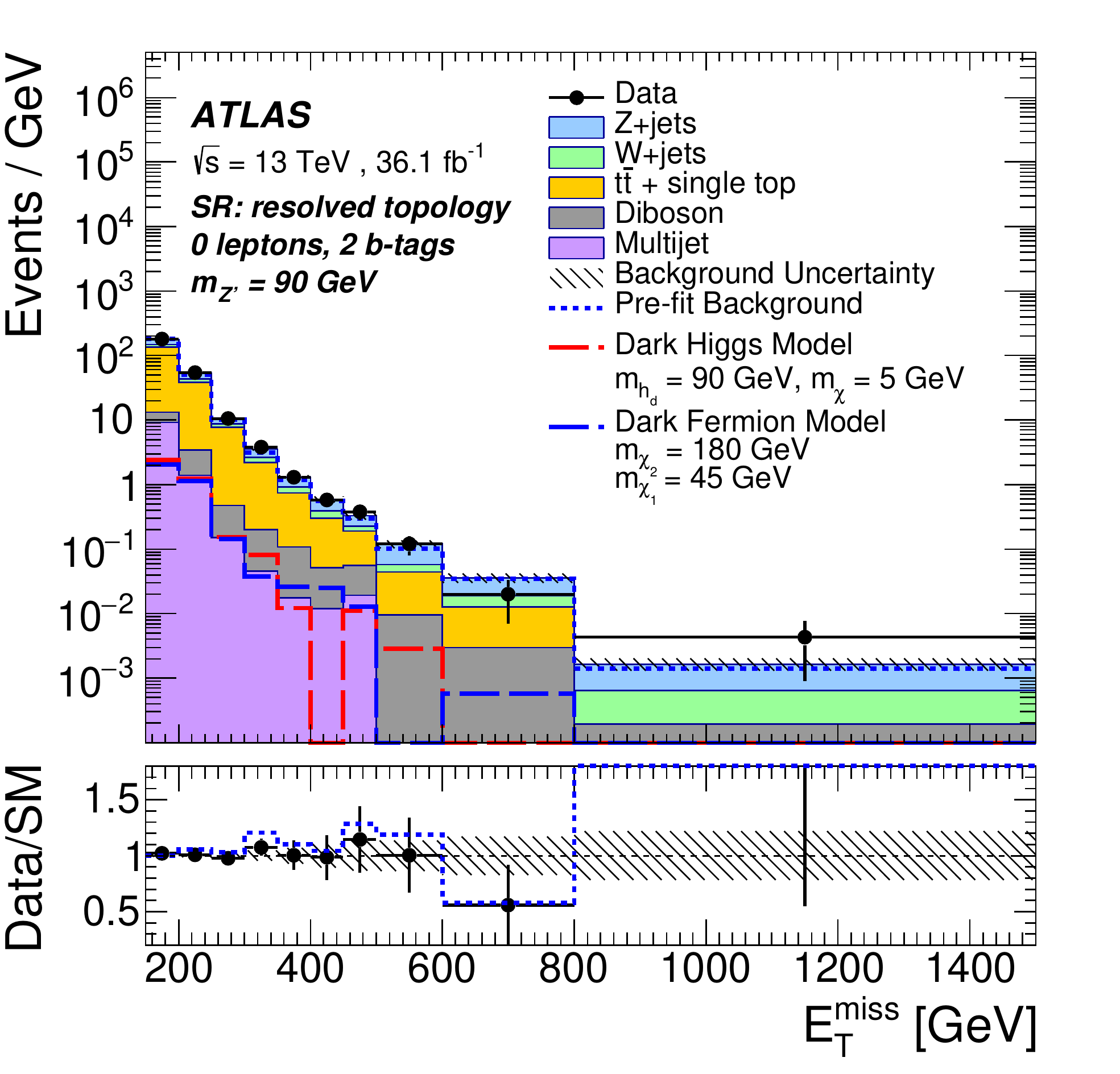}}
\subfigure[]{\includegraphics[width=0.39\textwidth]{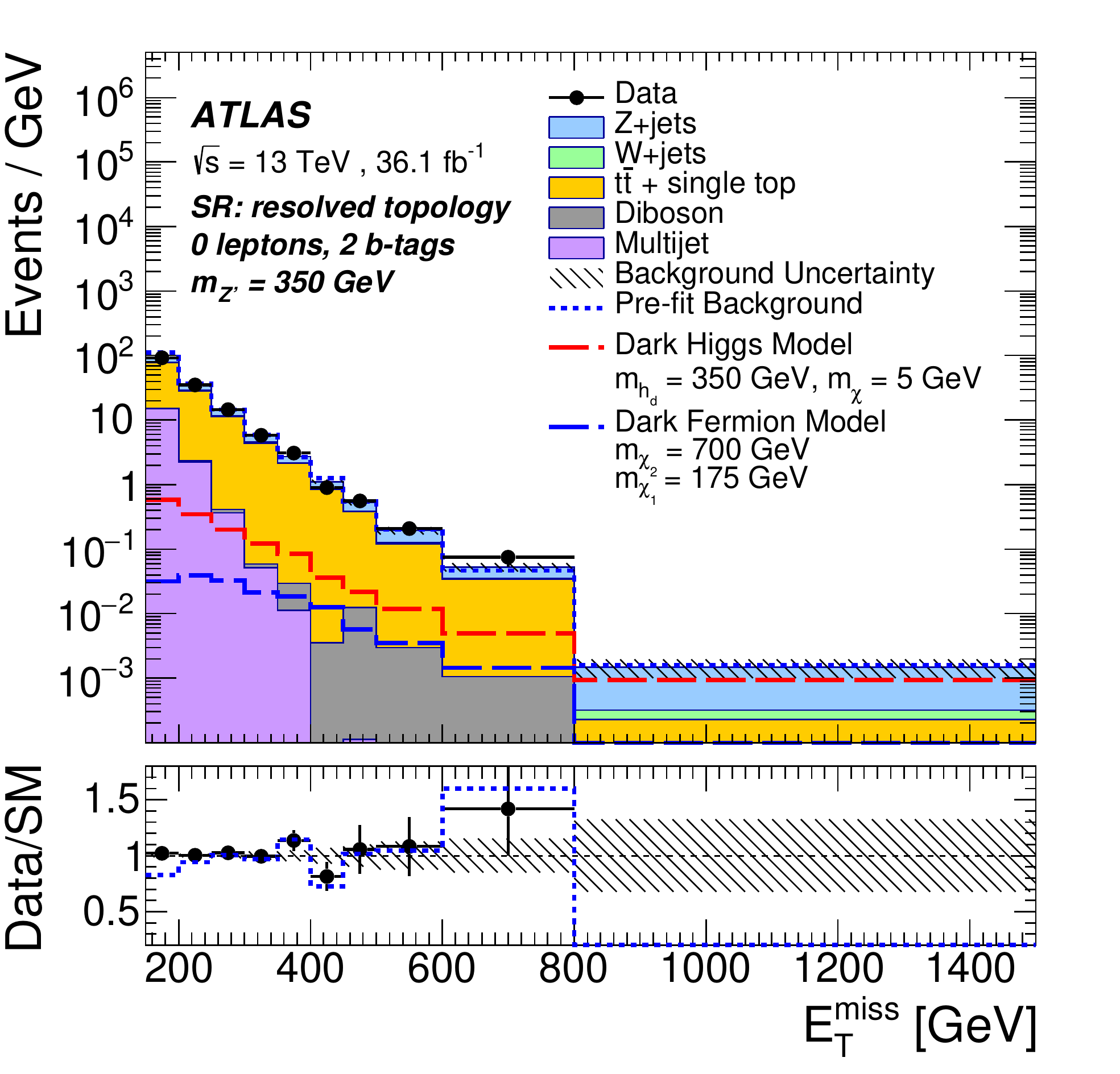}}
\vskip-0.4cm
\caption{
\label{fig:postfit_monozp_resolved}
The observed (dots) and expected (histograms) distribution of missing transverse momentum, $\met$, obtained with 36.1~\ifb of data at $\sqrt{s}=$~13~\TeV\ in the \monoZprime signal region with the resolved event topology after the profile likelihood fit (with $\mu = 0$), shown separately for the (a,b) $0b$, (c,d) $1b$ and (e,f) $2b$-tag event categories. On the left-hand side, the mediator mass of 90~\GeV\ and on the right-hand side of 350~\GeV\ is assumed. The total background contribution before the fit to data is shown as a dotted blue line. The hatched area represents the total background uncertainty. The expectations for the selected dark-Higgs (dashed red line) and dark-fermion (dashed blue line) signal points are shown for comparison. The inset at the bottom of each plot shows the ratio of the data to the total post-fit (dots) and pre-fit (dotted blue line) background expectation. 
}
\end{figure}

%%%%%%%%%%%%%%%%%
% mono Z' tables
%%%%%%%%%%%%%%%%%
%
%
Similarly, the observed and expected numbers of events passing the final \monoZprime selection are shown in Tables~\ref{tab:event_yields_monoZprime_90} and~\ref{tab:event_yields_monoZprime_300} for mediator masses $m_{Z'}$ of 90~\GeV\ and 350~\GeV\, respectively. The expected and observed numbers of background events for the $m_{Z'}$ hypothesis of 90~\GeV\ are similar to those from the \monoWZ search in all categories, except for the $2b$-tag category with resolved topology. There are about three times more events in that category for the \monoZprime search since no requirement on $\Delta R_{jj}$ is applied, as opposed to the strict requirement of $\Delta R_{jj}<$~1.25 employed in the \monoWZ search. The distributions of the missing transverse momentum in each \monoZprime signal region for these mediator masses are shown in Figures~\ref{fig:postfit_monozp_90_merged} and~\ref{fig:postfit_monozp_resolved}. 
\begin{table}[!t]
\begin{center}
\caption{
\label{tab:event_yields_monoZprime_90}
The expected and observed numbers of events for an integrated luminosity of 36.1~fb$^{-1}$ and $\sqrt{s}=$~13~\TeV, shown separately in each \monoZprime signal region category assuming $m_{Z'}=$~90~\GeV. The background yields and uncertainties are shown after the profile likelihood fit to the data (with $\mu = 0$). The quoted background uncertainties include both the statistical and systematic contributions, while the uncertainty in the signal is statistical only.  The uncertainties in the total background can be smaller than those in individual components due to anti-correlations of nuisance parameters.}
\resizebox{\linewidth}{!}{
\begin{tabular}{l cc cc c}
\hline\hline
& \multicolumn{5}{c}{Merged topology}\\
Process  &    \phantom{00}$0b$-HP  & \phantom{00}$0b$-LP & \phantom{00}$1b$-HP & \phantom{00}$1b$-LP & \phantom{00}$2b$\\
  \hline
 
Dark fermion, light sector  & 
\phantom{00}286 $\pm$ 54\phantom{0} & 
\phantom{00}125 $\pm$ 36\phantom{0} & 
\phantom{000}53 $\pm$ 23\phantom{0} & 
\phantom{000}26 $\pm$ 16\phantom{0} & 
\phantom{000}52 $\pm$ 23\phantom{0}\\
Dark fermion, heavy sector & 
\phantom{00}165 $\pm$ 18\phantom{0} & 
\phantom{000}71 $\pm$ 12\phantom{0} & 
\phantom{000}30.9 $\pm$ 7.7\phantom{00} & 
\phantom{000}18.6 $\pm$ 6.0\phantom{00} & 
\phantom{000}36.3 $\pm$ 8.4\phantom{00}\\
Dark Higgs, light sector & 
\phantom{00}253 $\pm$ 25\phantom{0} & 
\phantom{000}82 $\pm$ 14\phantom{0} & 
\phantom{000}37.7 $\pm$ 9.6\phantom{00} & 
\phantom{000}19.1 $\pm$ 6.9\phantom{00} & 
\phantom{0000}45 $\pm$ 11\phantom{00}\\
Dark Higgs, heavy sector &
\phantom{000}224 $\pm$ 14\phantom{00} & 
\phantom{000}75.9 $\pm$ 8.4\phantom{00} & 
\phantom{000}37.5 $\pm$ 5.9\phantom{00} & 
\phantom{000}21.2 $\pm$ 4.4\phantom{00} & 
\phantom{000}49.5 $\pm$ 6.8\phantom{00}\\
& & & & & \\

$W$+jets & 
\phantom{0}2960 $\pm$ 170\phantom{} & 
\phantom{0}5180 $\pm$ 280\phantom{} & 
\phantom{00}342 $\pm$ 52\phantom{0} & 
\phantom{00}680 $\pm$ 100\phantom{} & 
\phantom{00}120 $\pm$ 120\phantom{}\\
$Z$+jets & 
\phantom{0}4720 $\pm$ 190\phantom{} & 
\phantom{0}7990 $\pm$ 310\phantom{} & 
\phantom{00}628 $\pm$ 69\phantom{0} & 
\phantom{0}1280 $\pm$ 140\phantom{} & 
\phantom{00}265 $\pm$ 22\phantom{0}\\
$t\bar{t}$ & 
\phantom{00}780 $\pm$ 110\phantom{} & 
\phantom{00}440 $\pm$ 59\phantom{0} & 
\phantom{00}646 $\pm$ 59\phantom{0} & 
\phantom{00}434 $\pm$ 49\phantom{0} & 
\phantom{000}59 $\pm$ 19\phantom{0}\\
Single top-quark & 
\phantom{00}161 $\pm$ 15\phantom{0} & 
\phantom{00}113 $\pm$ 14\phantom{0} & 
\phantom{000}93 $\pm$ 10\phantom{0} & 
\phantom{000}94.1 $\pm$ 8.9\phantom{00} & 
\phantom{000}17.8 $\pm$ 2.8\phantom{00}\\
Diboson & 
\phantom{00}830 $\pm$ 130\phantom{} & 
\phantom{00}575 $\pm$ 95\phantom{0} & 
\phantom{00}129 $\pm$ 23\phantom{0} & 
\phantom{00}107 $\pm$ 18\phantom{0} & 
\phantom{000}61 $\pm$ 11\phantom{0}\\
Multijet & 
\phantom{0000}48 $\pm$ 41\phantom{00} & 
\phantom{000}21 $\pm$ 66\phantom{0} & 
\phantom{0000}1.2 $\pm$ 1.0\phantom{00} & 
\phantom{0000}5.4 $\pm$ 5.1\phantom{00} & 
\phantom{0000}0.52 $\pm$ 0.51\phantom{00}\\
\hline
Total background & 
\phantom{0}9498 $\pm$ 96\phantom{0} & 
\phantom{}14310 $\pm$ 120\phantom{} & 
\phantom{0}1840 $\pm$ 37\phantom{0} & 
\phantom{0}2600 $\pm$ 46\phantom{0} & 
\phantom{00}523 $\pm$ 19\phantom{0}\\
Data & 
\phantom{0}9516 \phantom{$\pm$ 000} & 
\phantom{}14282 \phantom{$\pm$ 000} & 
\phantom{0}1845 \phantom{$\pm$ 000} & 
\phantom{0}2628 \phantom{$\pm$ 000} & 
\phantom{00}534 \phantom{$\pm$ 000}\\
\hline\hline

& \multicolumn{5}{c}{Resolved topology}\\
Process  &   \phantom{00}$0b$ & &  \phantom{00}$1b$ & & \phantom{00}$2b$\\
  \hline
 
Dark fermion, light sector & 
\phantom{0}2060 $\pm$ 150\phantom{} & & 
\phantom{00}264 $\pm$ 52\phantom{00} & & 
\phantom{00}228 $\pm$ 55\phantom{0}\\
Dark fermion, heavy sector & 
\phantom{00}976 $\pm$ 44\phantom{0} & & 
\phantom{00}121 $\pm$ 15\phantom{00} & & 
\phantom{000}164 $\pm$ 18\phantom{00}\\
Dark Higgs, light sector  & 
\phantom{0}1206 $\pm$ 54\phantom{0} & & 
\phantom{00}135 $\pm$ 18\phantom{00} & & 
\phantom{00}197 $\pm$ 22\phantom{0}\\
Dark Higgs, heavy sector  & 
\phantom{00}953 $\pm$ 30\phantom{0} & & 
\phantom{00}112 $\pm$ 10\phantom{00} & & 
\phantom{000}146 $\pm$ 12\phantom{00}\\
& & & & & \\

$W$+jets  &             
\phantom{0}78400 $\pm$ 3400\phantom{} & &   
\phantom{0}4400 $\pm$ 690\phantom{} & &   
\phantom{0}1030 $\pm$  190\phantom{}\\
$Z$+jets  &            
\phantom{0}91700 $\pm$ 3800\phantom{} & &   
\phantom{0}6970 $\pm$ 690\phantom{} & &   
\phantom{0}2140 $\pm$ 210\phantom{}\\
$t\bar{t}$ &      
\phantom{}11170 $\pm$ 920\phantom{} & & 
\phantom{}10590 $\pm$ 530\phantom{} & & 
\phantom{0}7760 $\pm$ 230\phantom{}\\
Single top-quark & 
\phantom{0}1200 $\pm$  170\phantom{} & &  
\phantom{0}1006 $\pm$  74\phantom{0} & &  
\phantom{00}602 $\pm$  40\phantom{0}\\
Diboson &          
\phantom{0}6080 $\pm$  930\phantom{} & &   
\phantom{00}514 $\pm$  80\phantom{0} & &  
\phantom{00}337 $\pm$  55\phantom{0}\\
Multijet &        
\phantom{0}14700 $\pm$ 2500\phantom{} & &  
\phantom{0}1280 $\pm$ 540\phantom{} & &  
\phantom{00}540 $\pm$ 270\phantom{}\\
\hline
Total background & 
\phantom{}203990 $\pm$ 480\phantom{0}          & & 
\phantom{}24770 $\pm$ 220\phantom{}          & & 
\phantom{}12400 $\pm$ 110\phantom{}\\
Data &             
\phantom{}203991 \phantom{$\pm$ 0000} & & 
\phantom{}24783 \phantom{$\pm$ 000} & & 
\phantom{}12406 \phantom{$\pm$ 000}\\
\hline\hline
\end{tabular}
}
\end{center}
\end{table}
\begin{table}[!h]
\begin{center}
\caption{
\label{tab:event_yields_monoZprime_300}
The expected and observed numbers of events for an integrated luminosity of 36.1~fb$^{-1}$ and $\sqrt{s}=$~13~\TeV, shown separately in each \monoZprime signal region category assuming $m_{Z'}=$~350~\GeV. 
The background yields and uncertainties are shown after the profile likelihood fit to the data (with $\mu = 0$). The quoted background uncertainties include both the statistical and systematic contributions, while the uncertainty in the signal is statistical only.  The uncertainties in the total background can be smaller than those in individual components due to anti-correlations of nuisance parameters.}
\begin{tabular}{l ccc}
\hline\hline
& \multicolumn{3}{c}{Resolved topology}\\
Process  &         \phantom{00}$0b$& \phantom{00}$1b$& \phantom{00}$2b$\\
 \hline
Dark fermion, light sector &  
\phantom{00}655 $\pm$ 14\phantom{0} & 
\phantom{00}104.2 $\pm$ 5.8\phantom{00} & 
\phantom{000}89.5 $\pm$ 5.3\phantom{00}\\
Dark fermion, heavy sector & 
\phantom{000}70.79 $\pm$ 0.79\phantom{00} & 
\phantom{000}12.45 $\pm$ 0.33\phantom{00} & 
\phantom{0000}9.04 $\pm$ 0.28\phantom{00}\\
Dark Higgs, light sector   &  
\phantom{00}639 $\pm$ 13\phantom{0} & 
\phantom{000}96.7 $\pm$ 4.9\phantom{00} & 
\phantom{000}72.3 $\pm$ 4.3\phantom{00}\\
Dark Higgs, heavy sector   &  
\phantom{0}118.9 $\pm$ 1.4\phantom{0} & 
\phantom{00}19.62 $\pm$ 0.58\phantom{0} & 
\phantom{00}14.24 $\pm$ 0.50\phantom{0}\\
& & & \\

$W$+jets &               
\phantom{0}68300 $\pm$ 4300\phantom{} &  
\phantom{00}4270 $\pm$ 1100\phantom{} &   
\phantom{000}115 $\pm$ 84\phantom{00}\\
$Z$+jets &               
\phantom{0}72200 $\pm$ 3000\phantom{} &  
\phantom{00}7230 $\pm$ 800\phantom{0} &  
\phantom{00}1160 $\pm$ 110\phantom{0}\\

$t\bar{t}$ &        
\phantom{0}3900 $\pm$  460\phantom{} & 
\phantom{}10320 $\pm$  720\phantom{} & 
\phantom{0}4920 $\pm$  140\phantom{}\\

Single top-quark &   
\phantom{00}752 $\pm$   69\phantom{0} &  
\phantom{0}1530 $\pm$  110\phantom{} &  
\phantom{00}466 $\pm$  35\phantom{0}\\

Diboson &           
\phantom{0}2000 $\pm$  340\phantom{} &   
\phantom{00}282 $\pm$  47\phantom{0} &   
\phantom{000}14.6 $\pm$  2.8\phantom{00}\\

Multijet &         
\phantom{0}17100 $\pm$ 2300\phantom{} &  
\phantom{00}7870 $\pm$ 390\phantom{0} &  
\phantom{000}880 $\pm$ 140\phantom{0}\\
\hline
Total background & 
\phantom{}164310 $\pm$ 650\phantom{0} &           
\phantom{0}31520 $\pm$ 250\phantom{0} &           
\phantom{00}7567 $\pm$ 85\phantom{00}\\
Data &             
\phantom{}164386 \phantom{$\pm$ 0000}  & 
\phantom{}31465 \phantom{$\pm$ 000}  & 
\phantom{0}7597 \phantom{$\pm$ 000} \\
\hline\hline
\end{tabular}
\end{center}
\end{table}
%
%
%%%%%%%%%%%%%%%%%
% mono Z' plots
%%%%%%%%%%%%%%%%%
%
%

%%%%%%%%%%%%%%%%%%%%%%%%%%%%%%%%%%%%%
% Impact of systematic uncertainties
%%%%%%%%%%%%%%%%%%%%%%%%%%%%%%%%%%%%%
%
% 

The impact of the different sources of systematic uncertainty on the sensitivity of the \monoWZ and \monoZprime searches is estimated by means of fits of the signal-plus-background model to hypothetical data comprized of these signals (with signal strength $\mu=$~1) plus expected background contributions. The resulting uncertainties on the signal strength $\mu$ serve as a measure of the analysis sensitivity and are summarized in Table \ref{tab:systBreakdown_dmV_Vhad}.
\begin{table}[!htbp]
          \centering
          \caption{Breakdown of expected signal strength uncertainties for several \monoWZ and \monoZprime signal models, obtained for an integrated luminosity of 36.1~fb$^{-1}$ and $\sqrt{s}=$~13~\TeV. A dark matter mass of 1~\GeV\ is used for the two vector-mediator signals. Each systematic uncertainty contribution is determined from the quadratic difference between the total uncertainty and the uncertainty obtained by neglecting the systematic uncertainty source in question. Only the largest systematic uncertainties are shown.}
          \label{tab:systBreakdown_dmV_Vhad}
          \begin{tabular}{ l | c  c | c | c c}
          \hline\hline
          Source & \multicolumn{5}{c}{Uncertainty on $\mu=$1 [\%] } \\
          of uncertainty & \multicolumn{2}{c|}{Vector mediator, $m_{Z'}=$}& $H\to $invisible& \multicolumn{2}{c}{Dark fermion, $m_{Z'}=$}\\
	  & 200~\GeV & 600~\GeV& (\BHinv = 100\%)& 90~\GeV & 350~\GeV\\
          \hline

          Large-$R$ jets                 & \phantom{0}9  &           20  &  \phantom{.}17 &            23 & \phantom{0.}-- \\
          Small-R jets                   & \phantom{0}3  & \phantom{0}8  &  \phantom{0.}7 &            13 & \phantom{0.}7  \\
          Electrons                      & \phantom{0}4  & \phantom{0}9  &  \phantom{0.}6 &  \phantom{0}7 & \phantom{0.}6  \\
          Muons                          & \phantom{0}6  & \phantom{0}7  &  \phantom{0.}7 &            15 & \phantom{0}11 \\
          \met                           & \phantom{0}1  & \phantom{0}4  &  \phantom{0.}3 &  \phantom{0}4 & \phantom{0.}3  \\
          $b$-tagging (track jets)       & \phantom{0}4  & \phantom{0}4  &  \phantom{0.}4 &  \phantom{0}8 & \phantom{0.}-- \\
          $b$-tagging (small-$R$ jets)   & \phantom{0}2  & \phantom{0}4  &  \phantom{0.}2 &  \phantom{0}5 & \phantom{0.}5 \\
          Luminosity                     & \phantom{0}3  & \phantom{0}4  &  \phantom{0.}3 &  \phantom{0}4 & \phantom{0.}4  \\ 
          & \multicolumn{2}{c|}{}& & & \\

          Multijet normalization         & \phantom{0}7  &            11 & \phantom{.}11 &           13  &  \phantom{0.}6 \\
          Diboson normalization          & \phantom{0}5  &            11 & \phantom{0.}6 &  \phantom{0}3 &  \phantom{0.}1  \\
	      $Z$+jets normalization         & \phantom{0}5  &  \phantom{0}9 & \phantom{0.}4 &            15 &  \phantom{0.}9 \\
          $W$+jets normalization         & \phantom{0}3  &  \phantom{0}4 & \phantom{0.}2 &  \phantom{0}8 &  \phantom{0.}6  \\
          $t\bar{t}$ normalization       & \phantom{0}3  &  \phantom{0}1 &           0.3 &  \phantom{0}8 &  \phantom{0.}5 \\ 
          & \multicolumn{2}{c|}{}& & & \\

          Signal modelling                & \phantom{0}7 &  \phantom{0}9 & \phantom{.}20 &  \phantom{0}--&  \phantom{0.}-- \\ 
          $V$+jets modelling              & \phantom{0}4 &            10 & \phantom{0.}4 &  \phantom{0}7 &  \phantom{0}11 \\
          $t\bar{t}$ modelling            & \phantom{0}2 &  \phantom{0}4 & \phantom{0.}3 &            10 &  \phantom{0.}6  \\
          $V$+jets flavour composition    & \phantom{0}1 &  \phantom{0}3 & \phantom{0.}3 &  \phantom{0}4 &  \phantom{0.}2  \\
          Diboson modelling               & \phantom{0}1 &  \phantom{0}2 & \phantom{0.}2 &  \phantom{0}1 &            0.2  \\ 
          & \multicolumn{2}{c|}{}& & & \\

          Background MC stat.             &           10 &            18 & \phantom{.}14 &            20 &    \phantom{.}12 \\ 
          & \multicolumn{2}{c|}{}& & & \\

          \hline
          Total syst.                     &           21 &            40 & \phantom{.}38 &           45 &     \phantom{.}29 \\
          Data stat.                      & \phantom{0}7 &            21 & \phantom{0.}5 &           14 &     \phantom{.}12 \\
          \hline

          Total                           &           22 &            45 & \phantom{.}39 &           47 &     \phantom{.}32 \\
          \hline\hline
  \end{tabular}
\end{table}
Tests of the background-only versus the signal-plus-background hypothesis using a profile likelihood test statistic show no significant deviation from the SM background expectation for any of the signal mass points, in both the \monoWZ and \monoZprime searches. A modified frequentist method with the $\mathrm{CL_s}$ formalism~\cite{Read:2002hq} is used to set upper limits on the signal strength $\mu$ at 95\% confidence level for all signal models.

%%%%%%%%%%%%%%%%%%
% HInv results 
%%%%%%%%%%%%%%%%%%
%
%

\subsection{Constraints on invisible Higgs boson decays}

In the search for invisible Higgs boson decays, 
%upper limits at 95\% CL are derived on the product  $\sigma \times \BHinv$ of the Higgs boson production cross section and branching ratio for the decay into invisible particles, separately for each production mode. The observed (expected) limits of 1.5~(1.5)~pb, 2.1~(1.4)~pb and 123~(94)~pb are obtained for the $WH$, $ZH$ and $ggH$ production, respectively. 
%
an observed (expected) upper limit of 0.83 (0.58$^{\textrm{+0.23}}_{\textrm{-0.16}}$) is obtained at 95\% CL on the branching ratio \BHinv , assuming the SM production cross sections and combining the contributions from $VH$, $ggH$ and VBF production modes. The expected limit is a factor of about 1.5 better (while the observed is slightly worse) than the one reached by the previous analysis of Run~1 ATLAS data~\cite{HIGG-2014-07}.
%
% and comparable to the one from the analysis of the combined Run-1 and early Run-2 CMS data~\cite{CMS-HIG-16-016}. 
%The results are summarized in Table~\ref{tab:Hinv_results}.
%%
%\begin{table}[!htbp]
%          \centering
%          \begin{tabular}{l cccc}
%          \hline\hline
%
%\multicolumn{5}{c}{Exclusion limit at 95\% CL}\\
%Process& Observed& Expected& $-1\sigma$& $+1\sigma$\\
%\hline
%& \multicolumn{4}{l}{$\BHinv$}\\
%\hline
%$VH$+$ggH$+VBF& 0.83& 0.58& 0.42& 0.81\\
%\hline
%& \multicolumn{4}{l}{$\sigma \times \BHinv$ [pb] }\\
%\hline
%$WH$&  1.5& 1.5& 1.1& 2.1\\
%$ZH$&  2.1& 1.4& 1.0& 2.0\\
%$ggH$&   123&  94&  68& 131\\
%\hline
%
%\hline\hline
%\end{tabular}
%          \caption{Exclusion limits at 95\% CL on the branching ratio \BHinv and $\sigma \times \BHinv$ of invisible decays of the  Higgs boson produced via different production mechanisms, obtained for an integrated luminosity of 36.1~fb$^{-1}$ and $\sqrt{s}=$~13~\TeV. The limit on \BHinv is obtained assuming the SM Higgs boson production cross sections in each production mode. 
%          \label{tab:Hinv_results}}
%\end{table}
% 

%%%%%%%%%%%%%%%%%%
% mono-W/Z results 
%%%%%%%%%%%%%%%%%%
%
%
 
\subsection{Constraints on the simplified vector-mediator model}
 
\begin{figure}[!htbp]
\begin{center}
\subfigure[\label{fig:2dLimitPlots_a}]{\includegraphics[width=0.51\textwidth]{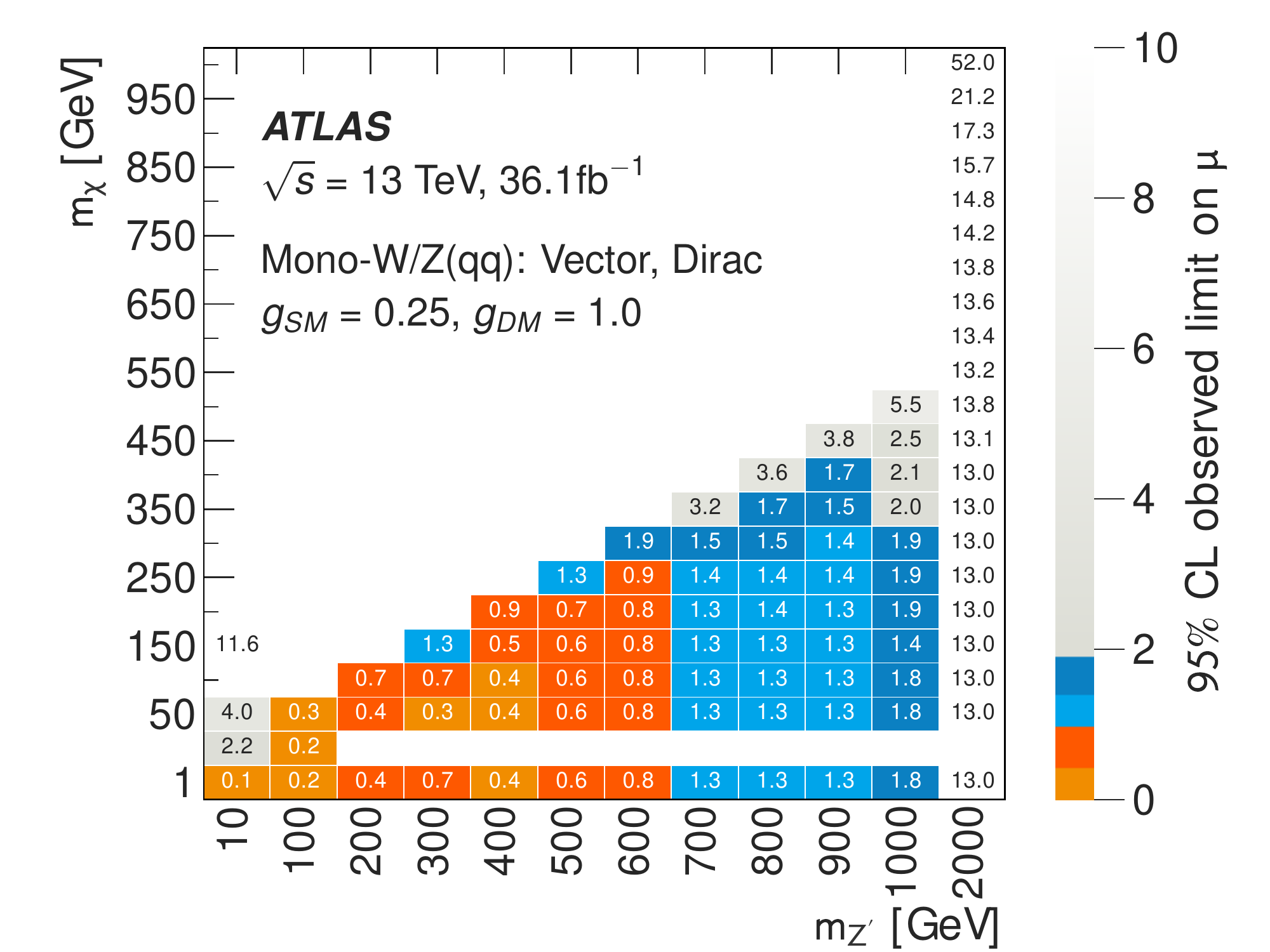}}
\subfigure[\label{fig:2dLimitPlots_b}]{\includegraphics[width=0.48\textwidth]{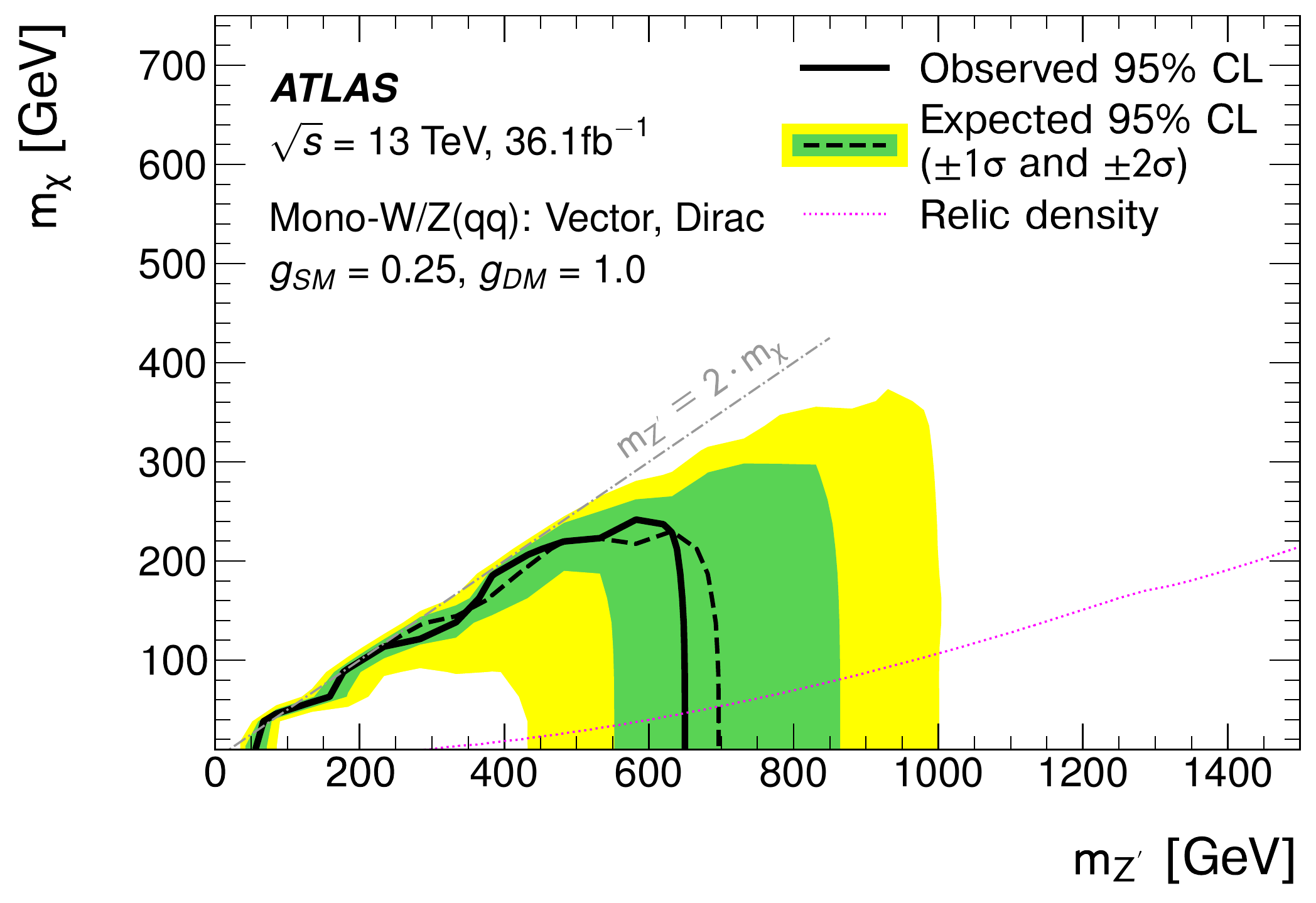}}
\caption{
\label{fig:2dLimitPlots} 
(a) Observed upper limits on the signal strength $\mu$ at 95\% CL in the grid of the DM and mediator particle masses, ($m_{\chi}$, $m_{Z'}$), for the combined mono-$W$  and mono-$Z$  search  in the simplified vector-mediator model with Dirac DM particles and couplings \gsm~=~0.25 and \gdm~=~1. There are no interpolated points and thus no limit values listed for the mass point ($m_{\chi}=$~100~\GeV, $m_{Z'}=$~10~\GeV) and in the parameter region ($m_{\chi}=$~10~\GeV, $m_{Z'}=$~200--2000~\GeV). (b) The corresponding exclusion contours at 95\% CL. The black solid (dashed) curve shows the observed (expected) limit. The dotted magenta curve corresponds to the set of points for which the expected relic density is consistent with the WMAP~\cite{WMAP} and Planck~\cite{Planck_relic} measurements ($\Omega h^2 = 0.12$), as computed with MadDM~\cite{MadDM}. The region below the curve corresponds to higher predicted relic abundance than these measurements.
}
\end{center}
\end{figure}

In the context of the \monoWZ simplified vector-mediator signal model, the exclusion limits on the signal strength are shown in Figure~\ref{fig:2dLimitPlots_a} and translated into limits on the dark matter and mediator masses (Figure~\ref{fig:2dLimitPlots_b}) for Dirac DM particles and couplings \gsm~=~0.25 and \gdm~=~1.  Since only a limited number of signal points were simulated, an interpolation procedure is employed to obtain the limits on the signal strength at other mass points in the ($m_{\chi}, m_{Z'}$) parameter plane. All signal processes with the same mediator mass $m_{Z'}$ and different $m_{\chi}$ values are assumed to have the same $(\mathcal{A} \times \varepsilon)_{\mathrm{total}}$ value as in the simulated sample with $m_{\chi}=$~1~\GeV. This was verified to be a reliable approximation for $m_{Z'}>2m_{\chi}$. Thus, the expected signal yield at a given mass point $(m_{Z'}, m_{\chi})$ only depends on the cross section $\sigma_{pp\to Z'\to \chi\chi}^{(m_{Z'}, m_{\chi})}$ at that mass point.  Under the narrow width approximation, this cross section can be expressed in terms of the cross section $\sigma_{pp\to Z'\to \chi\chi}^{(m_{Z'}, m_{\chi}=1~\mathrm{\GeV})}$  and the branching ratio $\mathcal{B}^{m_{\chi}=1~\mathrm{\GeV}}_{Z'\to \chi\chi}$ at the simulated mass 
point with $m_{\chi}=$~1~\GeV, 
\[
\sigma_{pp\to Z'\to \chi\chi}^{(m_{Z'}, m_{\chi})} = \sigma_{pp\to Z'\to \chi\chi}^{(m_{Z'}, m_{\chi}=1~\mathrm{\GeV})} \cdot \frac{\mathcal{B}^{m_{\chi}}_{Z'\to \chi\chi}}{\mathcal{B}^{m_{\chi}=1~\mathrm{\GeV}}_{Z'\to \chi\chi}}, 
\]
where the value of the branching ratio $\mathcal{B}^{m_{\chi}}_{Z'\to \chi\chi}$ is fully defined by the values of model parameters \gdm, \gsm, $m_{\chi}$ and $m_{Z'}$. For the given coupling choices, vector-mediator masses $m_{Z'}$ of up to 650~\GeV\ are excluded at 95\% CL for dark matter masses $m_{\chi}$ of up to 250~\GeV, agreeing well with the expected exclusion of $Z'$ masses of up to 700~\GeV\ for $m_{\chi}$ of up to 230~\GeV. The expected limits are improved by 15--30\%, depending on the DM mass, compared to the analysis presented in Ref.~\cite{EXOT-2015-08}.

%

%%%%%%%%%%%%%%%%%
% MILs
%%%%%%%%%%%%%%%%%
%
%
\subsection{{Mono-$W/Z$\xspace} constraints with reduced model dependence}

\begin{figure}[!htbp]
\centering
\includegraphics[width=0.45\textwidth]{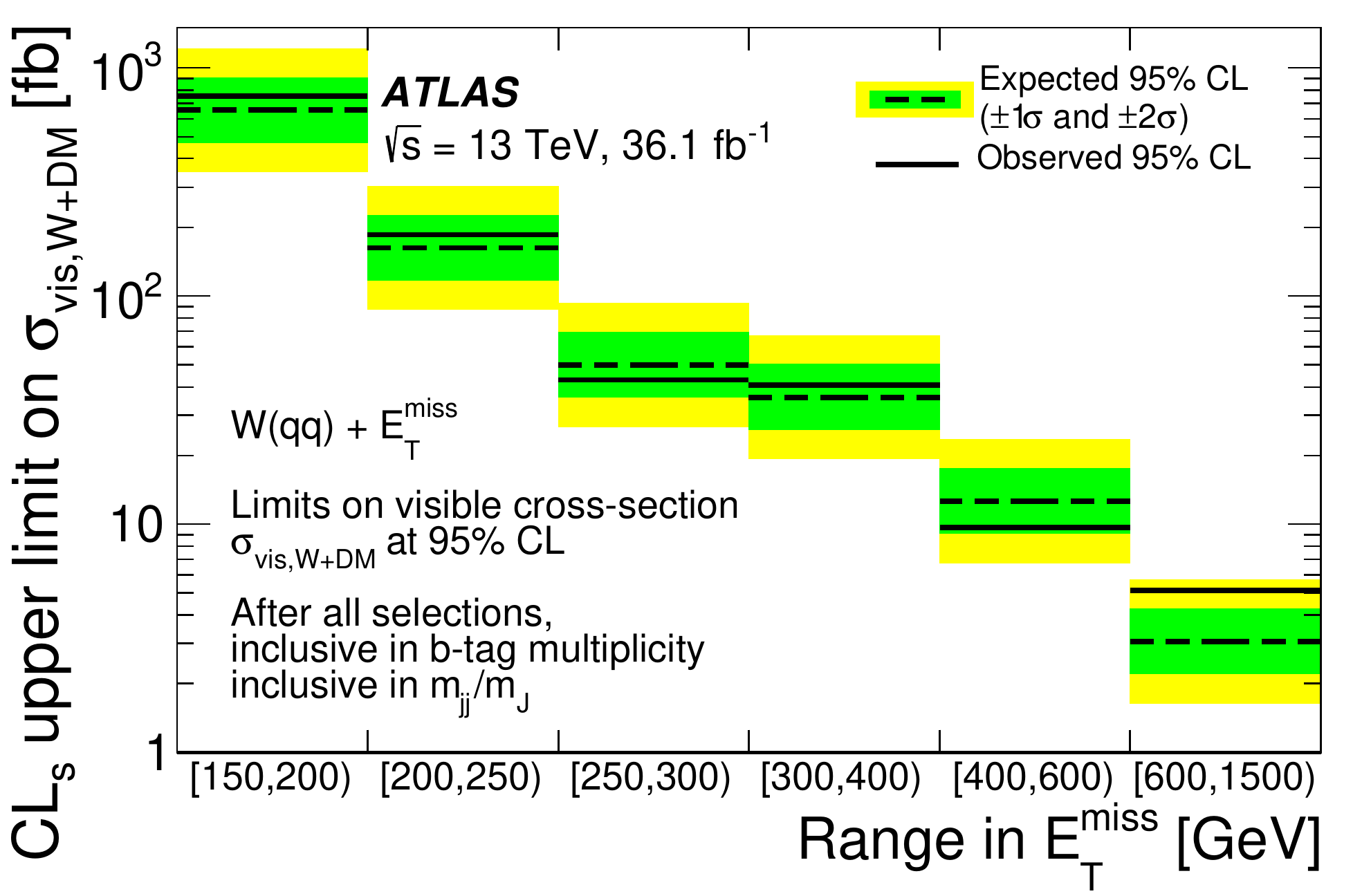}
\includegraphics[width=0.45\textwidth]{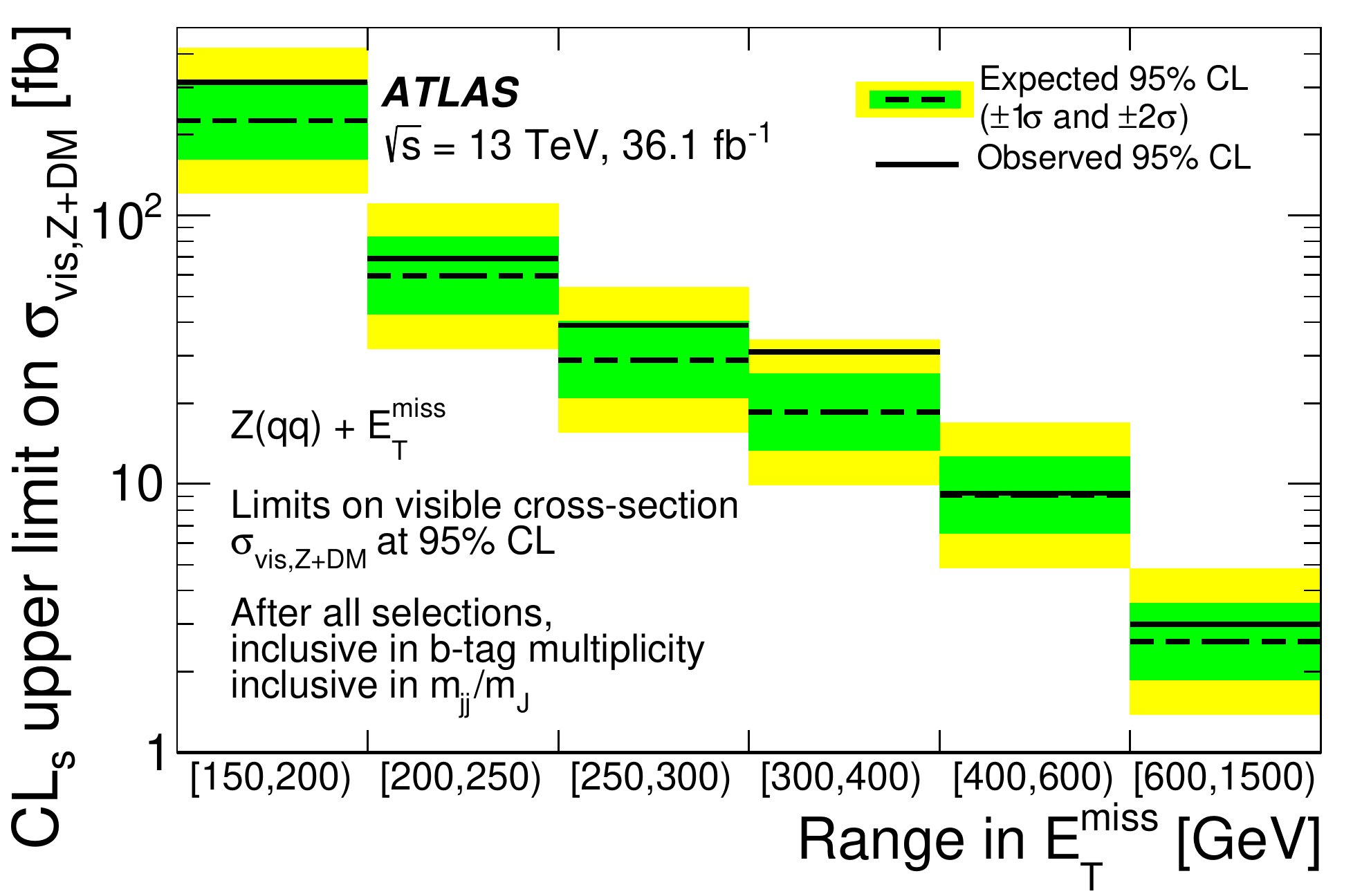}
\caption{
Upper limits at 95\% CL on the visible cross section \sigvisw~(left) and  \sigvisz~(right) in the six \met regions, after all selection requirements, but inclusive in the $b$-tag multiplicity and the $W/Z$ candidate mass $m_{jj}/m_J$. The observed limits (solid line) are consistent with the expectations under the SM-only hypothesis (dashed line) within uncertainties (filled bands).
\label{fig:mil}
}
\end{figure}

In addition to the interpretation of the \monoWZ search in terms of the simplified vector-mediator model and invisible Higgs boson decays, the analysis results are also expressed in terms of generic $\mathrm{CL_s}$ upper limits at 95\% CL on the allowed visible cross section \sigvis of potential \wdm or \zdm production. The limits on these two processes are evaluated separately to allow more flexibility in terms of possible reinterpretations, as new models might prefer one of these two final states. While the event selection and categorization is the same as described in Section~\ref{sec:selection}, i.e.\ including the $b$-tagging and mass window requirements, the exclusion limits are provided in the fiducial region that is defined by applying all signal region selection criteria except for the requirements on $m_{jj}$ or $m_J$ and the $b$-tagging multiplicity. With this definition, the exclusion limits  on \sigvis apply to any processes which are characterized by a generic back-to-back topology with a $W/Z$ boson recoiling against \met from weakly interacting particles such as DM. The limits on \sigvis are given as a function of the \met variable in order to avoid any additional model-dependent assumptions on the \met distribution. Hence, the \met~bins in the zero-lepton region are treated independently of each other in the statistical interpretation of the data. A reduced number of bins is used for $\met>300~\GeV$ to reduce the statistical uncertainty in the per-bin analysis. In all other aspects, the approach is identical to the \monoWZ analysis described above. The \monoWZ vector-mediator signal samples are used as a benchmark model to estimate the residual dependence of the \sigvis limits on the  kinematic properties of events within a given \met~range and on the $b$-tagging multiplicity. For this, a wide range of $(m_{Z'},m_\chi)$ model parameters that yield a sizeable contribution of at least 500 simulated events in a given \met range is considered. Corresponding variations of 15--50\% (25--50\%) in the expected limits on \sigvisw  (\sigvisz) are found. The weakest \sigvis limit is quoted in a given range of reconstructed \met in order to minimize the  dependence on a benchmark model. The observed and expected limits on \sigvis in each \met range are shown in Figure~\ref{fig:mil}, with the numerical values summarized in Tables~\ref{tab:sigvis_w} and~\ref{tab:sigvis_z}. As a general trend, the limits on \zdm production are somewhat stronger than those on \wdm since the former contributes significantly to the $2b$ category that has the highest sensitivity due to having the lowest SM background.
\begin{table}[!htbp]
          \centering
          \caption{The observed and expected exclusion limit at 95\% CL on \sigvis for \wdm production for an integrated luminosity of 36.1~fb$^{-1}$ and $\sqrt{s}=$~13~\TeV, together with the corresponding product of acceptance and efficiency (\aeps) for different regions of \met. 
          \label{tab:sigvis_w}}
          \begin{tabular}{ c | c c c c | c}
          \hline\hline

\met range& \multicolumn{4}{c|}{Upper limit at 95\% CL [fb]}& ~\\

[\GeV]& 
$\sigma_{\mathrm{vis}}^{\mathrm{obs}}$& 
$\sigma_{\mathrm{vis}}^{\mathrm{exp}}$& 
$-1\sigma$&  
$+1\sigma$& 
$\aeps$ \\

\hline
\multicolumn{6}{c}{$W$+DM, $W \to q'q$}\\
\hline

\lbrack150, \phantom{0}200]&             750 & \phantom{.}650&            470& \phantom{.}910& 20\% \\
\lbrack200, \phantom{0}250]&             185 & \phantom{.}163&            117& \phantom{.}226& 20\% \\
\lbrack250, \phantom{0}300]&   \phantom{0}43 & \phantom{0.}50&  \phantom{0}36& \phantom{0.}69& 30\% \\
\lbrack300, \phantom{0}400]&   \phantom{0}41 & \phantom{0.}36&  \phantom{0}26& \phantom{0.}50& 45\% \\
\lbrack400, \phantom{0}600]&   \phantom{.}9.7&           12.6& \phantom{.}9.1&           17.6& 55\% \\
\lbrack600,           1500]&  \phantom{.}5.1& \phantom{0}3.1& \phantom{.}2.2& \phantom{0}4.3& 55\% \\

\hline\hline
\end{tabular}
\end{table}
\begin{table}[!htbp]
          \centering
          \caption{The observed and expected exclusion limit at 95\% CL on \sigvis for \zdm production for an integrated luminosity of 36.1~fb$^{-1}$ and $\sqrt{s}=$~13~\TeV, together with the corresponding product of acceptance and efficiency (\aeps) for different regions of \met. 
          \label{tab:sigvis_z}}
          \begin{tabular}{ c | c c c c | c}
          \hline\hline

\met range& \multicolumn{4}{c|}{Upper limit at 95\% CL [fb]}& ~\\

[\GeV]& 
$\sigma_{\mathrm{vis}}^{\mathrm{obs}}$& 
$\sigma_{\mathrm{vis}}^{\mathrm{exp}}$& 
$-1\sigma$&  
$+1\sigma$& 
$\aeps$ \\

\hline
\multicolumn{6}{c}{$Z$+DM, $Z \to\qqbar$}\\
\hline

\lbrack150, \phantom{0}200]&  \phantom{.}313& \phantom{.}225& \phantom{.}162&  \phantom{.}314& 20\% \\
\lbrack200, \phantom{0}250]&  \phantom{0.}69& \phantom{0.}60& \phantom{0.}43&  \phantom{0.}83& 20\% \\
\lbrack250, \phantom{0}300]&  \phantom{0.}39& \phantom{0.}29& \phantom{0.}21&  \phantom{0.}40& 30\% \\
\lbrack300, \phantom{0}400]&            31.1&           18.5&           13.3&            25.7& 45\% \\
\lbrack400, \phantom{0}600]&  \phantom{0}9.2& \phantom{.}9.1& \phantom{.}6.5&            12.6& 50\% \\
\lbrack600,           1500]& \phantom{0}3.0& \phantom{.}2.6& \phantom{.}1.9&  \phantom{.}3.6& 55\% \\

\hline\hline
\end{tabular}
\end{table}

The observable \sigvis can be interpreted as
\begin{eqnarray*}
\label{eq:sigvisw}
\sigvisw(\met) &\equiv& \sigma_{\wdm}(\met) \times \br{W\to q'q} \times (\aeps)(\met)\quad\textrm{for}~\wdm~\textrm{events}\,,\\
\label{eq:sigvisz}
\sigvisz(\met) &\equiv& \sigma_{\zdm}(\met) \times \br{Z\to\qqbar} \times (\aeps)(\met)\quad\textrm{for}~\zdm~\textrm{events}\,,
\end{eqnarray*}
where $\sigma_{\wdm}$ ($\sigma_{\zdm}$) is the production cross section for \wdm (\zdm) events in a given \met range, $\br{W\to q'q}$  ($\br{Z\to\qqbar}$) is the branching ratio for the hadronic $W$ ($Z$) boson decay, and  $(\aeps)(\met)$ is the product of the kinematic acceptance and the experimental efficiency. This product represents the fraction of simulated \wzdm events in a given \met range at parton level\footnote{At parton level, \met is defined as the vector sum of momenta of neutrinos and DM particles in the transverse detector plane.} that fall into the same \met range at detector level after reconstruction, and pass the event selection criteria applied to determine \sigvis. To allow a generic interpretation, the requirements on $m_{jj}/m_J$ or $b$-tagging are not included in the latter. The product $(\aeps)(\met)$ in a given \met range has been evaluated for each simulated vector-mediator signal and the lowest of these values, rounded down in steps of 5\%, has been  taken for the limit calculation. 
The values obtained for each \met range are listed in Tables~\ref{tab:sigvis_w} and~\ref{tab:sigvis_z}.

%%%%%%%%%%%%%%%%%
% monoZ' results
%%%%%%%%%%%%%%%%%
%
%

\subsection{Constraints on \monoZprime models}

\begin{figure}[!htbp]
\centering
\subfigure[]{\includegraphics[width=0.49\textwidth]{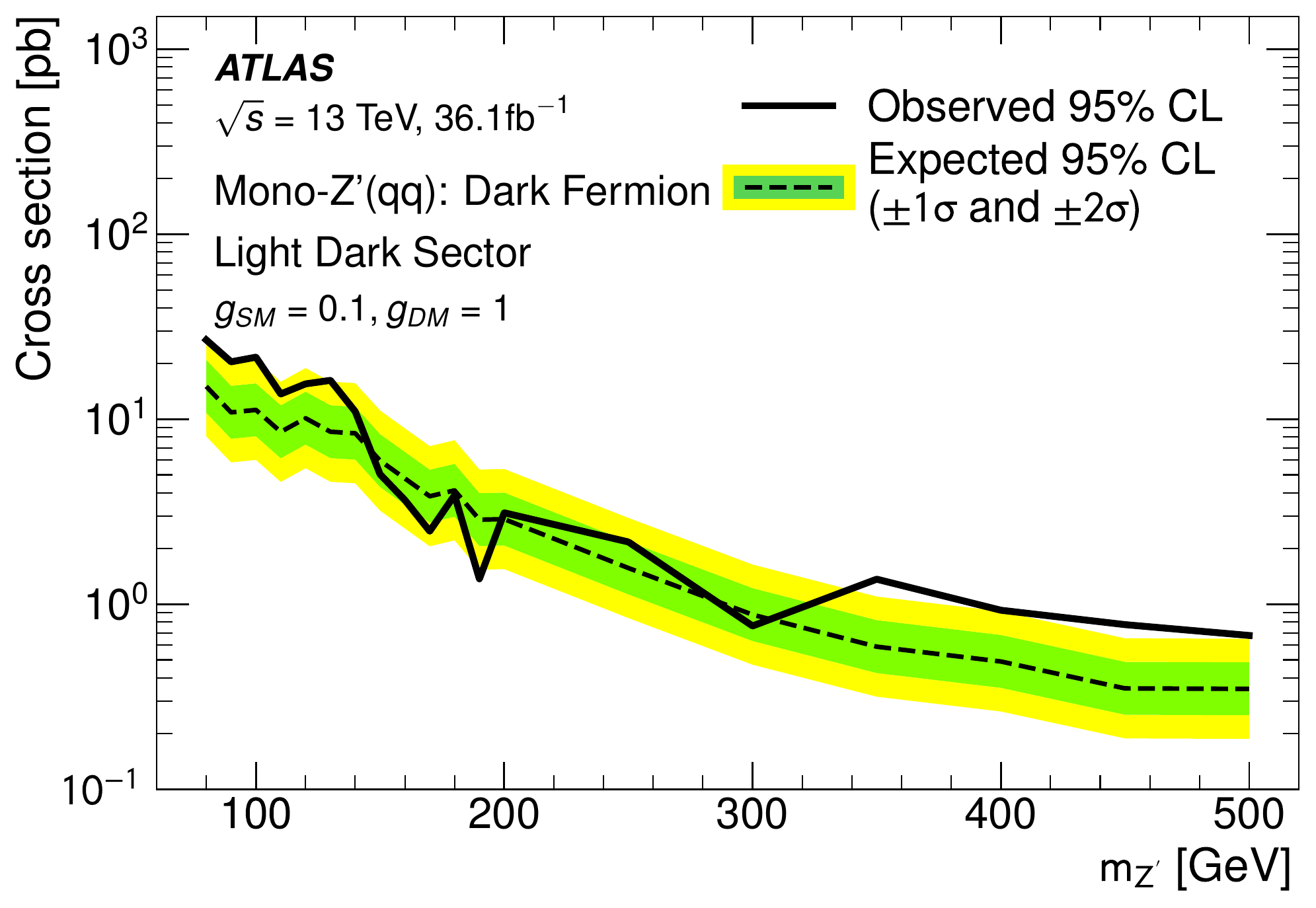}}
\subfigure[]{\includegraphics[width=0.49\textwidth]{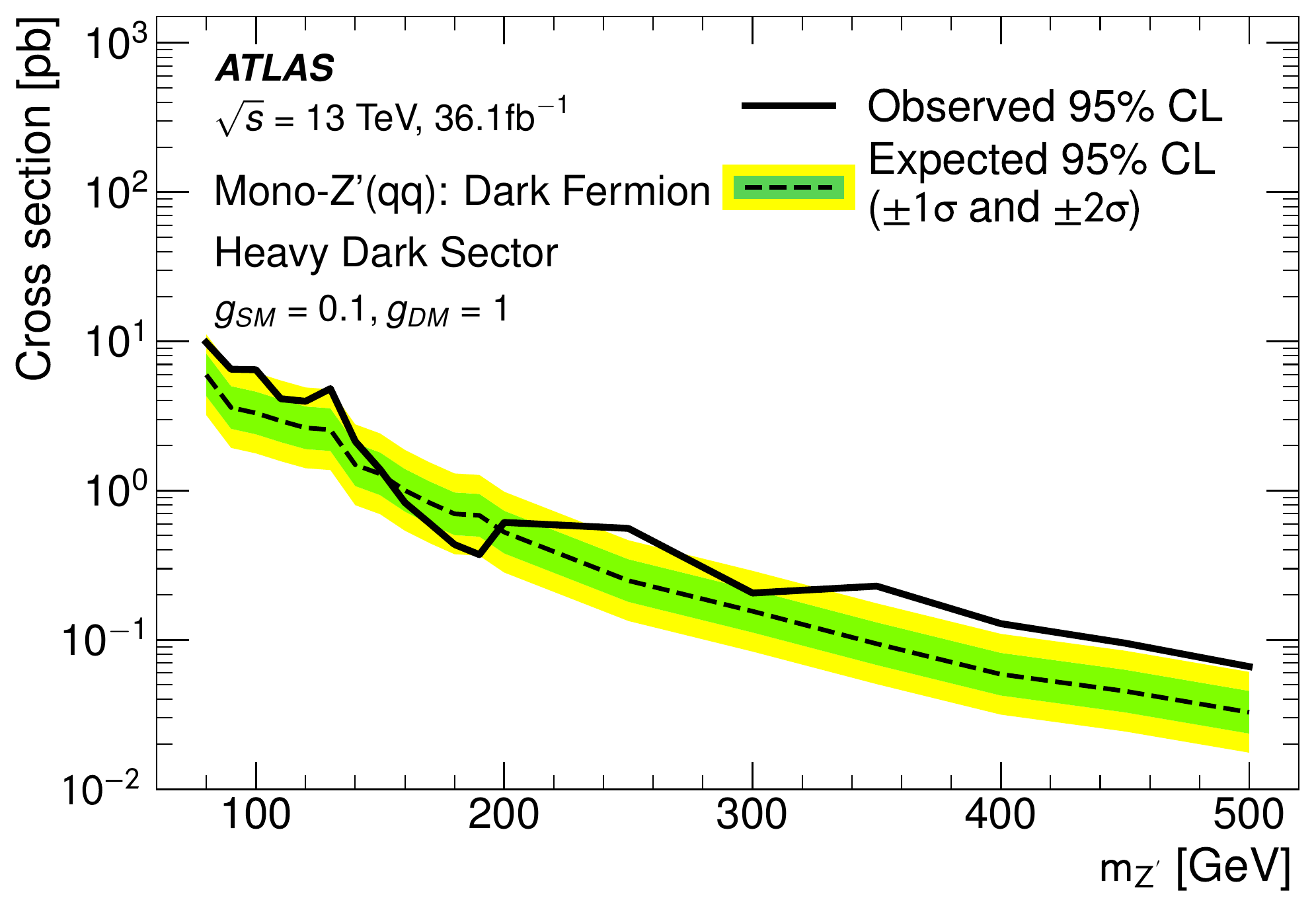}}
\subfigure[]{\includegraphics[width=0.49\textwidth]{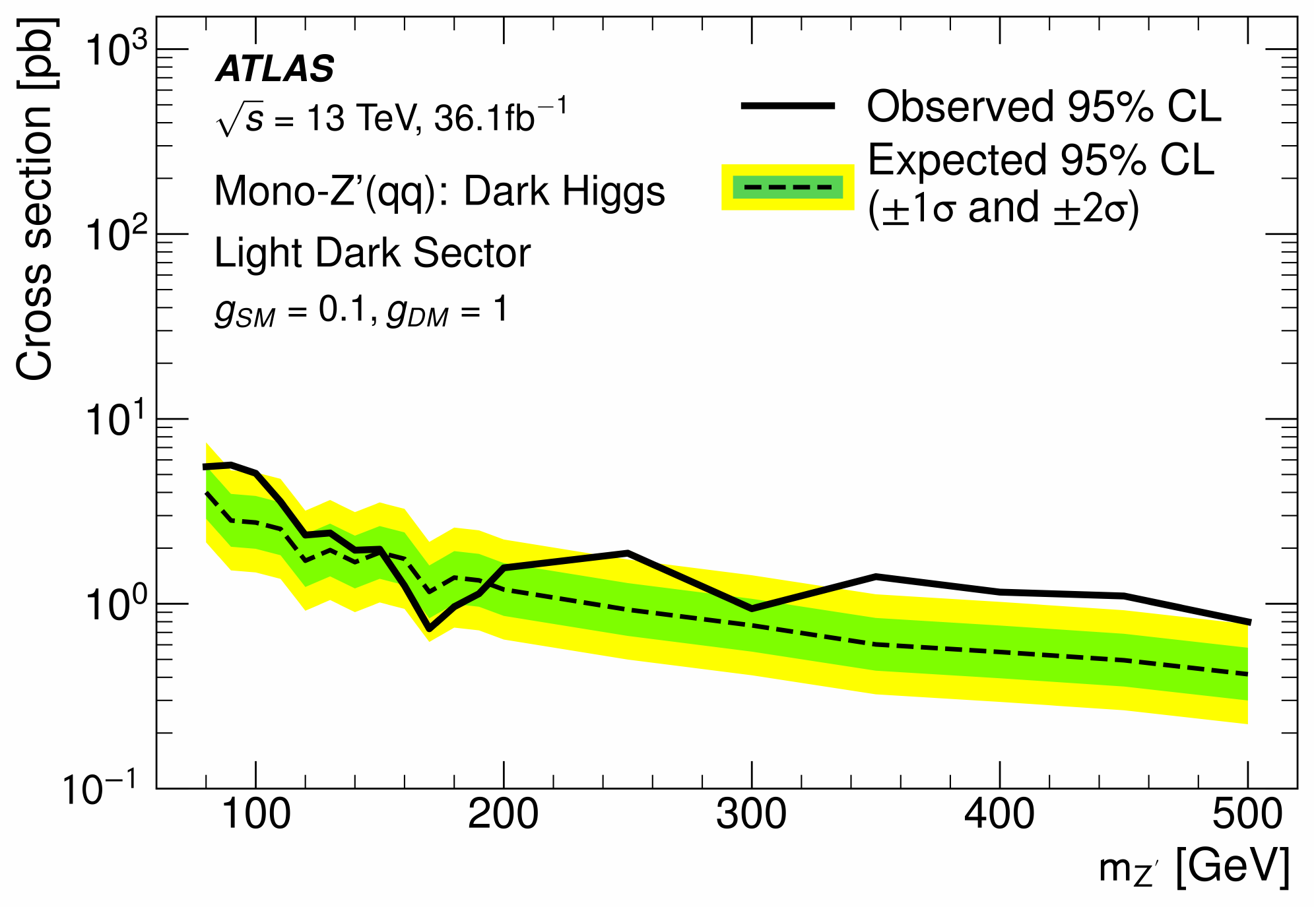}}
\subfigure[]{\includegraphics[width=0.49\textwidth]{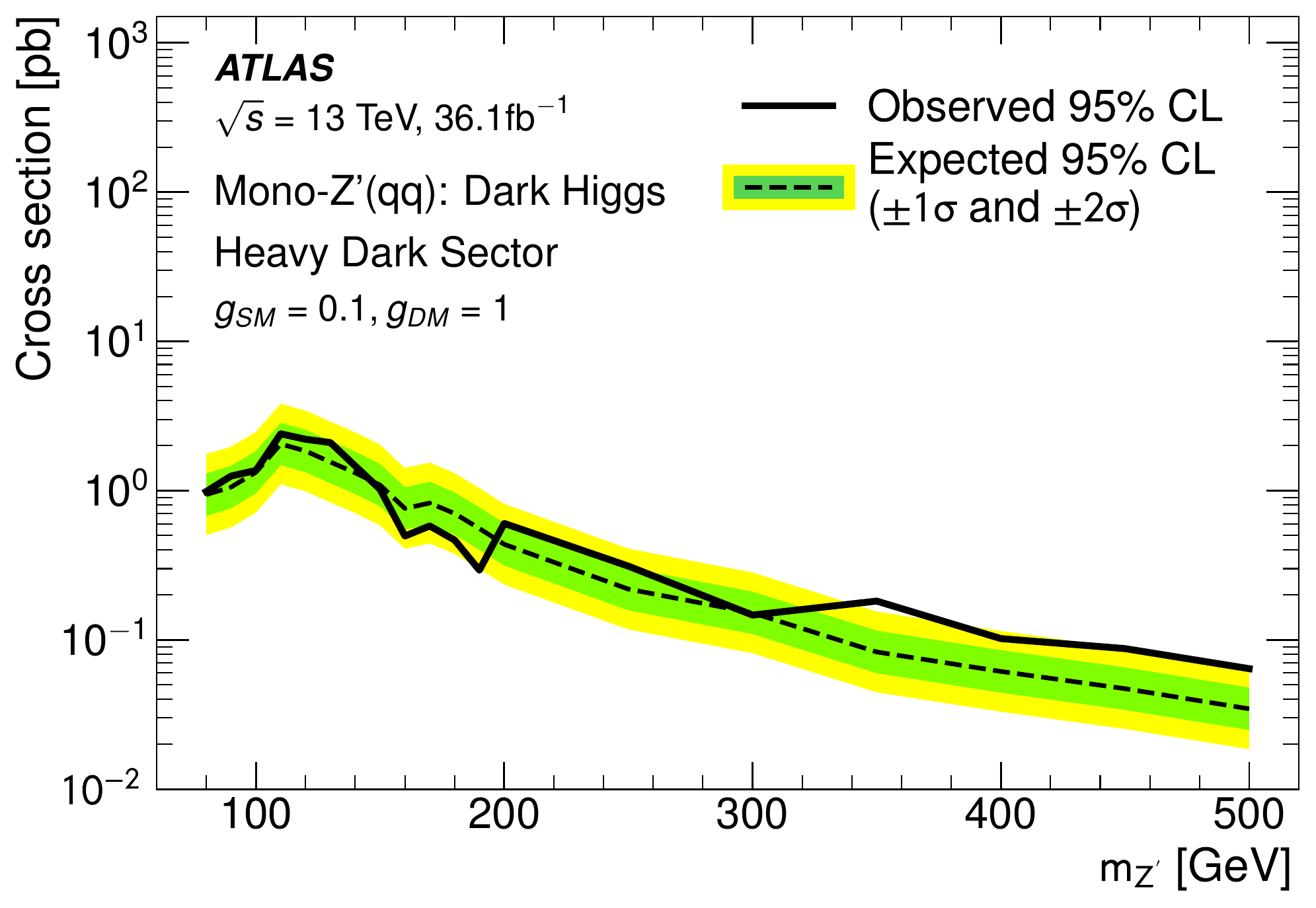}}
%\subfigure[]{\includegraphics[width=0.49\textwidth]{pfigures/results/monoZp/lv_light_2}}
%\subfigure[]{\includegraphics[width=0.49\textwidth]{pfigures/results/monoZp/lv_heavy_2}}
%\subfigure[]{\includegraphics[width=0.49\textwidth]{pfigures/results/monoZp/dh_light_2}}
%\subfigure[]{\includegraphics[width=0.49\textwidth]{pfigures/results/monoZp/dh_heavy_3}}
\caption{Upper limits at 95\% CL on the cross section times the branching ratio $\br{Z'\to q'q}$ in \monoZprime models as a function of the mediator mass, $m_{Z'}$, for the dark fermion model in the (a) light and (b) heavy dark-sector scenario, as well as the dark Higgs model in the (c) light and (d) heavy dark-sector scenario.
\label{fig:limits_monozp}}
\end{figure}

\begin{figure}[!htbp]
\centering
\subfigure[]{\includegraphics[width=0.49\textwidth]{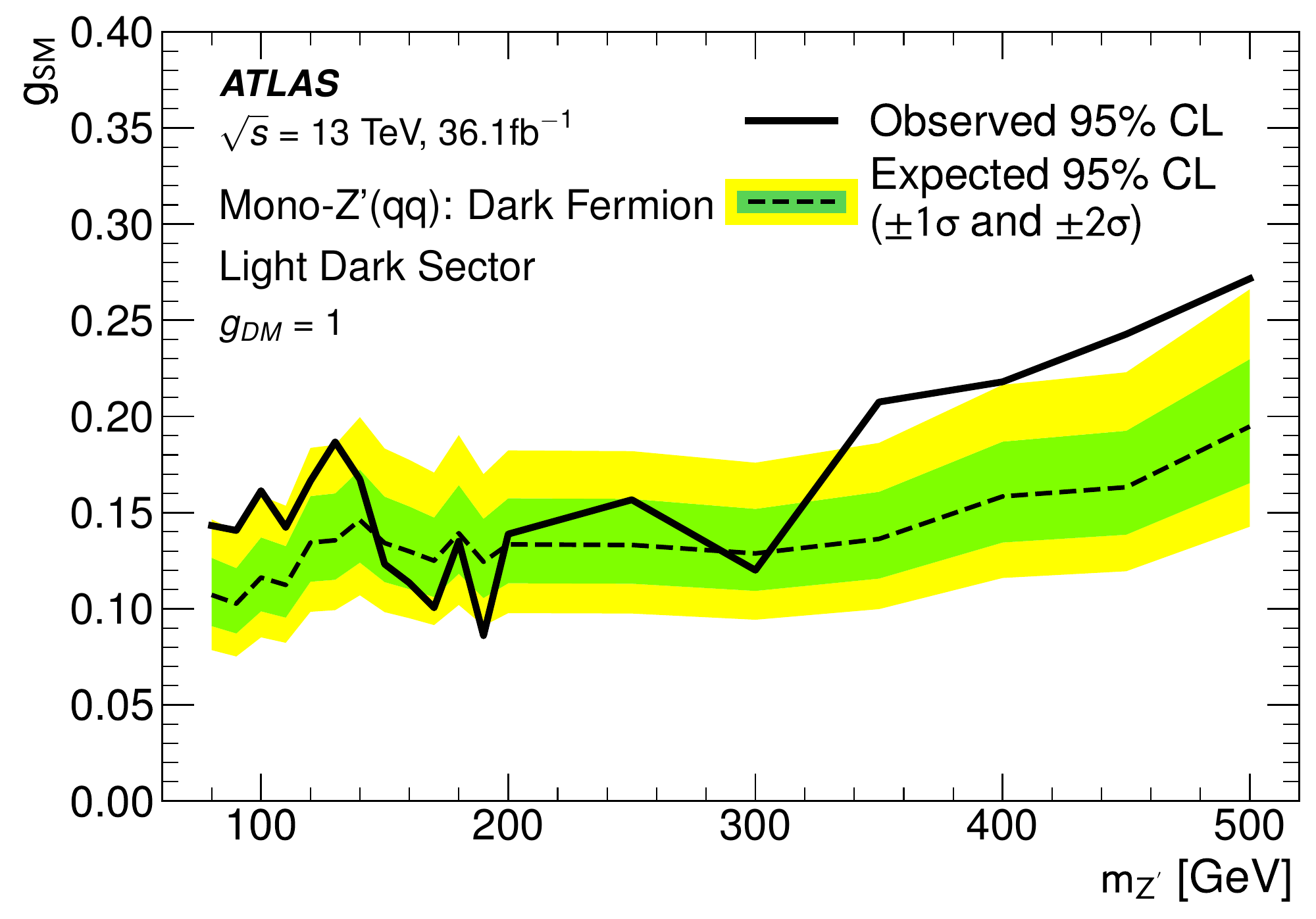}}
\subfigure[]{\includegraphics[width=0.49\textwidth]{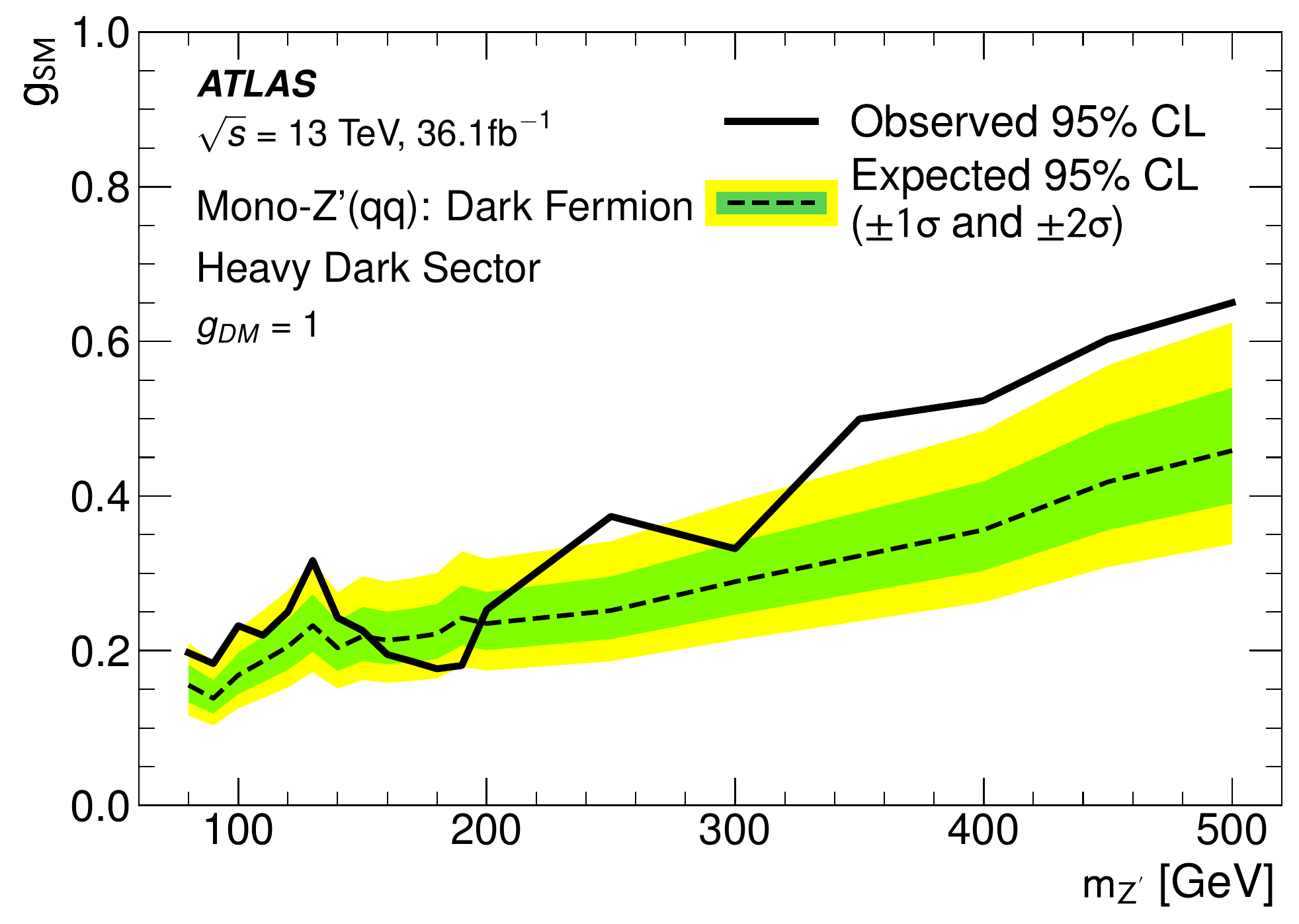}}
\subfigure[]{\includegraphics[width=0.49\textwidth]{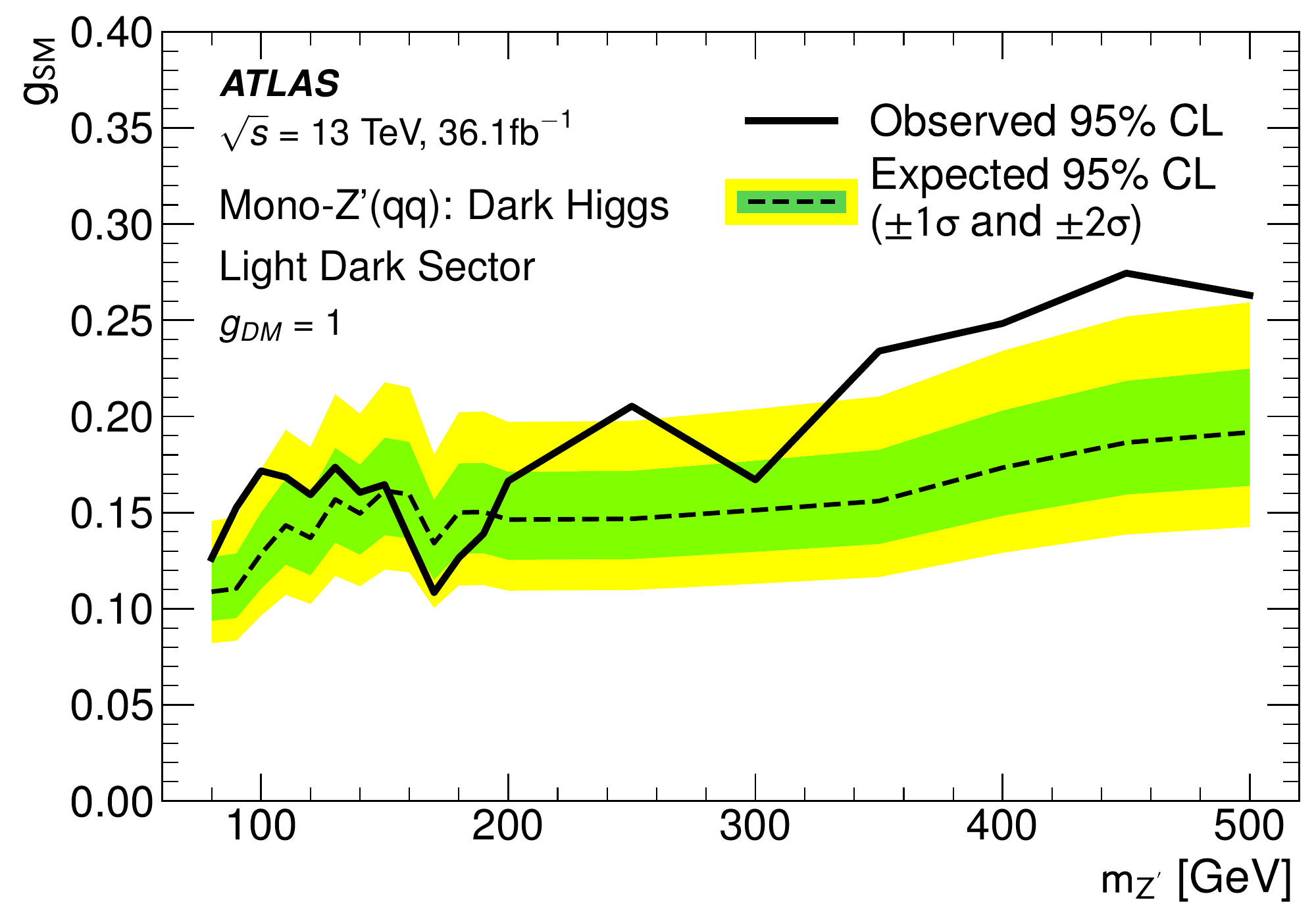}}
\subfigure[]{\includegraphics[width=0.49\textwidth]{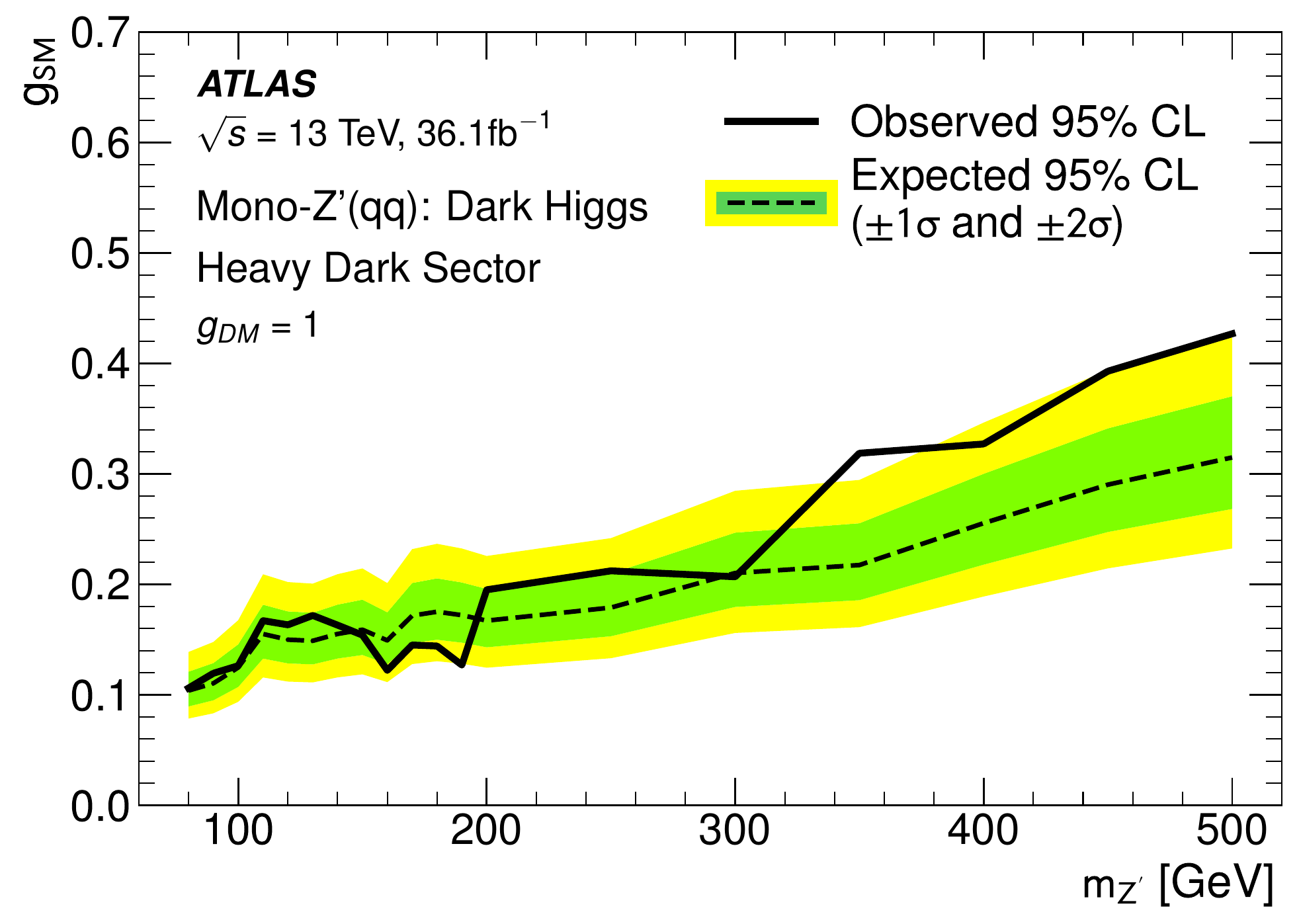}}
%\subfigure[]{\includegraphics[width=0.49\textwidth]{pfigures/results/monoZp/lv_light_couplings_2.pdf}}
%\subfigure[]{\includegraphics[width=0.49\textwidth]{pfigures/results/monoZp/lv_heavy_couplings_2.pdf}}
%\subfigure[]{\includegraphics[width=0.49\textwidth]{pfigures/results/monoZp/dh_light_couplings_2.pdf}}
%\subfigure[]{\includegraphics[width=0.49\textwidth]{pfigures/results/monoZp/dh_heavy_couplings_2.pdf}}
\caption{Upper limits at 95\% CL on the product of couplings \gsm\gdm in \monoZprime models as a function of the mediator mass  for the dark fermion model in the (a) light and (b) heavy dark-sector scenario, as well as the dark Higgs model in the (c) light and (d)  heavy dark-sector scenario.
\label{fig:limits_monozp_coupling}}
\end{figure}

For the \monoZprime models, the upper limits on the cross section times the branching ratio $\br{Z'\to q'q}$ at 95\% CL are shown in Figure~\ref{fig:limits_monozp} as a function of the mediator mass for both the dark-fermion and dark-Higgs models in the light and heavy dark-sector mass scenarios. The largest excess of the data above the expectation, corresponding to a local significance of 3$\sigma$, is observed for a hypothesized signal at $m_{Z'}=$~350~\GeV\ within the dark fermion model in the heavy dark-sector scenario. 
%The global significance of this excess is 1.9$\sigma$. 
Taking into account the look-elsewhere effect~\cite{LEE} with respect to the 19 overlapping mass windows examined in the \monoZprime search,  the excess corresponds to a global significance of 2.2$\sigma$.
%In addition, three excesses with a local (global) significance of 2.7$\sigma$ (1.6$\sigma$) are observed: two of these for the mass hypothesis $m_{Z'}=$~450~\GeV within the dark fermion and dark Higgs models in the light dark sector scenario, and one at 250~\GeV for the dark fermion model in the heavy scenario. All other measurement results agree with the SM prediction within 2.5 standard deviations. 
%Cross section exclusions limits are at the order of a few picobarn for $Z^\prime$ masses between 80 and 500~\GeV. 
Cross-section exclusion limits for the dark-fermion model (dark-Higgs model) in the light and the heavy dark-sector scenario are in the range of 0.68--27~pb and 0.066--9.8~pb (0.80--5.5~pb and 0.064--2.4~pb) respectively, for $Z^\prime$ masses between 80 and 500~\GeV. 
The corresponding observed and expected upper limits on the coupling \gsm are shown in Figure~\ref{fig:limits_monozp_coupling}, assuming \gdm~=~1.

%% file: summary.tex
A search for dark matter was performed in events having
a large-$R$ jet or a pair of small-$R$ jets compatible with a hadronic $W$ or $Z$ boson decay, and large $\met$. In addition, the as of yet unexplored hypothesis of a new vector boson $Z'$ produced in association with dark matter is considered. This search uses the ATLAS dataset corresponding to an integrated luminosity of 36.1 fb$^{-1}$ of $\sqrt{s}=13$~\TeV\ $pp$ collisions collected at the LHC in 2015 and 2016. It improves on previous searches by virtue of the larger dataset and further optimization of the selection criteria and signal region definitions. The results are in agreement with the SM predictions and are translated into exclusion limits on DM-pair production. 

Two simplified models are considered to describe DM production in the mono-$W/Z$ final state. For the simplified vector-mediator model in which the DM is produced via an $s$-channel exchange of a vector mediator $Z'$, masses $m_{Z'}$ of up to 650~\GeV\ are excluded for dark matter masses $m_{\chi}$ of up to 250~\GeV\ (assuming $\gsm=0.25$ and $\gdm=1.0$). This agrees well with the expected exclusion of $m_{Z'}$ values of up to 700~\GeV\ for $m_{\chi}$ of up to 230~\GeV. Limits are also placed on the visible cross section of non-SM events with large $\met$ and a $W$ or a $Z$ boson without extra model assumptions. In the search for invisible Higgs boson decays, an upper limit of 0.83 is observed at 95\% CL on the branching ratio \BHinv, while the corresponding expected limit is 0.58. 

Two additional signal models, for DM production in association with the non-SM vector boson $Z^\prime$, are considered. In the dark-fermion model, the intermediate $Z'$ boson couples to a heavier dark-sector fermion $\chi_2$ as well as the lighter DM candidate fermion $\chi_1$.  In the dark-Higgs model, a dark-sector Higgs boson which decays to a $\chi\chi$ pair is radiated from the $Z'$ boson. For coupling values of $\gsm = 0.1$ and $\gdm= 1.0$, two different choices of masses $m_{\chi_2}$ and $m_{h_{D}}$ of intermediate dark-sector particles are considered. Cross-section exclusion limits for the dark-fermion model in the light and heavy dark-sector scenarios are in the range of 0.68--27~pb and 0.066--9.8~pb respectively for $Z^\prime$ masses between 80 and 500~\GeV. The corresponding limits for the dark-Higgs model in the light and heavy dark-sector scenario are 0.80--5.5~pb and 0.064--2.4~pb, respectively.

%% file: acknowledgements/Acknowledgements.tex
% Acknowledgements for papers with collision data
% Version 14-Feb-2018

% Standard acknowledgements start here
%----------------------------------------------
We thank CERN for the very successful operation of the LHC, as well as the
support staff from our institutions without whom ATLAS could not be
operated efficiently.

We acknowledge the support of ANPCyT, Argentina; YerPhI, Armenia; ARC, Australia; BMWFW and FWF, Austria; ANAS, Azerbaijan; SSTC, Belarus; CNPq and FAPESP, Brazil; NSERC, NRC and CFI, Canada; CERN; CONICYT, Chile; CAS, MOST and NSFC, China; COLCIENCIAS, Colombia; MSMT CR, MPO CR and VSC CR, Czech Republic; DNRF and DNSRC, Denmark; IN2P3-CNRS, CEA-DRF/IRFU, France; SRNSFG, Georgia; BMBF, HGF, and MPG, Germany; GSRT, Greece; RGC, Hong Kong SAR, China; ISF, I-CORE and Benoziyo Center, Israel; INFN, Italy; MEXT and JSPS, Japan; CNRST, Morocco; NWO, Netherlands; RCN, Norway; MNiSW and NCN, Poland; FCT, Portugal; MNE/IFA, Romania; MES of Russia and NRC KI, Russian Federation; JINR; MESTD, Serbia; MSSR, Slovakia; ARRS and MIZ\v{S}, Slovenia; DST/NRF, South Africa; MINECO, Spain; SRC and Wallenberg Foundation, Sweden; SERI, SNSF and Cantons of Bern and Geneva, Switzerland; MOST, Taiwan; TAEK, Turkey; STFC, United Kingdom; DOE and NSF, United States of America. In addition, individual groups and members have received support from BCKDF, the Canada Council, CANARIE, CRC, Compute Canada, FQRNT, and the Ontario Innovation Trust, Canada; EPLANET, ERC, ERDF, FP7, Horizon 2020 and Marie Sk{\l}odowska-Curie Actions, European Union; Investissements d'Avenir Labex and Idex, ANR, R{\'e}gion Auvergne and Fondation Partager le Savoir, France; DFG and AvH Foundation, Germany; Herakleitos, Thales and Aristeia programmes co-financed by EU-ESF and the Greek NSRF; BSF, GIF and Minerva, Israel; BRF, Norway; CERCA Programme Generalitat de Catalunya, Generalitat Valenciana, Spain; the Royal Society and Leverhulme Trust, United Kingdom.

The crucial computing support from all WLCG partners is acknowledged gratefully, in particular from CERN, the ATLAS Tier-1 facilities at TRIUMF (Canada), NDGF (Denmark, Norway, Sweden), CC-IN2P3 (France), KIT/GridKA (Germany), INFN-CNAF (Italy), NL-T1 (Netherlands), PIC (Spain), ASGC (Taiwan), RAL (UK) and BNL (USA), the Tier-2 facilities worldwide and large non-WLCG resource providers. Major contributors of computing resources are listed in Ref.~\cite{ATL-GEN-PUB-2016-002}.
%----------------------------------------------

%% file: atlas_authlist.tex
% ATLAS Collaboration author list
% Reference date of EXOT-2016-23 is 2018-03-09
% Author list last updated on date 02-JUL-18
% Data extracted on 02-Jul-2018 for paper reference EXOT-2016-23
% at 2:17pm
 
\begin{flushleft}
{\Large The ATLAS Collaboration}

\bigskip

M.~Aaboud$^\textrm{\scriptsize 34d}$,    
G.~Aad$^\textrm{\scriptsize 99}$,    
B.~Abbott$^\textrm{\scriptsize 124}$,    
O.~Abdinov$^\textrm{\scriptsize 13,*}$,    
B.~Abeloos$^\textrm{\scriptsize 128}$,    
D.K.~Abhayasinghe$^\textrm{\scriptsize 91}$,    
S.H.~Abidi$^\textrm{\scriptsize 164}$,    
O.S.~AbouZeid$^\textrm{\scriptsize 39}$,    
N.L.~Abraham$^\textrm{\scriptsize 153}$,    
H.~Abramowicz$^\textrm{\scriptsize 158}$
% ...
\bigskip
\\

% ...
$^{13}$Institute of Physics, Azerbaijan Academy of Sciences, Baku; Azerbaijan.\\
% ...
$^{34}$$^{(a)}$Facult\'e des Sciences Ain Chock, R\'eseau Universitaire de Physique des Hautes Energies - Universit\'e Hassan II, Casablanca;$^{(b)}$Centre National de l'Energie des Sciences Techniques Nucleaires (CNESTEN), Rabat;$^{(c)}$Facult\'e des Sciences Semlalia, Universit\'e Cadi Ayyad, LPHEA-Marrakech;$^{(d)}$Facult\'e des Sciences, Universit\'e Mohamed Premier and LPTPM, Oujda;$^{(e)}$Facult\'e des sciences, Universit\'e Mohammed V, Rabat; Morocco.\\
% ...
$^{39}$Niels Bohr Institute, University of Copenhagen, Copenhagen; Denmark.\\
% ...
$^{91}$Department of Physics, Royal Holloway University of London, Egham; United Kingdom.\\
% ...
$^{99}$CPPM, Aix-Marseille Universit\'e, CNRS/IN2P3, Marseille; France.\\
% ...
$^{124}$Homer L. Dodge Department of Physics and Astronomy, University of Oklahoma, Norman OK; United States of America.\\
% ...
$^{128}$LAL, Universit\'e Paris-Sud, CNRS/IN2P3, Universit\'e Paris-Saclay, Orsay; France.\\
% ...
$^{153}$Department of Physics and Astronomy, University of Sussex, Brighton; United Kingdom.\\
% ...
$^{158}$Raymond and Beverly Sackler School of Physics and Astronomy, Tel Aviv University, Tel Aviv; Israel.\\
% ...
$^{164}$Department of Physics, University of Toronto, Toronto ON; Canada.\\
% ...
$^{a}$ Also at Borough of Manhattan Community College, City University of New York, NY; United States of America.\\
$^{b}$ Also at Centre for High Performance Computing, CSIR Campus, Rosebank, Cape Town; South Africa.\\
$^{c}$ Also at CERN, Geneva; Switzerland.\\
$^{d}$ Also at CPPM, Aix-Marseille Universit\'e, CNRS/IN2P3, Marseille; France.\\
$^{e}$ Also at D\'epartement de Physique Nucl\'eaire et Corpusculaire, Universit\'e de Gen\`eve, Gen\`eve; Switzerland.\\
% ...
$^{*}$ Deceased

\end{flushleft}

% Created with Glance <Atlas.Glance@cern.ch>